\def\draftversion{false}

\RequirePackage{ifthen}
\ifthenelse{\equal{\draftversion}{true}}{
  \documentclass[galley,aps,pra,10pt,amsmath,amssymb,
    superscriptaddress,nofootinbib,longbibliography]{revtex4-2}
}{
  \documentclass[twocolumn,aps,prb,10pt,amsmath,amssymb,
    superscriptaddress,nofootinbib,longbibliography]{revtex4-2}
}

\usepackage{graphicx}
\usepackage[usenames,dvipsnames]{color} 
\usepackage{bm} 
\usepackage{bbm} 
\usepackage{soul} 
\usepackage{mathtools}
\usepackage[makeroom]{cancel}
\usepackage{xspace}
\usepackage{float}
\usepackage{multirow}

\usepackage{hyperref}
\hypersetup{
    colorlinks=true,       
    linkcolor=blue,          
    citecolor=blue,       
    filecolor=magenta,      
    urlcolor=blue          
}

\ifthenelse{\equal{\draftversion}{true}}{
  \marginparwidth 2.7in
  \marginparsep 0.5in
  \newcounter{comm} 
  \def\commnext{\stepcounter{comm}}
  \def\commtext{{\bf\color{blue}[\arabic{comm}]}}
  \def\commmar{{\bf\color{blue}[\arabic{comm}]}}
  \def\dvm#1{\commnext\marginpar{\small DV\commmar: #1}\commtext}
  \def\aum#1{\commnext\marginpar{\small AU\commmar: #1}\commtext}
  \def\stm#1{\commnext\marginpar{\small ST\commmar: #1}\commtext}
  \def\ism#1{\commnext\marginpar{\small IS\commmar: #1}\commtext}
  \def\opm#1{\commnext\marginpar{\small OP\commmar: #1}\commtext}
  
  \newcommand{\seclab}[1]{\label{sec:#1}{\Red{\small\;\;[sec:~#1]}}}
  \newcommand{\eqlab}[1]{\Red{\hbox{\small\;\;[#1]}}\label{eq:#1}}
  \newcommand{\figlab}[1]{\Red{\hbox{\small\;\;[fig:~#1]}}\label{fig:#1}}
}{
  \def\dvm#1{}
  \def\aum#1{}
  \def\stm#1{}
  \def\ism#1{}
  \def\opm#1{}
  
  \newcommand{\eqlab}[1]{\label{eq:#1}}
  \newcommand{\seclab}[1]{\label{sec:#1}}
  \newcommand{\figlab}[1]{\label{fig:#1}}
  
}

\newcommand{\beq}{\begin{equation}}
\newcommand{\eeq}{\end{equation}}
\newcommand{\bea}{\begin{eqnarray}}
\newcommand{\eea}{\end{eqnarray}}
\newcommand{\nn}{\nonumber\\}
\newcommand{\eq}[1]{Eq.~(\ref{eq:#1})}
\newcommand{\Eq}[1]{Equation~(\ref{eq:#1})}
\newcommand{\eqs}[2]{Eqs.~(\ref{eq:#1}) and (\ref{eq:#2})}

\newcommand{\eqr}[2]{Eqs.~(\ref{eq:#1})-(\ref{eq:#2})}

\newcommand{\fref}[1]{Fig.~\ref{fig:#1}}
\newcommand{\Fref}[1]{Figure~\ref{fig:#1}}

\newcommand{\sref}[1]{Sec.~\ref{sec:#1}}
 
\newcommand{\sreffs}[2]{Secs.~\ref{sec:#1}-\ref{sec:#2}} 

\newcommand{\aref}[1]{Appendix~\ref{sec:#1}}
\newcommand{\Tref}[1]{Table~\ref{tab:#1}}

\newcommand{\ket}[1]{\vert#1\rangle}
\newcommand{\bra}[1]{\langle#1\vert}
\newcommand{\ip}[2]{\langle#1\vert#2\rangle}
\newcommand{\me}[3]{\langle#1\vert#2\vert#3\rangle}

\newcommand{\phm}{\phantom{-}}       
\newcommand{\q}{\textbf{q}}
\renewcommand{\k}{{\bf k}}
\renewcommand{\r}{{\bf r}}

\renewcommand{\b}{{\bf b}}

\newcommand{\bnk}{_{n\k}}

\newcommand{\0}{\mathbf{0}}
\renewcommand{\Re}{\mathrm{Re}}
\renewcommand{\Im}{\mathrm{Im}}
\newcommand{\w}{\omega}

\newcommand{\wln}{\w_{ln}}

\newcommand{\wnl}{\w_{nl}}
\newcommand{\fnl}{f_{nl}}

\newcommand{\mel}{m_{\rm e}}

\newcommand{\R}{{\bf R}}
\newcommand{\btau}{{\boldsymbol \tau}}
\newcommand{\W}{^{\rm W}}
 \renewcommand{\H}{^{\rm H}}
\newcommand{\I}{^{\rm I}}
\newcommand{\E}{^{\rm E}}
\newcommand{\X}{^{\rm X}}
\newcommand{\Acal}{\mathcal{A}}
\newcommand{\Bcal}{\mathcal{B}}
\newcommand{\Ccal}{\mathcal{C}}
\newcommand{\Dcal}{\mathcal{D}}
\newcommand{\Fcal}{\mathcal{F}}
\newcommand{\Hcal}{\mathcal{H}}
\newcommand{\Kcal}{\mathcal{K}}
\newcommand{\Tcal}{\mathcal{T}}
\newcommand{\Scal}{\mathcal{S}}
\newcommand{\Ucal}{\mathcal{U}}
\newcommand{\Vcal}{\mathcal{V}}
\newcommand{\Wcal}{\mathcal{W}}

\newcommand{\dd}{\mathfrak{d}}
\newcommand{\mm}{\mathfrak{m}}
\newcommand{\qq}{\mathfrak{q}}
\begin{document}


\title{Optical spatial dispersion via Wannier
      interpolation}

\author{Andrea Urru}
\thanks{These authors contributed equally to the manuscript.}
\affiliation{Department of Physics \& Astronomy, Rutgers University,
Piscataway, New Jersey 08854, USA}

\author{Ivo Souza}
\thanks{These authors contributed equally to the manuscript.}
\affiliation{Centro de F{\'i}sica de Materiales,
  Universidad del Pa{\'i}s Vasco, 20018 San Sebasti{\'a}n,
  Spain} \affiliation{Ikerbasque Foundation, 48013 Bilbao, Spain}

\author{\'Oscar Pozo Oca$\tilde{\text{n}}$a} \affiliation{Centro de
  F{\'i}sica de Materiales, Universidad del Pa{\'i}s Vasco, 20018 San
  Sebasti{\'a}n, Spain}

\author{Stepan S. Tsirkin} \affiliation{Centro de F{\'i}sica de
  Materiales, Universidad del Pa{\'i}s Vasco, 20018 San Sebasti{\'a}n,
  Spain} \affiliation{Ikerbasque Foundation, 48013 Bilbao, Spain}
\affiliation{Chair of Computational Condensed Matter Physics,
  Institute of Physics, École Polytechnique Fédérale de Lausanne
  (EPFL), CH-1015 Lausanne, Switzerland}

\author{David Vanderbilt}
\affiliation{
Department of Physics \& Astronomy, Rutgers University,
Piscataway, New Jersey 08854, USA}


\begin{abstract}
We present a numerical implementation, based on Wannier interpolation,
of a Kubo-Greenwood formalism for computing the spatially dispersive
optical conductivity in crystals at first order in the wave vector of
light. This approach is more efficient than direct
$\textit{ab initio}$ methods because, with less computational cost, it
allows for a much finer sampling of reciprocal space, resulting in
better resolved spectra. Moreover, Wannier interpolation avoids errors
arising from truncation of the sums over conduction bands when
evaluating the spatially dispersive optical matrix elements.  We
validate our method by computing the optical activity spectrum of
selected crystals, both polar (GaN) and chiral (trigonal Te, trigonal Se, and
$\alpha$-quartz), and comparing with existing literature.

\end{abstract}

\maketitle


\section{Introduction}
\seclab{intro}

When the wavelength of light is long compared to typical atomic
dimensions and bond lengths, the optical response of a medium can be
treated as local in space and, thus, does not depend on the wave
vector $\mathbf{q}$ of incident radiation. Beyond this approximation,
$\mathbf{q}$-dependent contributions to the optical response,
corresponding to spatially dispersive -- i.e., non-local -- effects,
must be taken into account~\cite{landau-84}.  Despite usually being
small corrections, spatial-dispersion terms describe phenomena that
are not captured by the $\mathbf{q}$-independent optical
response. Among them we mention natural optical activity (optical
activity henceforth), which stems from the first-order spatial
dispersion of the optical conductivity and shows up in acentric
systems. A well-known manifestation of optical activity is optical
rotation, i.e., the rotation of the plane of polarization of incident
linearly-polarized light as it travels through a chiral
medium~\cite{landau-84,barron-04,newnham-book05,malgrange-14,raab-05}.
Besides optical activity, optical spatial dispersion phenomena include
also magneto-optical effects such as gyrotropic
birefringence~\cite{raab-05,brown-jap63,hornreich-jap68,hornreich-pr68}
and non-reciprocal directional dichroism~\cite{kimura-prl24}, which
occur in acentric magnetic crystals (e.g., the antiferromagnet
Cr$_2$O$_3$).

Due to the fundamental interest in chiral molecules and their
relevance for industry, historically optical activity has been studied
more extensively in molecules than in solids. Starting a few decades
ago, several \textit{ab initio} approaches for computing optical
activity in molecules have been
developed~\cite{pedersen-cpl95,polavarapu-molphys97,yabana-pra99,varsano-pccp09,polavarapu-chirality02,autschbach-10,mattiat-helvchim21},
based on earlier molecular quantum theories. Until a few years ago,
not as many attempts to formulate an equivalent theory for periodic
solids had been
made~\cite{natori-jpsj75,zhong-prb93,malashevich-prb10}. More
recently, increased interest in chiral crystals has brought renewed
attention to optical activity and related phenomena. This has fostered
the development of first-principles theories of spatial dispersion in
bulk periodic systems~\cite{wang-prb23,pozo-scipost23,zabalo-prl23}.

Currently, the available \textit{ab initio} implementations of optical
spatial dispersion in solids follow one of two distinct approaches:
Kubo-Greenwood linear-response theory, or density-functional
perturbation theory (DFPT). Methods based on the Kubo
formulation~\cite{wang-prb23,pozo-scipost23} have a practical
advantage over the DFPT-based approaches \cite{zabalo-prl23} due to
their ease of implementation. Indeed, while Kubo linear response
theory can be implemented as a post-processing step after a
ground-state self-con\-sis\-tent first-principles calculation, DFPT
requires the self-consistent solution of a more involved Sternheimer
linear system, besides the ground-state calculation. Moreover, current
DFPT implementations tipically access only the zero-frequency limit of
optical activity, while the Kubo approach naturally accounts for the
frequency dependence of the response.  On the other hand, the DFPT
formulation carefully accounts for the induced variation of the
self-consistent potentials (Hartree and exchange-correlation), which
can be large in some cases~\cite{jonsson-prl96,zabalo-prl23}. Such
local-field effects are neglected in current implementations of the
Kubo formalism --~including the present one~-- which treat the
self-consistent potentials as a frozen quantities, neglecting their
response to the electromagnetic perturbation.

A Kubo approach for computing the spatially dispersive optical
conductivity has been recently proposed in two different ways in
Refs.~\cite{wang-prb23} and~\cite{pozo-scipost23}, building on an
earlier work~\cite{malashevich-prb10}. Both formulations take the
molecular multipole theory of
electromagnetism~\cite{barron-04,raab-05} as a basis, and recast the
multipolar terms in a way that is suitable for crystals as well.  One
drawback of using the Kubo approach is connected with the need to use
a sufficiently dense $\mathbf{k}$ mesh to resolve the features of
interest in optical spectra.  In the Kubo approach, this requires a
self-consistent solution at every point on this dense mesh, which is
computationally expensive. A second drawback is that the sum over
states in the formulation of Ref.~\cite{wang-prb23} has to be
truncated for practical purposes, thus leading to a possible band
truncation error.

Here, we elaborate on the formulation of Ref.~\cite{pozo-scipost23}
and present an {\it ab initio} implementation using Wannier
interpolation~\cite{wang-prb06,yates-prb07,lopez-prb12}.  This
technique is particularly well suited to evaluate the spatially
dispersive optical conductivity in the Kubo formalism, because it
allows for a very fine sampling of reciprocal space with a much lower
computational cost than a direct \textit{ab initio}
calculation. Furthermore, Wannier interpolation makes it possible to
recast the formalism avoiding sums over intermediate states when
evaluating the optical matrix elements at first order in $\mathbf{q}$,
so that it is expected to be less affected by band truncation
errors. We test our implementation by computing the optical activity
of both achiral GaN and chiral crystals Te, Se, and $\alpha$-SiO$_2$,
and by comparing our results with existing literature, both
theoretical~\cite{wang-prb23, zabalo-prl23} and
experimental~\cite{ades-josa75,fukuda-pssb75}.

The remainder of the manuscript is organized in the following way. In
\sreffs{theory}{wannier-interp} we present the underlying theoretical
formalism and the relevant details related to the implementation. We
start in \sref{theory} by recapitulating the general properties of the
spatially dispersive optical response and its expression in the Kubo
formalism. In \sref{sos} we discuss a sum-over-states approach for
evaluating the Kubo formula for the conductivity. Although this is not
the method implemented in this work, we find it ent to discuss it for
pedagogical reasons and to contrast it with our approach, which we
present in the following section. Specifically, in
\sref{wannier-basics} we introduce the Wannier interpolation scheme,
and in \sref{sigma-interp} we show how to use it to evaluate the
spatially dispersive optical conductivity. In \sref{tests} we present
our numerical results and validate our implementation by comparing
with previous literature. Finally, \sref{conclusions} contains a
summary and conclusions. We complement the manuscript with four
Appendices, containing analytical derivations and some numerical
tests.

\section{Optical spatial dispersion}
\seclab{theory}

\subsection{Phenomenology}
\seclab{OSD}

The optical response of a medium to a monochromatic electromagnetic
wave is described by the dielectric function $\epsilon$, which relates
the induction field $\mathbf{D}$ to the incident wave's electric field
$\mathbf{E}$. In the reciprocal variables $\omega$ and~$\mathbf{q}$
(frequency and wave vector), this relationship reads~\cite{landau-84}
\beq
D_{\alpha} (\omega, \mathbf{q}) =
\epsilon_{\alpha \beta} (\omega, \mathbf{q}) E_{\beta} (\omega, \mathbf{q}),
\eqlab{diel}
\eeq
where summation over repeated indices is implied, a convention that we
adopt henceforth.
In general, the response of a periodic medium to a perturbation with
wave vector $\mathbf{q}$ has contributions from all
$\mathbf{q}+\mathbf{G}$ wave vectors, with the
$\mathbf{G}\not=\mathbf{0}$ components resulting from variations of
the $\mathbf{D}$ field at spatial scales of the order of the bond
lengths. This fact underlies the presence of local-field corrections,
which are not included in this work, as mentioned in the Introduction.
The dielectric function can be Taylor-expanded as~\cite{landau-84}
\begin{equation}
\epsilon_{\alpha \beta} (\omega, \mathbf{q}) =
\epsilon_{\alpha \beta} (\omega, \0) +
i\eta_{\alpha \beta\gamma} (\omega) q_{\gamma} + \ldots.
\eqlab{diel-exp}
\end{equation}
Here $\epsilon_{\alpha \beta} (\omega, \0)$ is the dielectric function
in the infinite-wavelength limit, and
$\eta_{\alpha \beta\gamma} (\omega)$ describes spatial-dispersion
effects at first order in $\mathbf{q}$.

The $\eta_{\alpha \beta\gamma}$ tensor is odd under spatial inversion
$\mathcal{P}$, and thus vanishes identically in centrosymmetric
crystals. To analyze its behavior under time-reversal $\mathcal{T}$,
we split $\eta_{\alpha \beta\gamma}$ into symmetric (S) and
anti-symmetric (AS) parts under permutation of $\alpha$ and $\beta$,
\begin{equation}
\eta_{\alpha \beta\gamma}(\w) =
\eta^{\text{S}}_{\alpha \beta\gamma}(\w) +
\eta^{\text{AS}}_{\alpha \beta\gamma}(\w)\,.
\eqlab{eta-S-AS}
\end{equation}
As a consequence of Onsager's reciprocity relation,
$\eta^{\text{AS}}_{\alpha\beta\gamma}$ is $\mathcal{T}$-even while
$\eta^{\text{S}}_{\alpha\beta\gamma}$ is $\mathcal{T}$-odd.
$\eta^{\text{AS}}_{\alpha\beta\gamma}$ describes \textit{natural
  optical activity} (optical activity henceforth), which requires
broken $\mathcal{P}$ but not broken ${\mathcal T}$.
$\eta^{\text{S}}_{\alpha\beta\gamma}$ accounts for spatially
dispersive \textit{magneto-optical effects} in materials where both
${\mathcal P}$ and ${\mathcal T}$ symmetries are broken (but the
combined ${\mathcal P}{\mathcal T}$ may be present). The
$\mathcal{T}$-odd effects are \textit{nonreciprocal} in that one sees
distinct behaviors for forward vs.\ backward propagation of light. The
tensors $\eta^{\text{AS}}_{\alpha\beta\gamma}$ and
$\eta^{\text{S}}_{\alpha\beta\gamma}$ can be further decomposed as
\begin{subequations}
\begin{align}
\eta^{\text{S}}_{\alpha \beta\gamma}(\w) &=
\Re\, \eta^{\text{H}}_{\alpha \beta\gamma}(\w) +
i \Im\, \eta^{\text{AH}}_{\alpha \beta\gamma}(\w)\,, \\
\eta^{\text{AS}}_{\alpha \beta\gamma}(\w) &=
\Re\, \eta^{\text{AH}}_{\alpha \beta\gamma}(\w) +
i \Im\, \eta^{\text{H}}_{\alpha \beta\gamma}(\w)
 \eqlab{eta-H-AH}
 \end{align}
\end{subequations}
into hermitian (H) and anti-hermitian (AH) parts related to absorptive
and reactive phenomena, respectively.

As our main focus is on optical activity, we analyze
$\eta^{\text{AS}}_{\alpha\beta\gamma}$ in more detail. Being a rank-3
tensor antisymmetric in two indices, it has 9 independent components
that can be repackaged as a rank-2 \textit{gyration tensor} $G$ by
contracting with the Levi-Civita symbol
$\varepsilon_{\alpha \gamma \delta}$.  Following
Refs.~\cite{jerphagnon-jchemphys76,zabalo-prl23}, we define the
complex gyration tensor as
\begin{equation}
G_{\alpha \beta} (\omega) =
\frac{1}{2} \, \varepsilon_{\alpha \gamma \delta} \,
\eta^{\text{AS}}_{\gamma\delta\beta} (\omega)\,.
\eqlab{G}
\end{equation}
With this definition, $G$ has units of length and tends to a
real-valued constant in the $\w\rightarrow 0$
limit~\cite{zabalo-prl23}.\footnote{To obtain the dimensionless
  gyration tensor $g$ as defined in Ref.~\cite{landau-84}, multiply
  \eq{G} by a factor of $\w/c$. Note also that \eq{G} differs by a
  minus sign from the definition in
  Ref.~\cite{jerphagnon-jchemphys76}.}

It is useful to decompose $G$ into a trace piece, a traceless
symmetric part, and an antisymmetric
part~\cite{jerphagnon-jchemphys76}.  The first two form the symmetric
part of $G$, whose real and imaginary components account for optical
rotation and circular dichroism, respectively, of light propagating
along an optic axis.

Optical rotation --~the rotation of the plane of polarization of
linearly polarized incident light~-- results from a different speed of
propagation of left and right circular components. It is quantified by
the optical rotatory power, defined as the angle of rotation of the
plane of polarization of light per unit distance traveled inside the
material. Circular dichroism is the absorptive counterpart of optical
rotation, and corresponds to different absorption coefficients for
left and right circular components.  As a result, a linearly-polarized
incident wave develops an elliptical character as one of its circular
components decays more quickly than the other. For light not
propagating along an optic axis, linear birefringence typically
dominates, but the modes of propagation acquire some elliptical
character. The rotatory power $\rho$ and ellipticity $\theta$ are
obtained from the real and imaginary parts of $G$ as~\cite{landau-84}
\beq
\rho(\omega) + i \theta(\omega) = \frac{\omega^2}{2 c^2} \hat{q}_{\alpha} 
\left[
\text{Re}\, G_{\alpha \beta} (\omega) +
i \text{Im}\, G_{\alpha \beta} (\omega)
\right] 
\hat{q}_{\beta}
\eqlab{rot-ellip}
\eeq
with $\hat{\mathbf{q}}$ the unit vector along the propagation
direction~$\mathbf{q}$, assumed to be an optic axis.  As anticipated,
$\rho$ and $\theta$ only depend on the symmetric part of $G$. Since
$\Re\, G$ becomes constant at low frequencies, the rotatory power
decreases as $\w^2$, that is, the quantity
\beq
\bar \rho(\w) = \frac{\rho(\w)}{\w^2}
\eqlab{rho-bar}
\eeq
tends to a constant for $\w\rightarrow 0$.

The antisymmetric part of $G$ can be repackaged as a polar vector with
components
$G_{\alpha}=(1/2)\varepsilon_{\alpha\beta\gamma}G_{\beta\gamma}$. This
describes a situation in which the electric polarization vector of the
beam acquires a small longitudinal
component~\cite{jerphagnon-jchemphys76}. For example, light entering
along $x$ with polarization $z$ can propagate in the medium with a
slight elliptical polarization in the $x$-$z$ plane. By analogy with
\eq{rot-ellip}, one may quantify this ``polar optical activity'' via a
complex vector ${\bf d}(\w)$ with units of inverse length,
\beq
{\bf d}(\w) = \frac{\w^2}{2c^2} {\bf G}(\w)\,.
\eqlab{polar-NOA}
\eeq

Of the 21 acentric crystal classes, 18 allow for a nonzero gyration
tensor $G$. We identify as \textit{optically active} a material
belonging to any of these classes, according to a broad
definition~\cite{jerphagnon-jchemphys76}. The three acentric classes
that do not show optical activity are $\bar{6}$, $\bar{6}2m$, and
$\bar{4}3m$.  Among the 18 optically active crystal classes, three of
them ($3m$, $4mm$, and $6mm$) have a purely antisymmetric $G$
tensor. Materials that belong to these polar classes do not display
optical rotation or circular dichroism.  Nevertheless, we still
classify them here as optically active, while noting that some authors
adopt a more stringent definition that would exclude materials such as
these~\cite{halasyamani-chemmat98}. The other 15 classes, which do
support optical rotation, include 11 where the trace of $G$ is
nonzero. These \textit{chiral} or \textit{enantiomorphic} groups are
the only ones for which optical rotation is present even in
polycrystalline samples. Chiral materials can exist in two different
forms, called enantiomorphs, related to each other by a mirror
reflection. The two enantiomorphs show opposite (``dextrogyre'' and
``levogyre'') optical rotation. The remaining four classes ($\bar{4}$,
$\bar{4}2m$, $mm2$, and $m$) have a traceless $G$ tensor.  They still
show signs of optical rotation and circular
dichroism~\cite{malgrange-14}, but these effects are mixed with linear
birefringence, so that the solutions of the wave equation are
elliptically polarized in general.

\subsection{Microscopic theory}
\seclab{kubo}
%
In the Kubo linear-response formalism, one calculates the optical
conductivity tensor $\sigma$ relating the induced current density
$\mathbf{j}$ to the optical electric field $\mathbf{E}$,
\beq
j_{\alpha} (\w, \mathbf{q}) =
\sigma_{\alpha \beta} (\w, \mathbf{q}) E_{\beta} (\omega, \mathbf{q})\,.
\eqlab{sigma-w-q}
\eeq
The dielectric function and the optical conductivity are
related in CGS units by
\beq
\epsilon_{\alpha\beta}(\w,\mathbf{q}) =
\delta_{\alpha\beta} +
\frac{4\pi i}{\w}\sigma_{\alpha\beta}(\w,\mathbf{q})\,.
\eeq
Expanding the  latter in powers of $\q$ as
\beq
\sigma_{\alpha \beta} (\w, \mathbf{q}) =
\sigma_{\alpha \beta} (\w, \0) +
\sigma_{\alpha \beta\gamma} (\w) q_{\gamma} + \ldots
\eeq
and comparing with the expansion~\eqref{eq:diel-exp} of the former yields
\beq
\eta_{\alpha \beta\gamma}(\w) =
\frac{4\pi}{\w}\sigma_{\alpha \beta\gamma}(\w)\,,
\eqlab{eta-sigma}
\eeq
which combined with \eq{G} expresses the gyration tensor in
terms of the spatially dispersive conductivity.

The spatially dispersive optical conductivity evaluated from
Kubo linear response is given by the following expression,
equivalent to the one in Ref.~\cite{pozo-scipost23},
\begin{align}
&\sigma_{\alpha\beta\gamma}(\w) = \frac{ie^2}{\hbar} \sum_{ln} \int [d\k]
\Bigg\{
\frac{A_{\alpha,nl} \Tcal_{\beta\gamma,ln} + A_{\beta,ln} \Tcal_{\alpha\gamma,nl}}
{\wnl + \w + i \eta} \fnl\nn
&-
A_{\alpha,nl} A_{\beta,ln}
\left(
\frac{\w \bar f_{\gamma,nl} + \fnl \bar v_{\gamma,nl}}{\wnl + \w + i \eta} +
\frac{\wnl \fnl \bar v_{\gamma,nl}}{\left( \wnl + \w + i\eta \right)^2} \right)
\Bigg\}\nn
&+ \frac{e^2}{\hbar(\w + i\eta)} \sum_n \int [d\k]
\left(
f_{\alpha,n} \Tcal_{\beta\gamma,nn} - f_{\beta,n} \Tcal_{\alpha\gamma,nn} 
\right)\nn
&+ \frac{ie^2}{\hbar} \sum_n \int [d\k] g_{\alpha\beta,n} f_{\gamma,n}\nn &-
\frac{ie^2}{\hbar (\w + i\eta)^2}
\sum_n \int [d\k] f'_n v_{\alpha,n} v_{\beta,n} v_{\gamma,n}\,.
\eqlab{sigma-abc}
\end{align}
Here, $\int [d\k]=\int_{\rm BZ} d^3k/(2\pi)^3$, with the integral
spanning the first Brillouin zone (BZ), and $\k$ is implied in the
integrands.  $\wnl = w_n - \w_l$ and $ f_{nl} = f_n - f_l$ are
interband frequencies and differences in occupation factors,
respectively, and $\eta$ is a positive infinitesimal. We have also
defined
\begin{subequations}
\eqlab{defs}%
\begin{align}
v_{\alpha,n} &= \frac{\partial \w_n}{\partial k_\alpha}\,,
\eqlab{band-vel}
\\
f'_n &= \frac{\partial f_n}{\partial \w_n}\,,\\
f_{\alpha,n} &= \frac{\partial f_n}{\partial k_\alpha} = f'_n v_{\alpha,n}\,,\\
\bar f_{\alpha,nl} &=
\frac{1}{2} \left( f_{\alpha,n} + f_{\alpha,l} \right)\,,\\
\bar v_{\alpha,nl} &= \frac{1}{2} \left( v_{\alpha,n} + v_{\alpha,l} \right)\,.
\eqlab{v-bar}
\end{align}
\end{subequations}
Tensors $A_{\alpha,ln}$ and $\Tcal_{\alpha\beta,ln}$ containing
optical matrix elements will be introduced below, along with the
quantum-metric tensor $g_{\alpha\beta,n}$.

Every term in \eq{sigma-abc} except for the quantum-metric term has
either one or two powers of $\w+i\eta$ or $\wnl+\w+i\eta$ in the
denominator. Those terms with one power of frequency in the
denominator give $\delta$-like contributions to the optical
absorption, and those with two give $\delta'$-like contributions.

The first two lines of \eq{sigma-abc} contain the interband terms, and
the last three describe the intraband terms.  All intraband terms
contain derivatives $\partial f / \partial k$ or
$\partial f / \partial \omega$ and are thus
Fermi-surface-like. However, the reverse is not true: the term
containing $\bar f_{\gamma,nl}$ in the second line is interband and
Fermi-surface-like. In insulators and cold semiconductors only the
first line and the two terms containing $f_{nl}$ in the second line
survive, while in molecules only the first line survives, as all other
terms contain band velocities. The term in the last line, with a
product of three band velocities, is Drude-like in the sense of
depending only on the band structure.

All terms in the first two lines in \eq{sigma-abc} have parts that are
both symmetric and antisymmetric in $\alpha\beta$; the third line is
antisymmetric; and the fourth and the fifth are symmetric.  Since the
symmetric parts are time-odd, they vanish in the presence of
time-reversal symmetry.  Thus, only the third line and the
antisymmetric parts of the first two lines contribute in the case of
nonmagnetic materials. For a nonmagnetic or collinear-spin system
computed without spin-orbit coupling (SOC), the electron bands are all
pure spin up or down, so that matrices indexed by $ln$ are spin
diagonal and \eq{sigma-abc} can be evaluated separately in each spin
sector.

In \fref{fig1} we provide a schematic summary to help the
reader navigate \eq{sigma-abc} and classify all its
terms.

\begin{figure}[t]
\centering\includegraphics[width=0.6 \columnwidth]{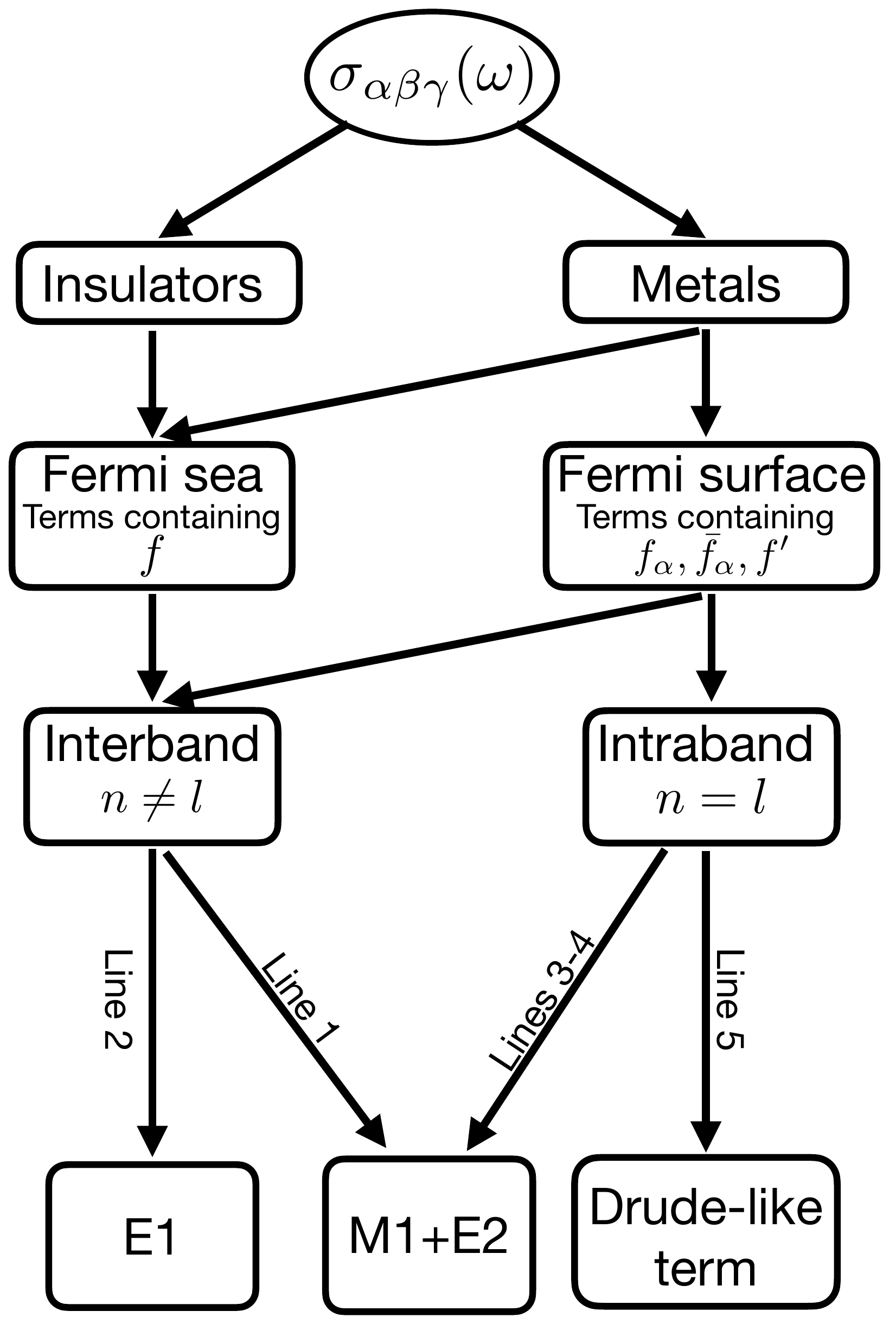}
\caption{Schematic chart of \eq{sigma-abc}. The contributions to
  $\sigma_{\alpha \beta\gamma}$ are classified at three different
  levels according to whether they are (i) Fermi sea or Fermi surface,
  (ii) interband or intraband, and (iii) electric dipole (E1) or
  magnetic dipole plus electric quadrupole (M1$+$E2). Terms containing
  the matrix $\mathcal{T}$ (see \fref{fig2} below) are of the M1$+$E2
  type, whereas the E1 terms do not contain $\mathcal{T}$. The
  quantum-metric term is classified as E2, whereas the Drude-like term
  in the last line of \eq{sigma-abc} falls outside the E1-M1-E2
  classification scheme.}
\figlab{fig1}
\end{figure}

The Hermitian matrices $A_\alpha$ and $\Tcal_{\alpha\beta}$ contain
the optical matrix elements at zeroth and first order in $\q$,
respectively.  $A_\alpha$ describes electric-dipole (E1) transitions,
while $\Tcal_{\alpha\beta}$ describes electric-quadrupole (E2) and
magnetic-dipole (M1) transitions via its symmetric and antisymmetric
parts, respectively (see \aref{multipoles}).  $A_\alpha$ takes the
form of a gauge-covariant Berry connection,
\beq
A_{\alpha,ln} = i\ip{u_l}{D_\alpha u_n}\,,
\eqlab{A}
\eeq
with
\beq
\ket{ D_\alpha u_n} =
\left(
1-\ket{u_n} \bra{u_n} 
\right) \ket{\partial_\alpha u_n}
\eqlab{cov-der}
\eeq
the covariant derivative of a cell-periodic Bloch state, and
$\partial_\alpha = \partial/\partial k_\alpha$ (here and throughout
this section, we assume nondegenerate bands).  While $A_{\alpha}$ is
purely orbital and off-diagonal ($A_{\alpha,nn}=0$),
$\Tcal_{\alpha\beta}$ has both orbital and spin parts, as well as
diagonal and off-diagonal parts. It is given by
\beq
\Tcal_{\alpha\beta,ln} = \frac{1}{2}
\left(
\Kcal_{\alpha\beta,ln} + \Kcal^*_{\alpha\beta,nl}
\right) -
\frac{g_{\rm s}}{2\mel}\epsilon_{\alpha\beta\gamma}\Scal_{\gamma,ln}
\,,
\eqlab{T}
\eeq
where in the first term the orbital matrix $\Kcal_{\alpha\beta}$ reads
\beq
\Kcal_{\alpha\beta,ln} =
\frac{1}{i\hbar} \bra{D_\alpha u_l}H-\epsilon_l\ket{D_\beta u_n} +
v_{\alpha,l} A_{\beta,ln}\,,
\eqlab{K}
\eeq
and in the second $g_{\rm s}\simeq 2$ is the spin $g$ factor and
\beq
\Scal_{\gamma,ln}
= \frac{\hbar}{2}\me{u_l}{\sigma_\gamma}{u_n}
\eqlab{S}
\eeq
is the spin matrix expressed in terms of Pauli matrices. The diagonal
part of $\Tcal_{\alpha\beta}$ appearing in the third line of
\eq{sigma-abc} is purely antisymmetric (M1), while the off-diagonal
part in the first line has both symmetric (E2) and antisymmetric (M1)
components. Finally, the quantity
\beq
g_{\alpha\beta,n} = \Re\,\ip{D_\alpha u_n}{D_\beta u_n}
\eqlab{g}
\eeq
in the fourth line of \eq{sigma-abc} is the single-band quantum
metric~\cite{provost80,marzari-prb97}.  It contributes a
frequency-independent and purely reactive intraband response of E2
character.

We have adopted a notation where lower-case letters ($f$, $g$, $v$)
are used for intraband quantities, and capital letters are used for
matrices in the band indices; calligraphic letters ($\Kcal$, $\Tcal$,
$\Scal$) are used for matrices with both interband and intraband
parts, and Roman letters ($A$) for purely interband
matrices. Intraband quantities can appear either as band vectors, like
$v_{\alpha,n}$ and $g_{\alpha\beta,n}$, or as matrices, in which case
they are automatically diagonal, e.g.,
$v_{\alpha,ln} = \delta_{ln} v_{\alpha,n}$.

The above quantities need to be computed on a sufficiently fine
$\bf k$ mesh to obtain converged results for the optical responses of
interest.  We assume this is being done as a post-processing step
after a self-consistent first-principles calculation. In the
sum-over-states approach to be summarized in \sref{sos}, it is
possible to compute the ingredients in \eqr{A}{g} directly, but with a
limitation on the number of contributing conduction bands. To prepare
for the Wannier interpolation approach to be introduced in
\sref{wannier-interp}, on the other hand, it is useful to define a set
of auxiliary tensors from which these matrices can be constructed in a
way that is free of the conduction-band limitation. These are
introduced next.

\subsection{Auxiliary matrices for Wannier interpolation}
\seclab{aux}

As we shall see in \sref{wannier-interp}, our Wannier interpolation
approach gives us convenient access to a set of auxiliary matrices
\begin{subequations}
\eqlab{A-C-D}%
\begin{align}
\Acal_{\alpha,ln} &= i\ip{u_l}{\partial_\alpha u_n} = \Acal^*_{\alpha,nl}\,,
\eqlab{Acal}\\
\Ccal_{\alpha\beta,ln} &=
\ip{\partial_\alpha u_l}{\partial_\beta u_n} = \Ccal^*_{\beta\alpha,nl}\,,
\eqlab{Ccal}\\
\Dcal_{\alpha\beta,ln} &=
\me{\partial_\alpha u_l}{H}{\partial_\beta u_n} = \Dcal^*_{\beta\alpha,nl}\,.
\eqlab{Dcal}
\end{align}
\end{subequations}
Here we show how the $\cal K_{\alpha\beta}$ and $g_{\alpha\beta}$
tensors of \eqs{K}{g} can be obtained from these ingredients.

The Berry connection matrix can be decomposed into intraband and
interband parts, $\Acal_\alpha=a_\alpha+A_\alpha$, with
\begin{subequations}
\begin{align}
a_{\alpha,ln} &= \delta_{ln}\Acal_{\alpha,ln} \,,\\
A_{\alpha,ln} &= (1-\delta_{ln})\Acal_{\alpha,ln} \,.
\end{align}
\eqlab{A-er-ra}
\end{subequations}
$A_\alpha$ is the same as in \eq{A}, and the covariant derivative in
\eq{cov-der} may now be written as
\beq
\ket{D_\alpha u_n} = \ket{\partial_\alpha u_n} +
i \ket{u_n} a_{\alpha,n}
\,,
\eqlab{cov-der-a}
\eeq
where we have converted $a_\alpha$ to band-vector form.  Inserting
this expression in \eqs{K}{g} and then using \eq{A-C-D} leads to
\begin{align}
\Kcal_{\alpha\beta,ln} &= \frac{1}{i\hbar}
\left( \Dcal_{\alpha\beta,ln} - \epsilon_l \Ccal_{\alpha\beta,ln} \right)\nn
&-
i\wln
A_{\alpha,ln}a_{\beta,n}
+ v_{\alpha,l}  A_{\beta,ln}
\eqlab{K-geom}
\end{align}
and
\beq
g_{\alpha\beta,n} = \Re\,\Ccal_{\alpha\beta,nn} -
a_{\alpha,n} a_{\beta,n}\,.
\eqlab{g-geom}
\eeq

Plugging \eq{K-geom} into \eq{T} for $\Tcal_{\alpha\beta}$ and
invoking the Hermiticity relations in \eq{A-C-D}, one finds that the
symmetric part of $\Dcal_{\alpha\beta}$ drops out, so that
$\Tcal_{\alpha\beta}$ and $\sigma_{\alpha\beta\gamma}(\w)$ only
involve the antisymmetric part. Instead, both the symmetric and
antisymmetric parts of $\Ccal_{\alpha\beta}$ contribute; separating
them as
\beq
\Ccal_{\alpha\beta} =
\frac{1}{2}\left( \Ccal_{\alpha\beta} + \Ccal_{\beta\alpha} \right) -
\frac{i}{2}\Fcal_{\alpha\beta}
\eqlab{C-decomp}
\eeq
where
\beq
\Fcal_{\alpha\beta} =
\partial_\alpha \Acal_\beta - \partial_\beta \Acal_\alpha
\eqlab{Fcal}
\eeq
is the Abelian Berry curvature matrix, \eq{K-geom} becomes
\begin{align}
\Kcal_{\alpha\beta}
&= \frac{1}{i\hbar}
\bigg[
\Dcal_{\alpha\beta} -
\frac{\epsilon}{2} \left( \Ccal_{\alpha\beta} +\Ccal_{\beta\alpha} \right) +
\frac{i\epsilon}{2}\Fcal_{\alpha\beta}\nn
&\qquad \; +
\epsilon A_\alpha a_\beta -
A_\alpha a_\beta \epsilon
\bigg] +
v_\alpha A_\beta\,,
\eqlab{K-geom-mat}
\end{align}
where $\epsilon$ is the diagonal band-energy matrix and all products
are matrix multiplications.

To recap, once the occupations $f$, band velocities $v_{\alpha}$ of
\eq{band-vel}, and tensors $\Acal_{\alpha}$, $\Ccal_{\alpha\beta}$,
and $\Dcal_{\alpha\beta}$ of \eq{A-C-D} are available at a given
$\bf k$ point, it is straightforward to compute, at that same point,
the various quantities introduced above.  The computation of these
tensor ingredients needed for the Wannier-interpolation approach will
be discussed in \sref{wannier-interp}.  \Fref{fig2} presents a
flowchart to summarize the important relationships to obtain the
building blocks of the E1, M1, and E2 contributions to the optical
conductivity in \eq{sigma-abc}.

\begin{figure}[t]
\centering\includegraphics[width=0.49\textwidth]{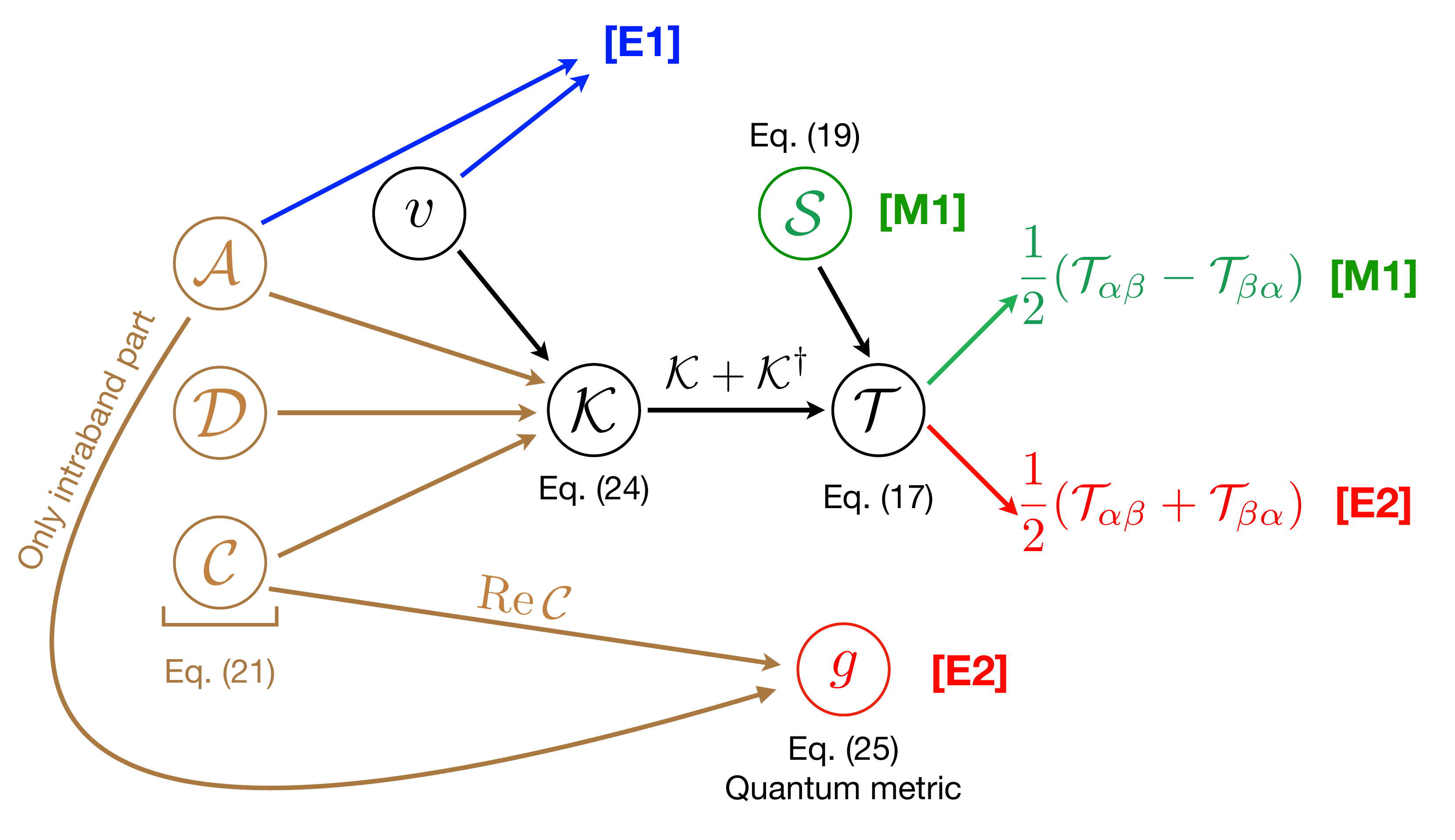}
\caption{Flowchart showing how the building blocks introduced in the
  main text are combined to obtain the electric dipole (E1, blue),
  magnetic dipole (M1, green) and electric quadrupole (E2, red)
  contributions to the spatially dispersive optical conductivity in
  \eq{sigma-abc}. Quantities needed exclusively for the Wannier
  interpolation approach are highlighted in brown; the sum-over-states
  approach to the Kubo formulation requires only the velocity matrix
  elements and the interband frequencies (see \sref{sos}).}
\figlab{fig2}
\end{figure}

\subsection{Summary}
\seclab{summary-end-II}

In this section, we have described the formalism and tensor
ingredients needed to compute the spatial-dispersion conductivity
tensor in \eq{sigma-abc}. In the remainder of this paper, we first
briefly describe a relatively straightforward sum-over-states approach
for computing this conductivity, although it is not the one we have
adopted.  We then provide the details of our Wannier-interpolation
approach and present some applications of this method.

\section{Sum-over-states approach}
\seclab{sos}

In this approach, one adopts a single $\bf k$ mesh for the entire
calculation and computes the needed matrix elements directly from the
Kohn-Sham wave functions on that mesh.  The details are as follows.

The spin matrices $\mathcal{S}_\gamma$ can be readily evaluated from
\eq{S}. As for the orbital tensors $A_\alpha$, $\Kcal_{\alpha\beta}$
and $g_{\alpha\beta}$, they are evaluated via the
sum-over-states expression for the covariant
derivative~\eqref{eq:cov-der},
\beq
\ket{D_\alpha u_n} = -\sum_{p\not= n}\, \ket{u_p}
\frac{\Vcal_{\alpha,pn}}{\w_{pn}}\,.
\eqlab{cov-der-sos}
\eeq
Here
\beq
\Vcal_{\alpha,pn} = \frac{1}{\hbar} \me{u_p}{(\partial_\alpha H)}{u_n}
\eqlab{V}
\eeq
is the velocity matrix, whose diagonal elements give the band
velocity~\eqref{eq:band-vel}:
\beq
v_{\alpha,n} = \Vcal_{\alpha,nn}\,.
\eqlab{vel}
\eeq
The orbital tensors can be written solely in terms of velocity matrix
elements and interband frequencies. This can be seen by inserting
\eq{cov-der-sos} in Eqs.~\eqref{eq:A}, \eqref{eq:K}, and~\eqref{eq:g},
which gives
\beq
A_{\alpha,ln} =
\begin{cases}
\frac{\Vcal_{\alpha,ln}}{i\wln} & \text{if $l \not= n$}\\
0                      & \text{if $l = n$}
\end{cases}\,,
\eqlab{A-sos}
\eeq
\beq
\Kcal_{\alpha\beta,ln} =
\left( \Vcal_\alpha A_\beta \right)_{ln}\,,
\eqlab{K-sos}
\eeq
and
\beq
g_{\alpha\beta,n} =
\Re \left( A_\alpha A_\beta \right)_{nn}\,.
\eqlab{g-sos}
\eeq

To obtain \eq{K-sos}, note the identity
\beq
\me{D_\alpha u_l}{H-\epsilon_l}{D_\beta u_n} =
i\hbar \sum_{p\not= l} \Vcal_{\alpha,lp} A_{\beta,pn}\,,
\eeq
which follows from \eqs{A}{cov-der-sos}.  Using that identity in
\eq{K} leads to
\begin{align}
\Kcal_{\alpha\beta,ln} &= \sum_{p\not= l}\Vcal_{\alpha,lp} A_{\beta,pn} +
v_{\alpha,l}A_{\beta,ln}\nn
&= \sum_p\,\Vcal_{\alpha,lp} A_{\beta,pn} -\Vcal_{\alpha,ll}A_{\beta,ln} +
v_{\alpha,l}A_{\beta,ln}\,.
\eqlab{Kcal2}
\end{align}
The last two terms cancel out, yielding \eq{K-sos}.

The sum-over-states evaluation of $\sigma_{\alpha\beta\gamma}(\w)$ was
implemented in Ref.~\cite{wang-prb23}.  The advantage of this type of
approach is the ease of implementation, especially if the {\it ab
  initio} code already provides the velocity matrix elements.  One
disadvantage is the need to truncate the implied summations over
intermediate states in \eqs{K-sos}{g-sos}. This can be traced back to
a band-truncation error in the covariant derivative of
\eq{cov-der-sos}, where index $p$ only runs over the conduction bands
carried in the first-principles calculation.  It also suffers from the
need to call the {\it ab initio} engine for every~$\k$ when performing
the BZ integral in \eq{sigma-abc}, which can become costly for
narrow-gap semiconductors and when evaluating Fermi-surface terms for
conductors.

\section{Wannier interpolation} 
\seclab{wannier-interp}

To overcome the difficulties mentioned above, we adopt an approach
based on Wannier interpolation. As we shall see, some terms in the
resulting expressions for $A_\alpha$, $\Kcal_{\alpha\beta}$ and
$g_{\alpha\beta}$ resemble the sum-over-states formulas in
\eqr{A-sos}{g-sos}, while others resemble the $k$-derivative formulas
in Eqs.~\eqref{eq:A}, \eqref{eq:g}, and~\eqref{eq:K-geom-mat}. This
approach typically cannot treat as many interband excitations as is
allowed by the sum-over-states approach. However, it is free from
band-truncation errors in the computation of optical matrix elements,
and it allows the optical properties to be calculated efficiently on a
fine interpolation mesh of $\bf k$ points.

\subsection{Basics of Wannier interpolation}
\seclab{wannier-basics}

The Wannier functions (WFs) are constructed as
\beq
\ket{\R j} = \frac{1}{N} \sum_\q\,
e^{-i\q\cdot(\R+\btau_j)} \ket{\psi\W_{j\q}}\qquad (j=1,\ldots,M)\,,
\eqlab{Rj}
\eeq
where
\beq
\ket{\psi\W_{j\q}} = \sum_n\,\ket{\psi_{n\q}}\Wcal_{nj}(\q)\,.
\eqlab{psi-W-q}
\eeq
The summation in \eq{Rj} is over the uniform Brillouin-zone mesh of
$N$ points $\q$ where the {\it ab initio} Bloch eigenstates
$\ket{\psi_{n\q}}$ were calculated, and the Wannier-gauge Bloch-like
states $\ket{\psi\W_{j\q}}$ are related to the Bloch eigenstates by a
unitary wannierization matrix $\Wcal(\q)$. That matrix is $M \times M$
without disentanglement~\cite{marzari-prb97}, and with disentanglement
it becomes $J(\q)\times M$ with $J(\q)\geq M$ at every
$\q$~\cite{souza-prb01}. Since in
\textsc{Wannier90}~\cite{pizzi-jpcm20} the Wannier centers
$\btau_j = \me{\0 j}{\r}{\0 j}$ are not included in the Fourier phase
factors, $\Wcal(\q)$ in \eq{psi-W-q} is related to
\textsc{Wannier90}'s wannierization matrix $\tilde \Wcal(\q)$ by
\beq
\Wcal_{nj}(\q) = \tilde \Wcal_{nj}(\q)\, e^{i\q\cdot \btau_j}\,.
\eqlab{W-matrix}
\eeq

Once the WFs have been constructed using a relatively coarse $\q$
grid, a Bloch basis is defined at each point $\k$ on the dense
interpolation grid as
\beq
\ket{\psi\W_{j\k}} = \sum_\R\, e^{i\k\cdot(\R+\btau_j)}\ket{\R j}\,.
\eqlab{u-W}
\eeq
Here the sum over $\R$ is restricted such that $\ket{\R j}$ lies
inside the Wigner-Seitz cell of the supercell conjugate to the coarse
$\q$ grid.\footnote{More precisely, we use the ``minimal-distance
  replica selection'' algorithm described in
  Refs.~\cite{pizzi-jpcm20,tsirkin-npjcm21}.}  The Hamiltonian matrix
in this basis reads
\beq
\Hcal\W_{ij}(\k) = \me{u\W_{i\k}}{H_\k}{u\W_{j\k}}\,,
\eqlab{H-W}
\eeq
where we have switched to cell-periodic states and defined
$H_\k = e^{-i\k\cdot \r} H e^{i\k\cdot\r}$.  To obtain the
interpolated band energies,\footnote{To avoid the heavy notations
  $\epsilon_{n\k}^{\rm H}$, $v^{\rm H}_{\alpha,n\k}$ and
  $\omega\H_{ln\k}$, we will denote the interpolated band energies,
  band velocities and interband frequencies by the same symbols
  $\epsilon_n$, $v_{\alpha,n}$ and $\wln$ as their {\it ab initio}
  counterparts. When referring to {\it ab initio} quantities, we will
  always include the subscript $\q$ for clarity; instead, we will
  often omit the subscript $\k$ from interpolated quantities.} the
matrix $\Hcal\W$ is diagonalized as
\beq
\Hcal\H_{ln}(\k) =
\left[
\Ucal^\dagger(\k) \Hcal\W(\k) \Ucal(\k)
\right]_{ln} =
\delta_{ln} \epsilon\bnk\,,
\eqlab{H-H}
\eeq
where $\Ucal(\k)$ is a unitary matrix and the H superscript stands for
Hamiltonian gauge. (When disentanglement is not used,
$\Ucal = \Wcal^\dagger$ on mesh points $\k=\q$.)

\Eq{H-H} can be brought to the same form as \eq{H-W} by defining
\beq
\ket{u\H\bnk} = \sum_j\, \ket{u\W_{j\k}}\Ucal_{jn}(\k)\,,
\eqlab{u}
\eeq
so that 
\beq 
\Hcal\H_{ln}(\k) = \me{u\H_{l\k}}{H_\k}{u\H_{n\k}} =
\delta_{ln}\epsilon\bnk\,.
\eqlab{H-H-b}
\eeq
Differentiating \eq{u} gives (from now on we drop $\k$)
\beq
\ket{\partial_\alpha u\H_n} =
\sum_j\,\ket{\partial_\alpha u\W_j}\Ucal_{jn} -
i\sum_l\,\ket{u\H_l}\Acal\I_{\alpha,ln}\,,
\eqlab{del-u}
\eeq
where
\beq
\Acal\I_\alpha = i \Ucal^\dagger \partial_\alpha \Ucal
\eqlab{Acal-I}
\eeq
is the Berry connection matrix constructed from derivatives of the
column vectors of $\Ucal$ (the eigenvectors of $\Hcal\W$).  We refer
to quantities of this kind, which can be expressed solely in terms of
the $M \times M$ Wannier Hamiltonian matrix $\Hcal\W(\k)$ and its
$M$-component column eigenvectors $\Ucal(\k)$, as \textit{internal}
(I). Instead, \textit{external}~(E) quantities are those that depend
on the character of the Wannier functions themselves relative to the
full Hilbert space.  Considering only the internal terms is equivalent
to treating the problem at a tight-binding level in the Wannier
subspace, with on-site energies, hoppings, and orbital centers as the
only model parameters. The external terms capture the extra embedding
information that is missed by the standard tight-binding
description. This partition becomes particularly clear and meaningful
when the intracell Wannier centers $\btau_j$ are included in the
Fourier phase factors, as done in \eqs{Rj}{u-W}.

Using \eq{del-u} in \eq{Acal} for $\Acal_\alpha$ yields the
Hamil\-to\-nian-gauge Berry connection~\cite{wang-prb06}
\beq
\Acal\H_\alpha = \Acal\I_\alpha + \Acal\E_\alpha\,,
\eqlab{Acal-H}
\eeq
where the external part reads
\beq
\Acal\E_\alpha = \Ucal^\dagger \Acal\W_\alpha \Ucal
\eqlab{Acal-E}
\eeq
with
\beq
\Acal\W_{\alpha,ij} = i\ip{u\W_i}{\partial_\alpha u\W_j}\,.
\eeq
More generally, for any Wannier-gauge matrix ${\cal O}\W$ we
define its external part to be
\beq
{\cal O}\E = \Ucal^\dagger {\cal O}\W \Ucal\,,
\eqlab{O-E}
\eeq
and when ${\cal O}\H={\cal O}\E$ that matrix is said to be gauge
covariant~\cite{wang-prb06}. In view of \eqs{H-H}{Acal-H}, we conclude
that $\Hcal$ is gauge covariant but $\Acal_{\alpha}$ is not.

We are now ready to evaluate the covariant
derivative~\eqref{eq:cov-der} of an interpolated Bloch state
$\ket{u\H_n}$.  Multiplying by $1-\ket{u\H_n}\bra{u\H_n}$ on both
sides of \eq{del-u} yields
\begin{align}
\ket{D_\alpha u\H_n} &= \sum_j\, \ket{\partial_\alpha u\W_j}\Ucal_{jn} +
i\ket{u\H_n}a\E_{\beta,n}\nn
&-i\sum_p\,\ket{u\H_p}A\I_{\alpha,pn}
\eqlab{cov-der-H}
\end{align}
where, by analogy with \eq{A-er-ra}, $a\E$ is the intraband part of
$\Acal\E$ and $A\I$ is the interband part of $\Acal\I$.  Note that
this expression for $\ket{D_\alpha u\H_n}$ does not suffer from the
same band-truncation error as was discussed above for the
sum-over-states approach: see \eq{cov-der-sos}.  Here, the covariant
derivative has the full accuracy of the plane-wave (or other
first-principles) basis, as do the various tensors computed from it.

\subsection{Application to \texorpdfstring{$\boldsymbol{\sigma_{\alpha\beta\gamma}(\w)}$}{sigma}}
\seclab{sigma-interp}

\subsubsection{Optical matrix elements}
\seclab{optical}

In \sref{sos}, the optical matrix elements entering \eq{sigma-abc} for
$\sigma_{\alpha\beta\gamma}$ were evaluated using the sum-over-states
formula of \eq{cov-der-sos} for $\ket{D_\alpha u_n}$.  In our approach
we evaluate the covariant derivatives using \eq{cov-der-H} instead.
Our immediate goal is to obtain the needed quantities such as
$A_\alpha$ of \eq{A} and ${\cal K}_{\alpha\beta}$ of \eq{K} on a
coarse $\bf q$ mesh. These will then be Wannier interpolated onto a
fine $\bf k$ mesh as described in detail later in
\sref{interpolation}.

The covariant derivatives $\ket{D_\alpha u_n}$ are not needed for the
Hamiltonian and spin matrices, which can be obtained directly from the
first-principles calculation as in \sref{sos}. For consistency with
what follows, however, we transform them to the Wannier gauge using
\eq{psi-W-q} to obtain
\begin{subequations}
\eqlab{H-S-W}%
\begin{align}
\Hcal\W_{ij} &= \me{u\W_i}{H}{u\W_j}\,,
\eqlab{H-W-repeat}\\
\Scal\W_{\alpha,ij} &= \me{u\W_i}{S_\alpha}{u\W_j}\,.
\eqlab{S-W}
\end{align}
\end{subequations}
We then also construct a set of Wannier-gauge orbital matrices
\begin{subequations}
\eqlab{W-matrices}%
\begin{align}
\Acal\W_{\alpha,ij} &= i\ip{u\W_i}{\partial_\alpha u\W_j}\,,
\eqlab{A-W}\\
\Bcal\W_{\alpha,ij} &= i\me{u\W_i}{H}{\partial_\alpha u\W_j}\,,
\eqlab{B-W}\\
\Ccal\W_{\alpha\beta,ij} &= \ip{\partial_\alpha u\W_i}{\partial_\beta u\W_j}\,,
\eqlab{C-W}\\
\Dcal\W_{\alpha\beta,ij} &= \me{\partial_\alpha u\W_i}{H}{\partial_\beta u\W_j}
\eqlab{D-W}
\end{align}
\end{subequations}
on the coarse mesh. The derivatives in these equations are evaluated
by finite differences on the coarse $\bf q$ mesh as described in
\sref{coarse-matrices}; this has to be done in the Wannier gauge to
enforce a smooth evolution of wave functions between neighboring
$\bf q$ points. After Wannier interpolation of the quantities in
\eqs{H-S-W}{W-matrices} onto the fine $\bf k$ mesh, it is relatively
straightforward to compute quantities such as those in \eqs{A}{K} in
the Hamiltonian gauge, as described below.

In \eq{W-matrices}, compared to \eq{A-C-D}, we have introduced an
additional matrix $\Bcal_\alpha$, which will be needed later in the
interpolation formula for $\Kcal_{\alpha\beta}$ if Wannier
disentanglement is used.  Recall from \eq{A-C-D} that $\Acal_\alpha$,
$\Ccal_{\alpha\beta}$, and $\Dcal_{\alpha\beta}$ satisfy Hermiticity
relations. Instead, $\Bcal_\alpha$ satisfies the generalized
Hermiticity constraint
\beq
\left(\Bcal\W_{\alpha,ji}\right)^* = \Bcal\W_{\alpha,ij} +
i\hbar\Vcal\W_{\alpha,ij} -
i\partial_\alpha \Hcal\W_{ij}
\eqlab{B-dagger}
\eeq
which follows from differentiating \eq{H-W} for $\Hcal\W$, with
$\Vcal_\alpha$ given by \eq{V}.  Following \eq{Fcal}, we also define
\beq
\Fcal\W_{\alpha\beta} =
\partial_\alpha \Acal\W_\beta - \partial_\beta \Acal\W_\alpha\,.
\eqlab{F-W}
\eeq

As a reminder, our goal is to obtain the quantities $A_\alpha$,
$\Kcal_{\alpha\beta}$, $g_{\alpha\beta}$, and $S_\alpha$ in the
Hamiltonian gauge from the above Wannier-gauge matrices.  Let us start
with the spin matrix in \eq{S}. Using \eq{u} we find
\beq
\Scal\H_\alpha = \Ucal^\dagger \Scal\W_\alpha \Ucal = \Scal\E_\alpha\,.
\eqlab{S-H}
\eeq
Thus, the $\Scal_\alpha$ matrix is gauge covariant, just like $\Hcal$.
But while $\Hcal\H$ in \eq{H-H} is diagonal by construction,
$\Scal\H_\alpha$ is generally nondiagonal in the presence of SOC.

The orbital matrices $A_\alpha$, $\Kcal_{\alpha\beta}$, and
$g_{\alpha\beta}$, on the other hand, are noncovariant (i.e., they are
not equal to their external parts). For example, inserting
\eq{cov-der-H} in \eq{A} for $A_\alpha$ gives
\begin{subequations}
\begin{align}
A\H_\alpha &= A\I_\alpha + A\E_\alpha\,,
\eqlab{A-H}\\
A\I_{\alpha,ln} &\equiv (1-\delta_{ln})\Acal\I_{\alpha,ln} =
\begin{cases}
\frac{\Vcal\I_{\alpha,ln}}{i\wln} & \text{if $l \not= n$}\\
0                      & \text{if $l = n$}
\end{cases}\,,
\eqlab{A-I}\\
A\E_{\alpha,ln} &\equiv (1-\delta_{ln})\Acal\E_{\alpha,ln} =
(1-\delta_{ln}) \left[ \Ucal^\dagger \Acal\W_\alpha \Ucal \right]_{ln}\,.
\eqlab{A-E}
\end{align}
\eqlab{A-decomp}%
\end{subequations}
In \eq{A-I} $\Vcal\I_\alpha$ is the internal velocity matrix
\beq
\Vcal\I_\alpha = \frac{1}{\hbar}
\Ucal^\dagger \left( \partial_\alpha \Hcal\W \right) \Ucal\,,
\eqlab{V-I}
\eeq
and the right-hand side of \eq{A-E} follows from \eq{Acal-E} for
$\Acal\E_\alpha$. Note that \eqs{A-I}{V-I} have the same form as
\eqs{A-sos}{V} for $A_\alpha$ and $\Vcal_\alpha$, except that they are
written in terms of internal quantities. Moreover, the interpolated
band velocities can be obtained from \eq{V-I} via the internal
analogue of \eq{vel},
\beq
v_{\alpha,n}=\Vcal\I_{\alpha,nn}\,.
\eqlab{v}
\eeq

Using \eq{cov-der-H} in \eq{K} for $\Kcal_{\alpha\beta}$ one finds
(see \aref{K-wann-deriv}) that it contains not only internal and
external terms, but also \textit{cross} (X) terms that mix internal
and external quantities,
\begin{subequations}
\begin{align}
\Kcal\H_{\alpha\beta} &= \Kcal\I_{\alpha\beta} +
\Kcal\E_{\alpha\beta} + \Kcal\X_{\alpha\beta}\,,
\eqlab{K-H}\\
\Kcal\I_{\alpha\beta} &= \Vcal\I_\alpha A\I_\beta\,,
\eqlab{K-I}\\
\Kcal\E_{\alpha\beta}
&= \frac{1}{i\hbar}
\bigg[
\Dcal\E_{\alpha\beta} -
\frac{\epsilon}{2} \left( \Ccal\E_{\alpha\beta} +\Ccal\E_{\beta\alpha} \right) +
\frac{i\epsilon}{2}\Fcal\E_{\alpha\beta}\nn
&\qquad \; +\epsilon A\E_\alpha a\E_\beta -
A\E_\alpha a\E_\beta \epsilon
\bigg] +
v_\alpha A\E_\beta\,,
\eqlab{K-E}\\
\Kcal\X_{\alpha\beta} &= \frac{1}{i\hbar}
\left[
A\I_\alpha B\E_\beta - \epsilon A\I_\alpha A\E_\beta +
\left( A\I_\beta B\E_\alpha - A\I_\beta A\E_\alpha \epsilon \right)^\dagger
\right]\,.
\eqlab{K-X}
\end{align}
\eqlab{K-decomp}%
\end{subequations}
Here $\epsilon=\Hcal\H$ and $v_\alpha$ are the diagonal band-energy
and band-velocity matrices, respectively, and $B\E_\alpha$ is the
off-diagonal part of ${\mathcal B}\E_\alpha$.  \Eq{K-I} is the
internal analogue of the sum-over-states formula~\eqref{eq:K-sos} for
$\Kcal_{\alpha\beta}$, and \eq{K-E} is the external analogue of the
$k$-derivative formula~\eqref{eq:K-geom-mat}.

When disentanglement is not used in the Wannier construction, \eq{K-X}
can be simplified using the relation
\beq
\Bcal\E_{\alpha,ln}=\epsilon_l\Acal\E_{\alpha,ln}
\eqlab{B-A-inner}
\eeq
from \aref{K-wann-deriv}.  This is an exact identity when the
interpolation point falls on the {\it ab initio} mesh and an excellent
approximation otherwise.  In this case, \eq{K-X} reduces to
\beq 
\Kcal\X_{\alpha\beta}=\frac{1}{i\hbar} \big[
A\I_\alpha \epsilon A\E_\beta - \epsilon A\I_\alpha A\E_\beta + ({\rm
  I}\leftrightarrow {\rm E})\big]\,, 
  \eqlab{B-removed} 
\eeq
which further highlights the parallelism with the sum-over-states
formula~\eqref{eq:K-sos} for $\Kcal_{\alpha\beta}$.  Of course, the
optical absorption spectra predicted from $\Kcal\H_{\alpha\beta,ln}$
can only cover the frequency range of transitions between wannierized
bands $l$ and $n$.

When disentanglement is used, we make the same assumption about the
range of computed spectra, but now insisting that bands $l$ and $n$
lie inside the frozen window.  In this case, \eq{K-decomp} is still
correct as written, but the replacement of \eq{K-X} by \eq{B-removed}
is no longer justified when the interior band index ($p$ in an
expression like $\sum_p A\I_{\alpha,lp}B\E_{\beta,pn}$) lies outside
the frozen window.  While \eq{K-X} could be used for both situations,
we find it convenient to use \eq{B-removed} when $p$ lies in the
frozen window and \eq{K-X} otherwise.\footnote{To avoid
  discontinuities, our implementation uses a smooth interpolation
  between evaluating $B\E_\alpha$ in the two different ways.}

Since $l$ and $n$ must lie inside the frozen window, the sums in the
Kubo formula~\eqref{eq:sigma-abc} for the optical conductivity are
restricted to a subset of the \textit{ab initio} bands. This leads to
a truncation error, inherent to the present methodology, which can be
controlled by enlarging the frozen window to capture the relevant
conduction states. This truncation error affects only the reactive
part of the response; the absorptive part is not affected because, as
noted in \sref{kubo}, it comes with $\delta(\omega_{nl} - \omega)$ and
$\delta'(\omega_{nl} - \omega)$ factors, which restrict the sums over
$l$ and $n$ to a small subset of states close to the optical
transition.

Finally, using \eq{cov-der-H} in \eq{g} gives for the quantum metric
\begin{subequations}
\begin{align}
g\H_{\alpha\beta,n} &=
g\I_{\alpha\beta,n} + g\E_{\alpha\beta,n} + g\X_{\alpha\beta,n}\,,
\eqlab{g-H}\\
g\I_{\alpha\beta,n} &=
\Re \left( A\I_\alpha A\I_\beta \right)_{nn}\,,
\eqlab{g-I}\\
g\E_{\alpha\beta,n} &= \Re\,\Ccal\E_{\alpha\beta,nn} -
a\E_{\alpha,n} a\E_{\beta,n}
\,,
\eqlab{g-E}\\
g\X_{\alpha\beta,n} &= \Re
\left( A\I_\alpha A\E_\beta + A\E_\alpha A\I_\beta \right)_{nn}\,.
\eqlab{g-X}
\end{align}
\eqlab{g-decomp}%
\end{subequations}
\Eq{g-I} is the internal counterpart of the sum-over-states
formula~\eqref{eq:g-sos} for $g_{\alpha\beta,n}$, and \eq{g-E} is the
external counterpart of the $k$-derivative formula~\eqref{eq:g-geom};
as in the case of $\Kcal_{\alpha\beta}$, the cross term is similar to
the internal term and to the sum-over-states expression.

In summary, the interpolated spin matrix is given by \eq{S-H}; of the
three orbital matrices entering \eq{sigma-abc}, $A_\alpha$
[\eq{A-decomp}] contains only internal and external terms in the
Wannier representation, while $\Kcal_{\alpha\beta}$ [\eq{K-decomp}]
and $g_{\alpha\beta}$ [\eq{g-decomp}] also contain cross terms. The
band energies and band velocities appearing in \eq{K-decomp} and
elsewhere in \eq{sigma-abc} are obtained from \eqs{H-H}{v},
respectively, with \eq{H-H} also providing the $\Ucal$ matrices.

The only remaining task is to compute the Wannier-gauge matrices in
\eqs{H-S-W}{W-matrices}. The derivatives with respect to wave vector
in \eq{W-matrices} will be evaluated by finite differences on the
coarse ${\bf q}$ mesh as described in the next section.  At this
point, if we were only interested in computing the responses on the
coarse mesh, we would essentially be done.  However, one of the main
advantages of the Wannier interpolation approach is that it now allows
these matrices to be interpolated efficiently onto a fine mesh for a
higher-resolution calculation of the optical response, as described in
\sref{interpolation}.

\subsubsection{Construction of Wannier-gauge matrices}
\seclab{coarse-matrices}

To interpolate the Wannier-gauge matrices $\mathcal{O}\W$, with
$\mathcal{O} = \Hcal, \Scal_\alpha, \Acal_\alpha, \Bcal_\alpha,
\Ccal_{\alpha\beta}, \Dcal_{\alpha\beta}$, onto a fine mesh of
$\mathbf{k}$ points, one needs first to compute $\mathcal{O}\W$ on the
coarse \textit{ab initio} $\mathbf{q}$ mesh. This is done starting
from \eq{H-S-W} for the Hamiltonian and spin matrices and from
\eq{W-matrices} for the orbital matrices, and using \eq{psi-W-q}. In
particular, the derivatives appearing in \eq{W-matrices} are computed
by finite differences on the $\mathbf{q}$ grid as~\cite{marzari-prb97}
\beq
\ket{\partial_{\alpha} u_{j \mathbf{q}}\W} \simeq
\sum_{\mathbf{b}} w_{b} b_{\alpha} \ket{u_{j \mathbf{q}+\mathbf{b}}\W},
\eeq
where $w_b$ are appropriately chosen weights, and the sum is performed
over shells of vectors $\mathbf{b}$ connecting a point~$\mathbf{q}$ of
the \textit{ab initio} grid to its neighbors. In this way, the
Wannier-gauge Hamiltonian and spin matrices in the coarse $\mathbf{q}$
mesh read
\begin{subequations}
\begin{align}
\Hcal\W(\q) &=
\Wcal^\dagger(\q) \mathbbm{H}(\q) \Wcal(\q)\,,
\eqlab{H-W-q-new}\\
\Scal\W_\alpha(\q) &=
\Wcal^\dagger(\q) \mathbbm{S}_\alpha(\q) \Wcal(\q)\,,
\eqlab{S-W-q-new}
\end{align}
\end{subequations}
whereas the orbital matrices take the form
\begin{widetext}
\begin{subequations}
\begin{align}
\Acal\W_{\alpha, ij}(\mathbf{q}) &\simeq i \sum_{\mathbf{b}} w_b b_{\alpha} \left[ \Wcal^\dagger(\q) {\mathbbm M}(\q,\q+\b) \Wcal(\q+\b) \right]_{ij}\,,\\
\Bcal\W_{\alpha, ij}(\mathbf{q}) &\simeq i \sum_{\mathbf{b}} w_b b_{\alpha} \left[ \Wcal^\dagger(\q) {\mathbbm H}(\q){\mathbbm M}(\q,\q+\b) \Wcal(\q+\b) \right]_{ij}\,,\\
\Ccal\W_{\alpha \beta, ij}(\mathbf{q}) &\simeq \sum_{\mathbf{b}, \mathbf{b}'} w_b w_{b'} b_\alpha b'_\beta \left[ \Wcal^\dagger(\q+\b) {\mathbbm M}(\q+\b,\q+\b') \Wcal(\q+\b') \right]_{ij}\,,\\
\Dcal \W_{\alpha \beta, ij}(\mathbf{q}) &\simeq \sum_{\mathbf{b}, \mathbf{b}'} w_b w_{b'} b_\alpha b'_\beta \left[ \Wcal^\dagger(\q+\b) {\mathbbm N}(\q+\b,\q+\b') \Wcal(\q+\b') \right]_{ij}\,.
\end{align}
\eqlab{R-q-new}
\end{subequations}
\end{widetext}
Here $\Wcal(\q)$ is related by \eq{W-matrix} to the $\tilde \Wcal(\q)$
matrix provided by \textsc{Wannier90}, and
\begin{subequations}
\begin{align}
\mathbbm{H}_{mn}(\q) &= \me{u_{m\q}}{H_\q}{u_{n\q}} = \delta_{mn} \epsilon_{n\q}\,,
\eqlab{H-q-new}\\
\mathbbm{S}_{\alpha,mn}(\q) &= \me{u_{m\q}}{S_\alpha}{u_{n\q}}\,, \\
{\mathbbm M}_{mn}(\q,\q+\b) &= \ip{u_{m\q}}{u_{n,\q+\b}}\,,
\eqlab{mmn-new}\\
{\mathbbm M}_{mn}(\q+\b,\q+\b') &= \ip{u_{m,\q+\b}}{u_{n,\q+\b'}}\,,
\eqlab{uIu-new}\\
{\mathbbm N}_{mn}(\q+\b,\q+\b') &= \me{u_{m,\q+\b}}{H_\q}{u_{n,\q+\b'}}\,,
\eqlab{uHu-new}
\end{align}
\eqlab{ab-initio-mat}
\end{subequations}
are matrices provided by the {\it ab initio} code through its
interface to \textsc{Wannier90}. This complete list of {\it ab initio}
matrices, which are needed to compute $\sigma_{\alpha\beta\gamma}$, is
also summarized in ~\Tref{matrices-new}.
\begin{table*}
\renewcommand{\arraystretch}{1.2}
\caption{Quantities needed from the {\it ab initio} code to evaluate
  \eq{sigma-abc} by Wannier interpolation. The filename convention
  follows \textsc{pw2wannier90}, the interface between the {\it ab
    initio} code \textsc{Quantum
    Espresso}~\cite{giannozzi-jpcm09,giannozzi-jpcm17} and
  \textsc{Wannier90}.}
\begin{tabular}{ccc}
{\it Ab initio} matrix & Needed for & Stored in\\
\hline
${\mathbbm H}(\q) =
\me{u_\q}{H_\q}{u_\q}$ & $\Hcal\W(\q), \Bcal\W_\alpha(\q)$ &
{\tt seedname.eig}\\
${\mathbbm S}_\alpha(\q) =
\me{u_\q}{S_\alpha}{u_\q}$ & $\Scal\W_\alpha(\q)$ &
{\tt seedname.spn}\\
${\mathbbm M}(\q,\q+\b) =
\ip{u_\q}{u_{\q+\b}}$ & $\Acal\W_\alpha(\q), \Bcal\W_\alpha(\q)$ &
{\tt seedname.mmn}\\
${\mathbbm M}(\q+\b,\q+\b') =
\ip{u_{\q+\b}}{u_{\q+\b'}}$ & $\Ccal\W_{\alpha\beta}(\q)$ &
{\tt seedname.uIu}\\
${\mathbbm N}(\q+\b,\q+\b') = \me{u_{\q+\b}}{H_\q}{u_{\q+\b'}}$ &
$\Dcal\W_{\alpha\beta}(\q)$ & {\tt seedname.uHu}\end{tabular} 
\label{tab:matrices-new}
\end{table*}

Compared to the ground-state bulk magnetization~\cite{lopez-prb12}, an
additional piece of information [\eq{uIu-new}], stored in the {\tt
  seedname.uIu} file, is required.  According to \eqs{T}{K-E}, that
information is only used for the symmetric (E2) part of the
$\Tcal_{\alpha\beta}$ matrix, not for the antisymmetric (M1) part.
Conversely, the information stored in {\tt seedname.uHu}
[\eq{uHu-new}] is only used for the antisymmetric part of
$\Tcal_{\alpha\beta}$; this is due to the fact that the symmetric part
of $\Dcal_{\alpha\beta}$ drops out from $\Tcal_{\alpha\beta}$, as
discussed in \sref{aux}.

\subsubsection{Wannier interpolation onto the fine mesh}
\seclab{interpolation}

Once the matrices on the $\mathbf{q}$ mesh are computed, they are
Fourier-interpolated onto a finer mesh in the following way. First, a
Fourier transform to real space is performed,
\beq
{\mathcal O}_{ij}(\R) = \frac{1}{N} \sum_\q\,
e^{-i\q \cdot (\R + \btau_j - \btau_i)} {\mathcal O}\W_{ij}(\q)\,,
\eqlab{O-R-new}
\eeq
then, the real space matrices are Fourier transformed to the fine
$\mathbf{k}$ mesh,
\beq
{\mathcal O}\W_{ij}(\k) =
\sum_\R\, e^{i\k \cdot (\R+\btau_j - \btau_i)} {\mathcal O}_{ij}(\R).
\eqlab{O-k-new}
\eeq
In these expressions,
$\mathcal{O}_{ij}(\R) = \me{\0 i}{\mathcal{O}}{\R j}$.  The real-space
matrix elements of $\Hcal$ and $\Scal_{\alpha}$ read
\begin{subequations}
\begin{align}
\Hcal_{ij}(\R) &= \me{\0 i}{H}{\R j}\,,
\eqlab{H-R-new}\\
\Scal_{\alpha,ij}(\R) &= \me{\0 i}{S_\alpha}{\R j}\,,
\eqlab{S-R-new}
\end{align}
\eqlab{H-S-R-new}%
\end{subequations}
whereas those of the orbital matrices are
\begin{subequations}
\begin{align}
\Acal_{\alpha,ij}(\R) &= \me{\0 i}{\left( r - R - \tau_j \right)_\alpha}{\R j}\,,
\eqlab{A-R-new}\\
\Bcal_{\alpha,ij}(\R) &= \me{\0 i}{H\left( r - R - \tau_j \right)_\alpha}{\R j}\,,
\eqlab{B-R-new}\\
\Ccal_{\alpha\beta,ij}(\R) &= \me{\0 i}{\left( r - \tau_i \right)_\alpha
                         \left( r - R - \tau_j \right)_\beta}{\R j}\,,
\eqlab{C-R-new}\\
\Dcal_{\alpha\beta,ij}(\R) &= \me{\0 i}{\left( r - \tau_i \right)_\alpha H
                         \left( r - R - \tau_j \right)_\beta}{\R j}\,.
\eqlab{D-R-new}
\end{align}
\eqlab{R-matrices-new}%
\end{subequations}

\Eq{O-k-new} concludes the Wannier interpolation. Once the
Wannier-gauge matrices are known on the fine mesh, they are used to
compute the optical matrix elements in the Hamiltonian gauge via the
interpolation formulas obtained in \sref{optical}. To this end, we
note that the internal velocity $\Vcal\I_\alpha$ is also needed,
particularly to compute the internal parts of the Berry connection
[\eq{A-I}] and of $\Kcal_{\alpha\beta}$ [\eq{K-I}]. To evaluate
$\Vcal\I_\alpha$ from \eq{V-I} one needs to differentiate
$\Hcal\W(\k)$, which can be easily done with the help of \eq{O-k-new},
\beq
\partial_\alpha \Hcal\W_{ij}(\k) =
i\sum_\R\, (R+\tau_j-\tau_i)_\alpha e^{i\k\cdot(\R+\btau_j-\btau_i)}
\Hcal_{ij}(\R)\,.
\eeq
In addition, to evaluate $\Kcal\E_{\alpha \beta}$ from \eq{K-E} one
also needs $\Fcal\W_{\alpha\beta}$. Starting from its definition in
\eq{F-W} and again using \eq{O-k-new}, one finds
\begin{align}
\Fcal\W_{\alpha\beta,ij}(\k) &=
i\sum_\R\, (R+\tau_j-\tau_i)_\alpha e^{i\k\cdot(\R+\btau_j-\btau_i)}
 \Acal_{\beta,ij}(\R)\nn
 &- \left( \alpha \leftrightarrow \beta \right)\,.
\eqlab{F-W-fourier}
\end{align}
Once all the necessary ingredients are assembled,
$\sigma_{\alpha\beta\gamma}$ is computed using \eq{sigma-abc}.

The way we have presented the interpolation of orbital matrices
follows Ref.~\cite{wang-prb06}, except for the inclusion of the
Wannier centers in the Fourier phase factors.  In this formulation,
the real-space orbital matrices contain, in each factor depending on
$\mathbf{r}$, the position of only one of the two Wannier centers
involved, either $\boldsymbol{\tau}_i$ or
$\mathbf{R} + \boldsymbol{\tau}_j$: see \eq{R-matrices-new}.

Recently, an improved procedure was
introduced~\cite{lihm-unpublished}, whereby the Wannier centers
$\boldsymbol{\tau}_i$ and $\mathbf{R} + \boldsymbol{\tau}_j$ are
replaced by their average
$(\mathbf{R} + \boldsymbol{\tau}_i + \boldsymbol{\tau}_j)/2$.  This
formulation preserves symmetries better, has a faster convergence with
respect to the size of the coarse $\mathbf{q}$ grid, and preserves the
generalized hermiticity conditions in \eqs{A-C-D}{B-dagger}, which in
the standard approach~\cite{wang-prb06} are only satisfied in the
limit of a dense $\q$ grid. Our implementation uses this revised
procedure, which we discuss in \aref{jml-scheme} for completeness.

\section{Results}
\seclab{tests}

In this section we present our results for the optical activity of
selected materials, and compare with previous works. The workflow is
summarized in \fref{fig3}, and a more complete schematic is provided
in the Supplementary Material.

We start with an \textit{ab initio} ground-state calculation performed
using \textsc{Quantum Espresso}, which provides the Kohn-Sham Bloch
eigenstates and energy eigenvalues on the coarse $\q$ mesh. This is
followed by the wannierization of the low-energy \textit{ab initio}
electronic structure. In this step, the \textit{ab initio} matrices
listed in \Tref{matrices-new} are computed with \textsc{pw2wannier90},
and the wannierization matrices $\tilde W(\q)$ in \eq{W-matrix} are
subsequently generated by \textsc{Wannier90}.  The final step --~the
Wannier interpolation of $\sigma_{\alpha\beta\gamma}(\w)$~-- is
carried out using the \textsc{WannierBerri}
code~\cite{tsirkin-npjcm21}, where the methodology presented in this
work was implemented.

All the materials studied are either semiconductors or insulators,
thus only the Fermi-sea terms in \eq{sigma-abc} and in \fref{fig1}
contribute at zero electronic temperature. Moreover, since all the
materials studied are nonmagnetic, the spin contribution to
$\sigma_{\alpha \beta\gamma}$ vanishes in the absence of SOC. With
SOC, the spin terms are found to be negligible.

\begin{figure*}[t]
\centering\includegraphics[width=1.5\columnwidth]{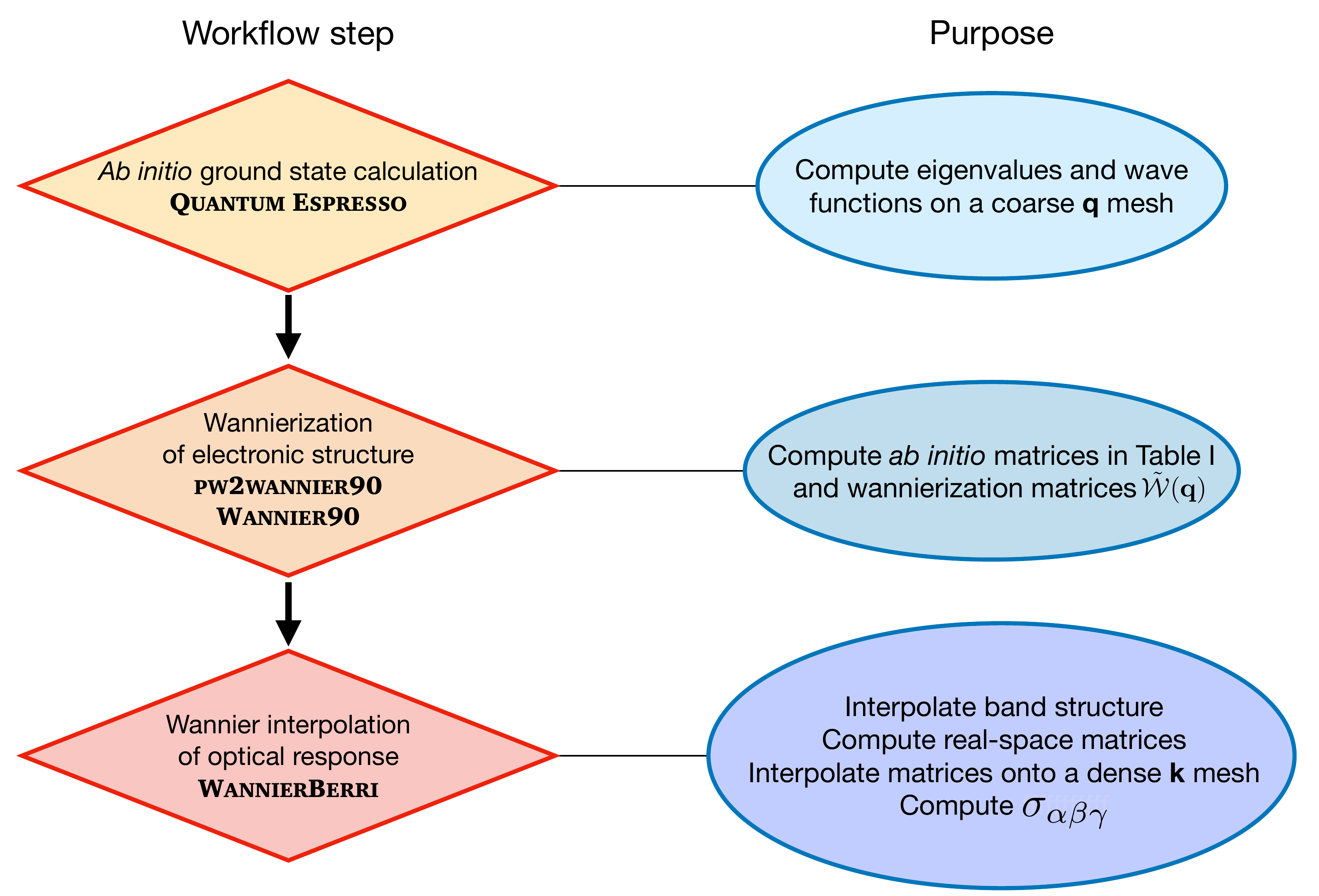}
\caption{Workflow of the computational approach used for evaluating
  the spatially dispersive optical conductivity.}
\figlab{fig3}
\end{figure*}

\subsection{Computational details}
\seclab{comp-det}

\subsubsection{Density-functional theory calculations}

Density-functional theory (DFT) electronic-structure calculations are
performed for wurtzite GaN, trigonal Se and Te, and $\alpha$-SiO$_2$
using \textsc{Quantum Espresso}.  The local density approximation
(LDA) is employed for GaN, and the generalized gradient approximation
(GGA) for Se and $\alpha$-SiO$_2$.  LDA calculations use the
Perdew-Zunger~\cite{perdew-prb81} parametrization of the
exchange-correlation energy, while GGA calculations use the
Perdew-Burke-Ernzerhof (PBE) scheme~\cite{perdew-prl96}. In the case
of Te, exact Fock exchange is added on top of the PBE
exchange-correlation energy, adopting a
Heyd-Scuseria-Ernzerhof~\cite{heyd-jcp03} hybrid functional.

The ions are described via scalar-relativistic and fully-relativistic
optimized norm-conserving Vanderbilt
pseudopotentials~\cite{hamann-prb13}. Specifically, for GaN, Se, and
$\alpha$-SiO$_2$ we use scalar-relativistic pseudopotentials (valence
configurations: $3d^{10} 4s^2 4p^1$ for Ga, $2s^2 2p^3$ for N,
$3d^{10} 4s^2 4p^4$ for Se, $3s^2 3p^2$ for Si, and $2s^2 2p^4$ for
O), whereas for Te we use both scalar-relativistic and
full-relativistic pseudopotentials (valence configuration
$4d^{10} 5s^2 5p^4$). The pseudo-wavefunctions are expanded in
plane-wave basis sets with kinetic energy cut-offs of 80 Ry for GaN,
60 Ry for Te, and 100 Ry for both Se and $\alpha$-SiO$_2$.

The BZ is sampled using $\Gamma$-centered Monkhorst-Pack
meshes~\cite{monkhorst-prb76}. The mesh size is $6 \times 6 \times 4$
for GaN, $6 \times 6 \times 6$ for Te and $\alpha$-SiO$_2$, and
$8 \times 8 \times 8$ for Se.

\subsubsection{Construction of  Wannier functions}
\seclab{wannier-constr}

For the construction of Wannier functions using \textsc{Wannier90}, we
choose atom-centered $s$ and $p$ trial orbitals for each atomic
species in each material studied; the one exception is oxygen in
$\alpha$-SiO$_2$, for which we use $p$ orbitals only.  These choices
correspond to the dominant orbital character of the highest occupied
and lowest unoccupied bands. The disentanglement
procedure~\cite{souza-prb01} is used when necessary. Details about the
ranges of the frozen and disentanglement energy windows adopted for
each material are provided in the Supplementary Material.

The resulting Wannier functions reproduce the low-energy \textit{ab
  initio} electronic structure around the Fermi level.  As discussed
in \sref{optical}, this allows to compute the absorptive part of the
low-frequency optical conductivity $\sigma_{\alpha\beta\gamma}(\w)$
without any band-truncation errors. On the other hand, the reactive
part of the calculated optical response does suffer from such errors.

The Wannier functions generated by \textsc{Wannier90} slightly break
crystal symmetries, because symmetry is not enforced in the
wannierization algorithm.  To ensure that the calculated
$\sigma_{\alpha\beta\gamma}$ tensor complies with the point-group
symmetry, two seperate symmetrization procedures are applied during
Wannier interpolation, as described in \aref{symm}.

When studying in \aref{trunc-error} the effect of band-truncation
errors on the rotatory power of Se, we need to wannierize a number of
empty states that is larger than those accessible by using only $s$
and $p$ trial orbitals. To that end, we exploit the ``Selected Columns
of the Density Matrix'' (SCDM) method \cite{damle-jctc15,
  damle-mms18}, as implemented in
\textsc{Wannier90}~\cite{vitale-npjcm20}. This method is used as a
guide to identify the centers of the Wannier functions that span the
additional conduction bands. We interpret those Wannier functions as
interstitial orbitals, for which $s$-like trial orbitals are used. A
more detailed explanation is provided in the Supplementary Material.

\subsection{Polar optical activity of wurtzite GaN}

Wurtzite is a polar structure with point group $6mm$, whose gyration
tensor is purely antisymmetric and hence shows neither optical
rotation nor circular dichroism (\sref{OSD}). With the polar axis
along $z$, the nonzero tensor components are $G_{xy}=-G_{yx} (=G_z)$,
which map onto $d_z$ via \eq{polar-NOA}.  We evaluate as a function of
frequency its absorptive (imaginary) part, and compare with a recent
Kubo-formula calculation~\cite{wang-prb23} that used a sum-over-states
approach (\sref{sos}).

\Fref{fig4} shows the calculated spectrum from 1 to 3~eV (solid black
line), broken down into E1, M1, and E2 terms (solid colored
lines). This decomposition, depicted schematically in \fref{fig1} and
detailed in \aref{multipoles}, is somewhat different from that in
Ref.~\cite{wang-prb23}, which is also shown in \fref{fig4} as
calculated in the present work (dashed lines).  The spectrum in
\fref{fig4}(a) is obtained using the same broadening parameter
$\eta=0.1$~eV as in Ref.~\cite{wang-prb23}, with an interpolation mesh
of $198 \times 198 \times 132$ $\mathbf{k}$ points.  Both the relative
peak positions and their heights are in good quantitative agreement
with that work.

The M1 and E2 terms cancel to a large extent in \fref{fig4}(a), so
that $\Im\, d_z(\w)$ follows closely the E1
term~\cite{wang-prb23}. However, E2 has a sharper rise compared to M1,
which makes the cancellation incomplete for frequencies right below
the gap, pushing the resonance peak to lower frequencies by about
$0.05$ eV. This can be better appreciated by reducing the broadening
parameter to $\eta = 0.04$~eV and sampling the BZ on a denser
$300 \times 300 \times 200$ mesh to ensure convergence.  The resulting
spectrum in \fref{fig4}(b) reveals an additional E2 feature close to
the band-gap frequency.
\color{black}

\begin{figure}[t]
\centering\includegraphics[width=0.9\columnwidth]{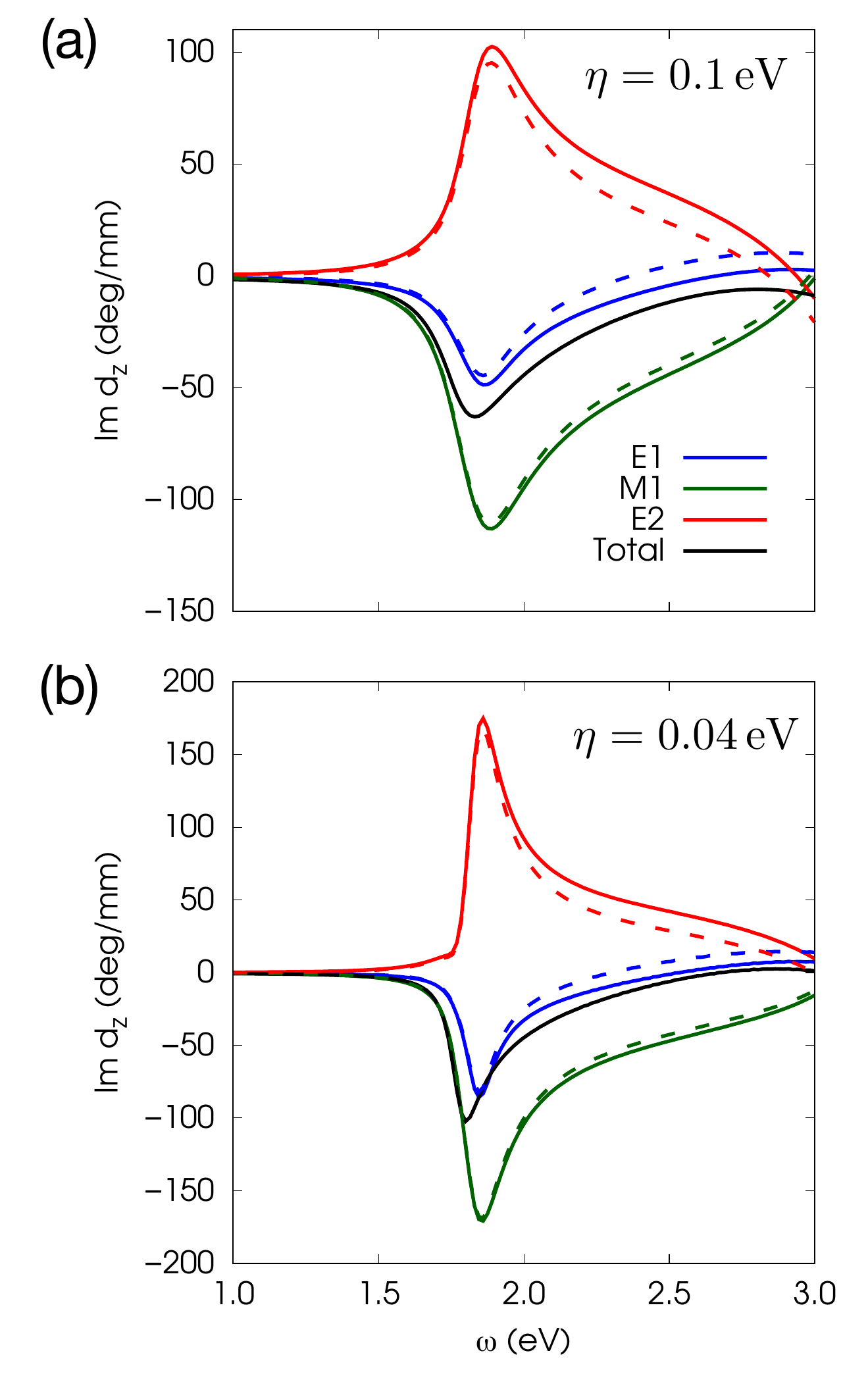}
\caption{Absorptive part of the polar optical activity spectrum of
  GaN, calculated with two different broadening parameters $\eta$ and
  broken down into E1, M1, and E2 contributions.  Dashed lines indicate
  the decomposition used in Ref.~\cite{wang-prb23} (where the E1 terms
  were denoted as ``band dispersion''), while solid lines indicate the
  revised decomposition presented in this work.}
\figlab{fig4}
\end{figure}

\subsection{\texorpdfstring{Rotatory power of  tellurium,  selenium, and $\alpha$-quartz}{Te, Se, alpha-quartz}
}
\seclab{RP}

Recently, the rotatory power of several chiral materials was computed
using both Kubo~\cite{wang-prb23} and DFPT~\cite{zabalo-prl23}
methods.  The latter is at present restricted to the low-frequency
limit, but it includes local-field effects that are instead neglected
in the former. As mentioned earlier, such effects were found to play a
substantial role in the optical rotation of some
materials~\cite{jonsson-prl96,zabalo-prl23}.

Here, we compute the rotatory power of trigonal Te, trigonal Se, and
$\alpha$-SiO$_2$.  The results for Te are compared with the Kubo
calculations of Ref.~\cite{wang-prb23} and with experimental
measurements~\cite{ades-josa75, fukuda-pssb75}. In the case of Se and
$\alpha$-SiO$_2$, we compare with the DFPT results of
Ref.~\cite{zabalo-prl23}.

All three materials are chiral, with their left- and right-handed
enantiomorphs crystallizing in the $P3_221$ and $P3_121$ space groups,
respectively (point group $32$). The gyration tensor is diagonal, with
$G_{xx} = G_{yy} \ne G_{zz}$.  According to \eq{rot-ellip}, the
rotatory power for light propagating along the trigonal $z$ axis is
given by
\beq
\rho(\omega) = \frac{\omega^2}{2c^2} \Re\, G_{zz} (\omega).
\eeq
The calculations reported below were carried out for the
left-handed enantiomorphs.

\subsubsection{Tellurium}

\Fref{fig5}(a) shows the rotatory-power spectrum of tellurium,
calculated both with and without SOC using a
$200 \times 200 \times 150$ interpolation mesh, together with
experimental data from Refs.~\cite{ades-josa75,fukuda-pssb75}.  Due to
the large atomic number of Te, SOC plays a substantial role. The
spectrum computed without SOC overestimates the experimental one,
while that obtained with SOC underestimates it.

\begin{figure}
\centering\includegraphics[width=\columnwidth]{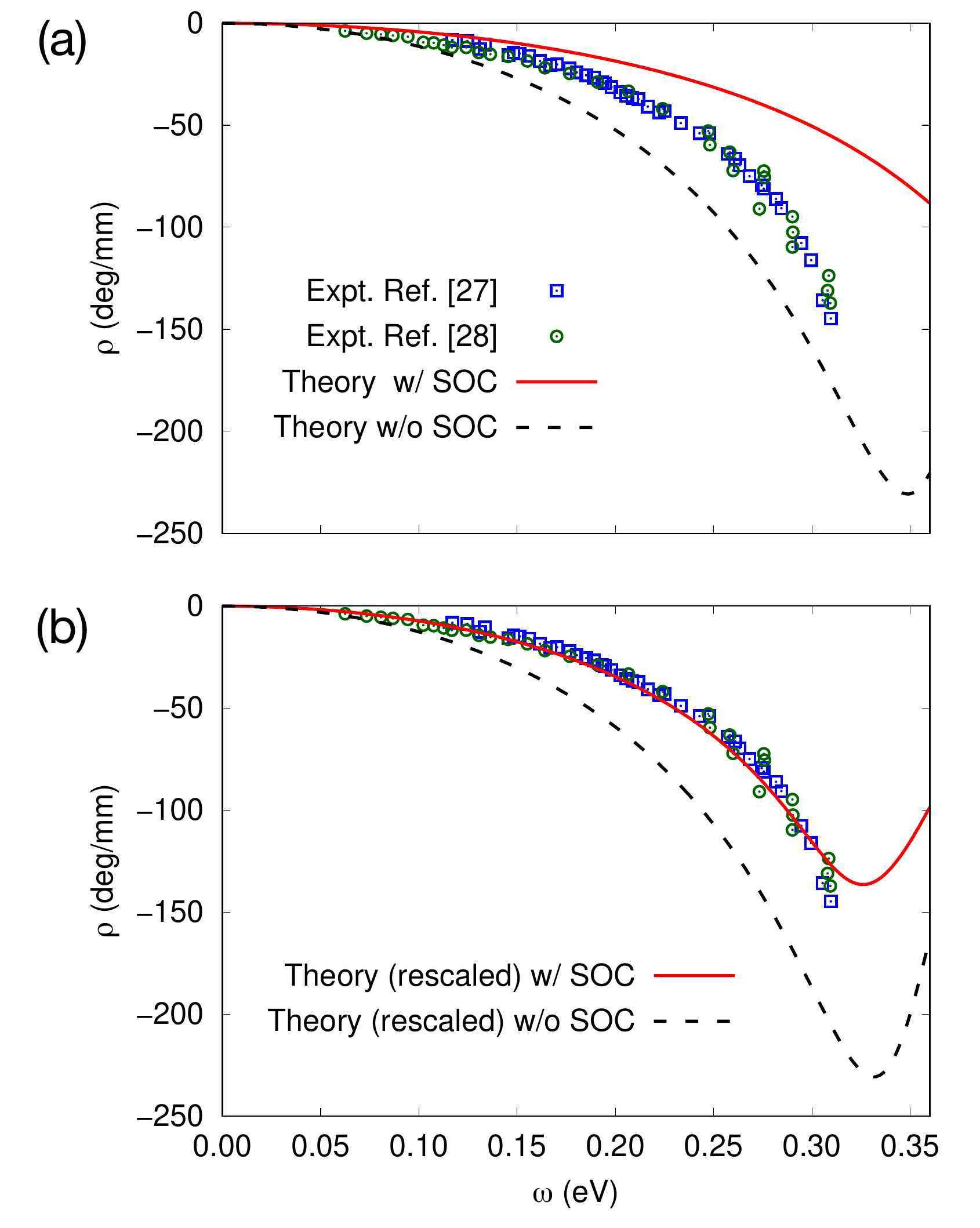}
\caption{Calculated and experimental rotatory power of left-handed
  trigonal Te for light propagating along the trigonal axis. (a) Raw
  calculated data obtained with a broadening parameter of
  $\eta = 25$~meV, compared to experimental data. (b) Same data as in
  (a), with a rescaled frequency axis for the theoretical results, to
  align the \textit{ab initio} energy gap with the experimental one.}
\figlab{fig5}
\end{figure}

The band gap is important in determining $\rho(\w)$. Since reactive
and absorptive responses satisfy Kramers-Kr\"onig relations,
$\rho(\w)$ shows a resonance at the gap frequency where circular
dichroism sets in.  This, in turn, affects the rate at which
$\rho(\w)$ increases at subgap frequencies.  An accurate band gap is
therefore important to obtain a spectrum in good agreement with
experiment. For this reason, our calculations were performed with a
hybrid functional (see \sref{comp-det}), which typically provides
improved band gaps compared to LDA and PBE. This is particularly true
for Te, where hybrid-functional calculations with (without) SOC give a
band gap of 0.42~eV (0.342~eV), to be compared with the experimental
value of 0.323~eV~\cite{anzin-pssa77}. Instead, the PBE band gap is
0.015 eV~\cite{wang-prb23}, 20 times smaller than experiment.

Even though hybrid functionals provide a substantially improved
electronic structure, the band gap obtained with SOC is still $0.1$ eV
larger than experiment. This pushes up the $\rho(\w)$ resonance
resulting in a slower increase at subgap frequencies, which may
explain the underestimation of the experimental spectrum. To test this
conjecture, in \fref{fig5}(b) we renormalize the frequency axis to
align the resonance frequency with the experimental band gap. This
{\it ad hoc} correction substantially improves the results obtained
with SOC.

\subsubsection{\texorpdfstring{Selenium and $\alpha$-quartz}{Se and alpha-quartz}}
\seclab{se-quartz}

The rotatory power of both elemental Se (isostructural to Te) and
$\alpha$-SiO$_2$ is strongly affected by local-field
effects~\cite{zabalo-prl23,jonsson-prl96}.  The $\omega\rightarrow 0$
limit of the quantity $\bar{\rho} (\w) = \rho(\w)/\w^2$ introduced in
\eq{rho-bar} was computed in Ref.~\cite{zabalo-prl23} using a DFPT
approach.  It was found that the removal of local-field effects
reduces $\bar{\rho} (0)$ by factors of 4 and 7 for Se and
$\alpha$-SiO$_2$, respectively.

\begin{figure}
\centering\includegraphics[width=\columnwidth]{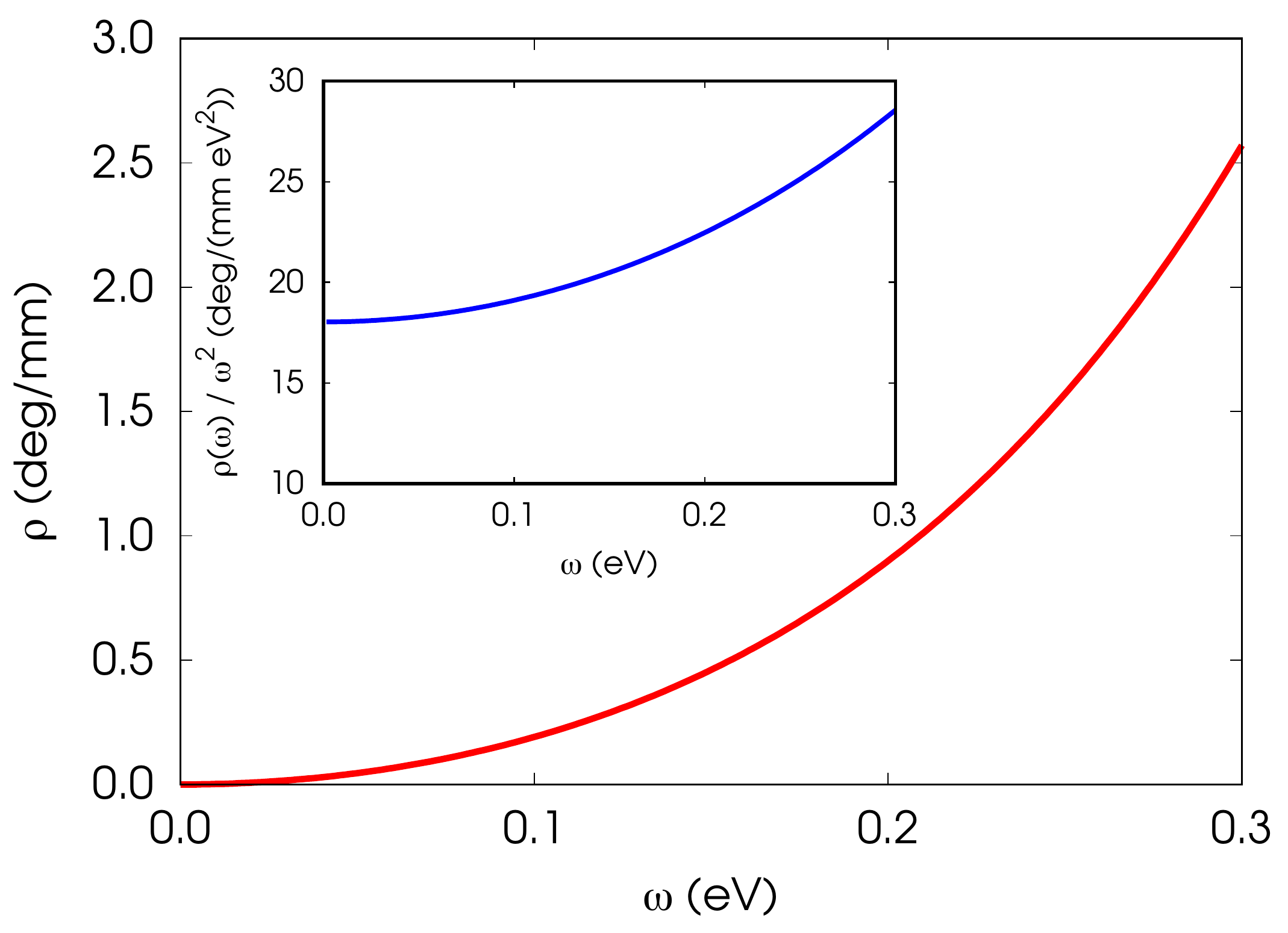}
\caption{Calculated rotatory power of trigonal Se for light
propagating along the trigonal axis. Inset: rotatory power divided
by $\omega^2$, which tends to a nonzero constant at low
frequencies.}
\figlab{fig6}
\end{figure}

\Fref{fig6} shows our results for the rotatory power of Se, obtained
with a $50 \times 50 \times 36$ interpolation mesh, much coarser than
for Te due to the larger band gap ($\approx 0.96$~eV).  The quantity
$\bar{\rho}(\w)$ is plotted in the inset, and its value at $\w=0$ is
in agreement with that obtained in Ref.~\cite{zabalo-prl23} without
self-consistent fields, see \Tref{local-field}.\footnote{In
  Ref.~\cite{zabalo-prl23}, the quantity $\bar\rho(0)$ was calculated
  for right-handed crystals. To compare with our results for
  left-handed crystals, in \Tref{local-field} we flip the sign of the
  $\bar\rho(0)$ values reported in that work.}  To calculate the
rotatory power of $\alpha$-SiO$_2$, we use a
$100 \times 100 \times 100$ interpolation grid. Also in this case, the
value of $\bar\rho(0)$ is in reasonable agreement with the one
obtained in Ref.~\cite{zabalo-prl23} without self-consistent terms.
\def\phm{\phantom{-}}
\begin{table}
\centering
\begin{ruledtabular}
\caption{Static limit of
  $\bar{\rho} (\omega) = \rho(\omega) / \omega^2$ in comparison with
  the results of Ref.~\cite{zabalo-prl23} without and with local
  fields (LF).}
\begin{tabular}{lccc}
& This work & 
  Ref.~\cite{zabalo-prl23} w/o LF &
  Ref.~\cite{zabalo-prl23} w/ LF \\
\hline
Se & $\phantom{+} 18.0$ & $\phantom{+} 17.8$ & $\phantom{+} 74.5$ \\
$\alpha$-SiO$_2$ & $-1.1$ & $- 0.7$ & $- 4.9$
\end{tabular}
\end{ruledtabular}
\label{tab:local-field}
\end{table}

Since optical rotation is a reactive response, its calculation with
our methodology is affected by band-truncation errors (see
\sref{optical}).  It is therefore important to assess the convergence
of $\rho(\w)$ with respect to the number of conduction bands spanned
by the Wannier basis. In \aref{trunc-error} we present such a
convergence test for Se.

\section{Summary and outlook}
\seclab{conclusions}

Wannier interpolation is a fast and accurate technique for computing
$\k$-dependent quantities across the BZ, using Wannier functions as a
tight-binding-like representation of the low-energy \textit{ab initio}
electronic structure. The method is widely used to evaluate the
$k$-space Berry curvature and its BZ integral -- the intrinsic
anomalous Hall conductivity~\cite{wang-prb06}. Other applications
include computing $\sigma_{\alpha\beta}(\w)$ (the optical conductivity
in the E1 approximation)~\cite{yates-prb07,assmann-cpc2016}, and the
ground-state orbital magnetization~\cite{lopez-prb12,hanke-prb16}. For
a comprehensive review, see Ref.~\cite{marrazzo-rmp24}.

In this work, we developed and implemented a Wannier interpolation
scheme to evaluate the Kubo-Greenwood formula~\eqref{eq:sigma-abc} for
$\sigma_{\alpha\beta\gamma}(\w)$, the optical conductivity of bulk
crystals at first order in the wavevector of light. This quantity
contains not only E1 matrix elements as in the case of the Berry
curvature and $\sigma_{\alpha\beta}(\w)$, but also M1 and E2 matrix
elements. The latter are trickier to evaluate, as they involve virtual
interband transitions; by expressing them as $k$ derivatives, the
intermediate band summations were circumvented, avoiding a source of
truncation errors that was present in a prior \textit{ab initio}
implementation~\cite{wang-prb23}.

The proposed method is particularly well suited to compute the
absorptive part of $\sigma_{\alpha\beta\gamma}(\w)$ (e.g., circular
dichroism) in the energy range covered by the Wannier basis, which is
done without introducing any band-truncation errors. On the other
hand, such errors do occur for the reactive response (e.g., rotatory
power), which is effectively evaluated as a truncated Kramers-Kr\"onig
transform of the absorptive response. Methods based on
DFPT~\cite{zabalo-prl23} seem better suited to obtain the reactive
response with high accuracy, as they completely avoid band summations
and automatically account for local-field effects. Nevertheless, our
tests indicate that the proposed method can give semiquantitative
results for the rotatory power with a manageable number of Wannier
functions.

The method was applied to the optical activity of insulators and
semiconductors at zero temperature, for which the Fermi-surface terms
in \eq{sigma-abc} --~both intraband and interband~-- do not play any
role. In metals and doped semiconductors such terms do contribute to
the optical activity, yielding in the static limit the kinetic
magnetoelectric effect (the bulk version of the Edelstein effect at
surfaces)~\cite{pozo-scipost23}.  The low-frequency optical activity
of conductors arising from those terms is an interesting topic for
future studies employing the present method. Another promising
direction is to consider the time-odd (symmetric) part of
$\sigma_{\alpha\beta\gamma}(\w)$; that part becomes nonzero in
acentric antiferromagnetic materials, giving rise to characteristic
optical and static responses.

\section*{Acknowledgments}

The authors thank Massimiliano Stengel, Seung-Ju Hong, and Cheol-Hwan
Park for valuable discussions, and Xiaoming Wang for sharing
computational data relative to Ref.~\cite{wang-prb23}. This work has
been supported by the NSF Grant No. DMR-2421895. AU acknowledges
support from the Abrahams Postdoctoral Fellowship of the Center for
Materials Theory at Rutgers University. IS, OPO and SST acknowledge
support from Grant No. PID2021-129035NB-I00 funded by
MCIN/AEI/10.13039/501100011033 and by ERDF/EU, and from the IKUR
Strategy under the collaboration agreement between the Ikerbasque
Foundation and the Material Physics Center on behalf of the Department
of Universities, Science and Innovation of the Basque Government.
Computational resources were provided by the Beowulf cluster at the
Department of Physics and Astronomy at Rutgers University.

\section*{Data Availability}

The data that support the findings of this article are openly available \cite{urru_2025_15236403}.

\appendix

\section{Multipole-like expansion of the optical matrix elements in
  crystals}
\seclab{multipoles}

In this appendix we expand to first order in $\q$ the combination of
optical matrix elements entering the Kubo formula for the optical
conductivity $\sigma_{\alpha\beta}(\w,\q)$, and express the result in
a multipole-like form.  We leave out spin terms which are
straightforward, and focus on the more subtle orbital terms.

\subsection{Finite systems}

Before considering bulk crystals, let us work out the standard
multipole expansion for finite systems.  We start from the
paramagnetic current operator written in Fourier space, and expand it
to first order in $\q$,
\begin{align}
 J_\alpha(\q) &= -\frac{|e|}{2}
\left(
 v_\alpha e^{-i{\bf q} \cdot  {\bf r}} +
e^{-i{\bf q} \cdot  {\bf r}}  v_\alpha
\right)\nn
&\simeq -|e| v_\alpha + \frac{i|e|}{2} q_\beta
\left(  v_\alpha  r_\beta +  r_\beta  v_\alpha \right)\,.
\end{align}
Next, we take its matrix elements in the eigenstate basis,
\begin{align}
\me{l}{ J_\alpha(\q)}{n} &\simeq -|e| \me{l}{ v_\alpha}{n} +
\frac{i|e|}{2}q_\beta\sum_p\,
\left(
\me{l}{ v_\alpha}{p} \me{p}{ r_\beta}{n} \right. \nonumber \\ & \left. +
\me{l}{ r_\beta}{p} \me{p}{ v_\alpha}{n}
\right)\nonumber \\
&= i|e|\w_{nl} \me{l}{ r_\alpha}{n} + \frac{|e|}{2} q_\beta \sum_p\,
\left(
\w_{pl} \me{l}{ r_\alpha}{p} \me{p}{ r_\beta}{n} \right. \nonumber \\ & \left. +
\w_{np} \me{l}{ r_\beta}{p} \me{p}{ r_\alpha}{n}
\right)\,,
\eqlab{J-exp}
\end{align}
where the second equality follows from the relation
\beq
\Vcal_{\alpha,ln} \equiv
\me{l}{ v_\alpha}{n} =
i\w_{ln} \me{l}{r_\alpha}{n}\,.
\eqlab{V-at}
\eeq
Writing the antisymmetric part of the $\q$-linear term according to
the first line of \eq{J-exp} and the symmetric part according to the
second, one finds
\begin{align}
\me{l}{ J_\alpha(\q)}{n} &\simeq -i\w_{nl} d_{\alpha,ln}\nn
&+i\left(  m_{\alpha\beta,ln} +
\frac{i}{2} \w_{nl}q_{\alpha\beta,ln} \right) q_\beta\,,
\eqlab{J-mult}
\end{align}
where
\begin{align}
d_{\alpha,ln} &= -|e|\me{l}{ r_\alpha}{n}\,,
\eqlab{d-fin}\\
m_{\alpha\beta,ln} &= -\frac{|e|}{2}\me{l}{ r_\alpha  v_\beta - 
  r_\beta  v_\alpha}{n}
\,,
\eqlab{m-fin}\\
q_{\alpha\beta,ln} &= -|e|\me{l}{ r_\alpha  r_\beta}{n}
\eqlab{q-fin}
\end{align}
are the E1, M1, and E2 matrix elements, respectively.  \Eq{J-mult} is
analogous to the Fourier-transformed multipole expansion of a current
distribution~\cite{melrose-91,raab-05,malashevich-prb10},
\beq
J_\alpha(\w,\q) \simeq -i\w d_\alpha(\w) +
i\left[ m_{\alpha\beta}(\w) +
\frac{i}{2}\w q_{\alpha\beta}(\w) \right] q_\beta\,.
\eqlab{J-mult-phenom}
\eeq

The matrix elements of the current operator enter the Kubo formula in
the combination
\beq
\me{n}{ J_\alpha(\q)}{l} \me{l}{ J_\beta(-\q)}{n}
\eqlab{JJ}
\eeq
which, in view of \eq{J-mult}, can be written at first order in
$\mathbf{q}$ as
\begin{align}
\me{n}{ J_\alpha(\q)}{l}& \me{l}{ J_\beta(-\q)}{n} \simeq
e^2{\mathcal V}_{\alpha,nl} {\mathcal V}_{\beta,ln}\nn
&+ ie^2
\left(
{\mathcal V}_{\alpha,nl}{\mathcal T}_{\beta\gamma,ln} -
{\mathcal V}_{\beta,ln}{\mathcal T}_{\alpha\gamma,nl}
\right)
q_\gamma\,,
\eqlab{JJ-exp}
\end{align}
where the velocity matrix $\Vcal_\alpha$ is related to the E1 matrix
by \eq{V-at}, and the antisymmetric and symmetric parts of
$\Tcal_{\alpha\beta}$ give the M1 and E2 matrices, respectively. All
in all, we have
\begin{subequations}
\begin{align}
\wln d_{\alpha,ln} &= i|e| \Vcal_{\alpha,ln}\,,
\eqlab{d}\\
m_{\alpha\beta,ln} &= \frac{|e|}{2}
\left( {\mathcal T}_{\alpha\beta,ln} - {\mathcal T}_{\beta\alpha,ln} \right)\,,
\eqlab{m}\\
\wln q_{\alpha\beta,ln} &= i|e|
\left( {\mathcal T}_{\alpha\beta,ln} + {\mathcal T}_{\beta\alpha,ln} \right)\,.
\eqlab{q}
\end{align}
\eqlab{d-m-q}
\end{subequations}

\subsection{Bulk crystals}

When the expansion~\eqref{eq:JJ-exp} is carried out in the Bloch
eigenstate basis, one faces the problem that $\k$ derivatives of
cell-periodic Bloch states do not transform covariantly under gauge
transformations.  By introducing covariant derivatives each term in
the expansion becomes gauge covariant~\cite{pozo-scipost23}, and one
finds that the matrix $\Tcal_{\alpha\beta}$ appearing in \eq{JJ-exp}
is given by \eq{T}.\footnote{The tensor $B_{\alpha\beta}$ in
  Ref.~\cite{pozo-scipost23} is equal to $\Tcal_{\alpha\beta}$ in
  \eq{T} without the $v_\alpha A_\beta$ term in \eq{K} for
  $\Kcal_{\alpha\beta}$. That accidental omission explains the missing
  terms in the expressions for $\mm_{\alpha\beta}$ and
  $\qq_{\alpha\beta}$ in Ref.~\cite{pozo-scipost23}: see
  \eqr{m-bloch}{q-sos} below and related discussion.  Nevertheless,
  those terms were correctly included, although classified as E1
  (``band dispersive'') contributions, in the expression for
  $\sigma_{\alpha\beta\gamma}(\w)$ given in that work, which is
  consistent with \eq{sigma-abc}.}  Using \eq{K} for
$\Kcal_{\alpha\beta}$ and comparing with \eqs{A-sos}{d-m-q} leads to
\begin{align}
\dd_{\alpha,ln} &= -i|e| \ip{u_l}{D_\alpha u_n} = -|e|A_{\alpha,ln}\,,
\eqlab{d-bloch}
\\
\mm_{\alpha\beta,ln} &= \frac{|e|}{2i\hbar}
\me{D_\alpha u_l}{ H - \bar \epsilon_{ln}}{D_\beta u_n}\nn
&+\frac{|e|}{2} \bar v_{\alpha,ln} A_{\beta,ln} -
(\alpha \leftrightarrow \beta)\,,
\eqlab{m-bloch}
\\
\wln \qq_{\alpha\beta,ln} &= -\frac{|e|}{2} \w_{ln}
\ip{D_\alpha u_l}{D_\beta u_n}\nn
&+i|e| \bar v_{\alpha,ln} A_{\beta,ln} +
(\alpha \leftrightarrow \beta)\,,
\eqlab{q-bloch}
\end{align}
with $\bar \epsilon_{ln} = \left( \epsilon_l + \epsilon_n \right)/2$
and $\bar v_{\alpha,ln}$ given by \eq{v-bar}.  We denote these
multipole-like bulk tensors as $\dd$, $\mm$, and $\qq$ to distinguish
them from the standard multipoles $d$, $m$, and $q$ in
\eqr{d-fin}{q-fin}, which become ill-defined for extended systems.

The expressions above for $\mm_{\alpha\beta}$ and $\qq_{\alpha\beta}$
differ from the ones in Ref.~\cite{pozo-scipost23} by the presence of
the off-diagonal $\bar v_\alpha A_\beta$ terms.  If \eq{K-sos} is used
for $\Kcal_{\alpha\beta}$ instead of \eq{K}, the following
sum-over-states formulas are obtained,
\begin{align}
\mm_{\alpha\beta,ln} &= \frac{i|e|}{4} \sum_{p\not= l}\,
\frac{\Vcal_{\alpha,lp} \Vcal_{\beta,pn} - (\alpha\leftrightarrow\beta)}{\w_{lp}}\nn
&-\frac{i|e|}{4} \sum_{p\not= n}\,
\frac{\Vcal_{\alpha,lp} \Vcal_{\beta,pn} - (\alpha\leftrightarrow\beta)}{\w_{pn}}
\eqlab{m-sos}
\end{align}
and
\begin{align}
\wln \qq_{\alpha\beta,ln} &=
\frac{|e|}{2} \sum_{p\not= l}\,
\frac{\Vcal_{\alpha,lp} \Vcal_{\beta,pn} +
(\alpha \leftrightarrow \beta)}{\w_{lp}}\nn
&+
\frac{|e|}{2} \sum_{p\not= n}\,
\frac{\Vcal_{\alpha,lp} \Vcal_{\beta,pn} +
\alpha \leftrightarrow \beta)}{\w_{pn}} 
\eqlab{q-sos}\,.
\end{align}
In both equations, only the terms with vanishing denominators are
excluded. Instead, in the sum-over-state formulas of
Ref.~\cite{pozo-scipost23} an additional nonsingular term is excluded
from each summation when $l\not=n$.\footnote{To compare \eq{q-sos}
with Eq.~(24c) of Ref.~\cite{pozo-scipost23}, multiply the latter by
$\wln$ and then use $\wln/(\w_{lp} \w_{pn})=1/\w_{lp} + 1/\w_{pn}$.}

Interband orbital moments are needed to evaluate the orbital Hall
conductivity, and the expression used in the recent
literature~\cite{pezo-prb22,gobel-prl24} agrees with the
sum-over-states formula of Ref.~\cite{pozo-scipost23}.  The impact on
the orbital Hall conductivity of the additional off-diagonal term in
\eqs{m-bloch}{m-sos} remains to be studied. As it contains the band
velocity, that term is ``itinerant'' (it vanishes for crystals
composed of nonoverlapping units); its contribution should therefore
be enhanced in crystals with strongly dispersive bands.

In closing, we note that a different expression for the interband
orbital moment has been proposed within the semiclassical wavepacket
formalism~\cite{gao-prl19}.  That expression is non-Hermitian, and its
Hermitian part almost agrees with \eq{m-bloch}, the only difference
being that the itinerant $\bar v_\alpha A_\beta$ term is twice as
large.  Upon integrating over the BZ and performing an integration by
parts, that term becomes an interband version of the Berry-curvature
contribution to the ground-state orbital magnetization of a non-Chern
insulator, which reads~\cite{xiao-prl05}
\beq
M_{\alpha\beta} =
\sum_n^{\rm occ} \int [d\k]
\left(
\mm_{\alpha\beta,nn}
-\frac{|e|}{\hbar}\epsilon_n\Fcal_{\alpha\beta,nn}
\right)\,,
\eeq
with $\Fcal_{\alpha\beta}$ given by \eq{Fcal} and $\mm_{\alpha\beta}$
by \eq{m-bloch} ($\Fcal_{\alpha\beta,nn}$ and $\mm_{\alpha\beta,nn}$
are the Berry curvature and the intrinsic orbital moment of a Bloch
state, respectively).

\section{\texorpdfstring{Derivation of \eq{K-decomp} for $\Kcal\H_{\alpha \beta}$}{K-der}}
\seclab{K-wann-deriv}

We want to evaluate by Wannier interpolation the matrix
$\Kcal_{\alpha\beta}$ defined by \eq{K}. Denoting its first term as
$(1/i\hbar)Z_{\alpha\beta}$ and expanding the covariant derivatives
according to \eq{cov-der-H}, we find
\begin{widetext}
\begin{align}
Z\H_{\alpha\beta,ln} &= \sum_{ij}\,
\Ucal^\dagger_{li}
\me{\partial_\alpha u\W_i}{H-\epsilon_l}{\partial_\beta u\W_j} \Ucal_{jn}
+i\sum_i\,\Ucal^\dagger_{li}
\me{\partial_\alpha u\W_i}{H-\epsilon_l}{u\W_n} a\E_{\beta,n}\nn
&-i\sum_{pj}\, \Ucal^\dagger_{lj}
\me{\partial_\alpha u\W_j}{H-\epsilon_l}{u\H_p} A\I_{\beta,pn}
-i\sum_j\, a\E_{\alpha,l}
\me{u\H_l}{H-\epsilon_l}{\partial_\beta u\W_j}\Ucal_{jn}\nn
&+i\sum_{pj}\,A\I_{\alpha,lp}
\me{u\H_p}{H-\epsilon_l}{\partial_\beta u\W_j}\Ucal_{jn}
+(\epsilon_l-\epsilon_n)A\I_{\alpha,ln}
  a\E_{\beta,n}
\nn
&
+\sum_p\,(\epsilon_p-\epsilon_l)A\I_{\alpha,lp}A\I_{\beta,pn}
\,,
\eqlab{Z-intp1}
\end{align}
\end{widetext}
where \eq{H-H-b} was used to eliminate two terms and to simplify two
other terms.

Next, use \eq{u} for the Hamiltonian-gauge Bloch states together with
the definitions in \eq{W-matrices} and with the
prescription~\eqref{eq:O-E} for obtaining external objects.
\Eq{Z-intp1} then becomes
\begin{align}
Z\H_{\alpha\beta,ln} = \Dcal\E_{\alpha\beta,ln} -
\epsilon_l \Ccal\E_{\alpha\beta,ln} &-
\left[
\left( \Bcal\E_{\alpha,nl} \right)^* -
\epsilon_l \Acal\E_{\alpha,ln}
\right]
a\E_{\beta,n}\nn
&+\sum_p\,
\left[
\left( \Bcal\E_{\alpha,pl} \right)^* - \epsilon_l \Acal\E_{\alpha,lp}
\right] A\I_{\beta,pn}\nn
&-a\E_{\alpha,l}
\left(
\Bcal\E_{\beta,ln} - \epsilon_l \Acal\E_{\beta,ln}
\right)\nn
&+\sum_p\,A\I_{\alpha,lp}
\left(
\Bcal\E_{\beta,pn} - \epsilon_l \Acal\E_{\beta,pn}
\right)\nn
&+(\epsilon_l-\epsilon_n)A\I_{\alpha,ln}a\E_{\beta,n}\nn
&+\sum_p\,\left(\epsilon_p - \epsilon_l\right)A\I_{\alpha,lp}A\I_{\beta,pn}\,.
\eqlab{Z-interp2}
\end{align}

To proceed further, assume that both $\epsilon_l$ and $\epsilon_n$
--~but not necessarily $\epsilon_p$~-- lie inside the frozen window,
and use
\beq
\Bcal\E_{\alpha,np} = \epsilon_n\Acal\E_{\alpha,np}\,,\;
\forall p\text{ and }
\epsilon_n \in \text{ frozen window}\,.
\eqlab{B-A}
\eeq
This relation, which follows from
$H\ket{u\H_n}=\epsilon_n\ket{u\H_n}$, holds exactly when the
interpolation point $\k$ falls on the {\it ab initio} mesh ($\k=\q$),
and it holds to an excellent approximation otherwise; when
$\epsilon_n$ lies outside the frozen window $\ket{u\H_n}$ ceases to be
an eigenstate of $H$, and as a result \eq{B-A} does not hold even
approximately.

Under the stated assumption, \eq{B-A} can be used in the first and
third lines of \eq{Z-interp2}; it makes the third line vanish, and it
turns the quantity inside square brackets in the first line into
$(\epsilon_n - \epsilon_l) A\E_{\alpha,ln}$, with $A\E_\alpha$ the
off-diagonal part of $\Acal\E_\alpha$.  Separating $Z\H_{\alpha\beta}$
into internal, external and cross terms we find
\begin{align}
Z\I_{\alpha\beta,ln} &=
\sum_p\,\left(\epsilon_p - \epsilon_l\right)A\I_{\alpha,lp}A\I_{\beta,pn}\,,
\eqlab{Z-I}\\
Z\E_{\alpha\beta,ln} &=
\Dcal\E_{\alpha\beta,ln} - \epsilon_l \Ccal\E_{\alpha\beta,ln} +
(\epsilon_l-\epsilon_n)A\E_{\alpha,ln}a\E_{\beta,n}\,,
\eqlab{Z-E}\\
Z\X_{\alpha\beta,ln} &=
\sum_p\,A\I_{\alpha,lp}
\left(
\Bcal\E_{\beta,pn} - \epsilon_l \Acal\E_{\beta,pn}
\right)\nn
&+
\sum_p\,
\left[
\left( \Bcal\E_{\alpha,pl} \right)^* - \epsilon_l \Acal\E_{\alpha,lp}
\right] A\I_{\beta,pn}\nn
&+
(\epsilon_l-\epsilon_n)A\I_{\alpha,ln}a\E_{\beta,n}\,.
\eqlab{Z-X}
\end{align}

The matrix $\Kcal\H_{\alpha\beta}$ defined by \eq{K} can be similarly
decomposed.  The internal part of its first term is given by \eq{Z-I}
divided by $i\hbar$, and to evaluate its second term we use
\eqs{A-decomp}{v} for the interband Berry connection and for the band
velocity, respectively.  Collecting all internal terms we obtain
[compare with \eq{Kcal2}]
\begin{align}
\Kcal\I_{\alpha\beta,ln} &= \frac{1}{i\hbar}
\sum_p\,\left(\epsilon_p - \epsilon_l\right)A\I_{\alpha,lp}A\I_{\beta,pn} +
\Vcal\I_{\alpha,ll} A\I_{\beta,ln}\nn
&=\sum_{p\not= l}\,\Vcal\I_{\alpha,lp} A\I_{\beta,pn} +
\Vcal\I_{\alpha,ll} A\I_{\beta,ln}\nn
&= \sum_p\, \Vcal\I_{\alpha,lp} A\I_{\beta,pn}\,,
\end{align}
which corresponds to \eq{K-I}.

To evaluate the external part of $\Kcal\H_{\alpha\beta}$, use
\eqs{Z-E}{A-decomp} in \eq{K}; after converting band energies and band
velocities to matrix form,
we obtain \eq{K-E}.

Finally, to evaluate the cross term
$\Kcal\X_{\alpha\beta}=(1/i\hbar)Z\X_{\alpha\beta}$ split
$\Acal\E_\beta$ and $\Bcal\E_\beta$ into interband and intraband parts
as
\begin{align}
\Acal\E_{\beta,pn} &= A\E_{\beta,pn} + \delta_{pn} a\E_{\beta,n}\,,\\ 
\Bcal\E_{\beta,pn} &=
B\E_{\beta,pn} + \delta_{pn} \epsilon_n a\E_{\beta,n}
\end{align}
(the second relation is only valid for $\epsilon_n$ inside the frozen
window).  Inserting these expressions in \eq{Z-X} gives
\begin{align}
Z\X_{\alpha\beta,ln} &=
\sum_p\,A\I_{\alpha,lp}
\left(
B\E_{\beta,pn} - \epsilon_l A\E_{\beta,pn}
\right) \nonumber \\ &+
\sum_p\,
\left[
\left( B\E_{\alpha,pl} \right)^* - \epsilon_l A\E_{\alpha,lp}
\right] A\I_{\beta,pn}\nonumber \\ 
&+\epsilon_n A\I_{\alpha,ln} a\E_{\beta,n} -
\epsilon_l A\I_{\alpha,ln} a\E_{\beta,n} \nonumber \\ & +
\epsilon_l a\E_{\alpha,l} A\I_{\beta,ln} -
\epsilon_l a\E_{\alpha,l} A\I_{\beta,ln}+
\left( \epsilon_l - \epsilon_n \right) A\I_{\alpha,ln} a\E_{\beta,n}\,.
\nn
\end{align}
The terms in the third and fourth line cancel out, and dividing the
first two lines by $i\hbar$ we arrive at \eq{K-X}.

\section{Improved scheme for evaluating the real-space orbital matrices}
\seclab{jml-scheme}

Here we describe the improved procedure to evaluate the real-space
matrices \cite{lihm-unpublished}, mentioned in \sref{interpolation},
that is used in our implementation. The idea is to introduce
recentered real-space matrices where the Wannier centers $\btau_i$ and
$\R+\btau_j$ in \eq{R-matrices-new} are both replaced by their average
\beq
\bar \r_{ij}(\R) = \frac{1}{2} \left( \R + \btau_j + \btau_i \right)\,,
\eeq
resulting in
\begin{subequations}
\begin{align}
\bar \Acal_{\alpha,ij}(\R) &=
\me{\0 i}{\left[ r - \bar r_{ij}(\R) \right]_\alpha}{\R j}\,,
\eqlab{A-bar-new}\\
\bar \Bcal_{\alpha,ij}(\R) &=
\me{\0 i}{H\left[ r - \bar r_{ij}(\R) \right]_\alpha}{\R j}\,,
\eqlab{B-bar-new}\\
\bar \Ccal_{\alpha\beta,ij}(\R) &=
\me{\0 i}{\left[ r - \bar r_{ij}(\R) \right]_\alpha
          \left[ r - \bar r_{ij}(\R) \right]_\beta}{\R j}\,,
\eqlab{C-bar-new}\\
\bar \Dcal_{\alpha\beta,ij}(\R) &=
\me{\0 i}{\left[ r - \bar r_{ij}(\R) \right]_\alpha H
          \left[ r - \bar r_{ij}(\R) \right]_\beta}{\R j}\,.
\eqlab{D-bar-new}
\end{align}
\eqlab{recentered-matrices-new}%
\end{subequations}
Once these recentered matrices have been computed as described below,
the original ones in \eq{R-matrices-new} are recovered using
\begin{subequations}
\begin{align}
\Acal_{\alpha,ij}(\R) &= \bar \Acal_{\alpha,ij}(\R)\,,
\eqlab{A-conv-new}\\
\Bcal_{\alpha,ij}(\R) &= \bar \Bcal_{\alpha,ij}(\R) -
d_{\alpha,ij}(\R) \Hcal_{ij}(\R)\,,
\eqlab{B-conv-new}\\
\Ccal_{\alpha\beta,ij}(\R) &= \bar \Ccal_{\alpha\beta,ij}(\R) +
d_{\alpha,ij}(\R) \bar \Acal_{\beta,ij}(\R)\nn
&\qquad \qquad \quad \; \, - d_{\beta,ij}(\R) \bar \Acal_{\alpha,ij}(\R) \,,
\eqlab{C-conv-new}\\
\Dcal_{\alpha\beta,ij}(\R) &= \bar \Dcal_{\alpha\beta,ij}(\R) +
d_{\alpha,ij}(\R) \bar \Bcal_{\beta,ij}(\R)\nn
&\qquad \qquad \quad \; \; -
d_{\beta,ij}(\R) \bar \Bcal_{\alpha,ij}(\R)\nn
&\qquad \qquad \quad \;\; - d_{\alpha,ij}(\R) d_{\beta,ij}(\R) \Hcal_{ij}(\R)\,,
\eqlab{D-conv-new}
\end{align}
\eqlab{conv-new}%
\end{subequations}
where
\beq
{\bf d}_{ij}(\R) = \frac{1}{2} \left( \R + \btau_j - \btau_i \right)\,.
\eeq
Note that the last term in \eq{D-conv-new} is symmetric in
$\alpha\beta$; since only the antisymmetric part of
$\Dcal_{\alpha\beta}$ contributes to $\sigma_{\alpha\beta\gamma}$,
that term drops out.

The next step is to evaluate the recentered matrices via the
approximate identity
\beq
\r - \r_0 \simeq i\sum_\b\, w_b \b e^{-i\b\cdot(\r-\r_0)}\,,
\eqlab{r-approx-new}
\eeq
which follows from expanding the exponential to first order, and then
using the relation
$\sum_\b\, w_b b_\alpha b_\beta =
\delta_{\alpha\beta}$~\cite{marzari-prb97} (the $\b$ vectors connect
neighboring points on the $\q$-grid, and $w_b$ are weight factors).
\Eq{r-approx-new} provides an approximate representation of the
position operator on a periodic supercell commensurate with the
Wannier functions given by \eq{Rj}. As that representation is most
accurate near $\r_0$~\cite{stengel-prb06}, the choice
$\r_0=\bar \r_{ij}(\R)$ (halfway between the Wannier centers
$\me{\0 i}{\r}{\0 i}$ and $\me{\R j}{\r}{\R j}$) is optimal for the
purpose of evaluating orbital matrix elements between $\ket{\0 i}$ and
$\ket{\R j}$.

To proceed, write
\begin{subequations}
\begin{align}
\left( \r - \r_0 \right) \ket{\R j} &\simeq
i\sum_\b\, w_b \b e^{-i\b\cdot(\r-\r_0)} \ket{\R j}\,,
\eqlab{r-ket-new}\\
\bra{\0 i} \left( \r - \r_0 \right) &\simeq
-i\sum_b\, w_b \b \bra{\0 i} e^{i\b\cdot(\r-\r_0)}\,,
\eqlab{bra-r-new}
\end{align}
\end{subequations}
plug these expressions into \eq{recentered-matrices-new} setting
$\r_0 = \bar \r_{ij}(\R)$, and then use \eqs{Rj}{psi-W-q} to express
the WFs in terms of the {\it ab initio} Bloch eigenfunctions.
Invoking the orthonormality relation
\beq
\ip{\psi_\q}{\chi_{\q'}} = N\delta_{\q,\q'}\ip{u_\q}{v_\q}
\eeq
between Bloch-like states $\psi_\q(\r)=e^{i\q\cdot\r}u_\q(\r)$ and
$\chi_{\q'}(\r)=e^{i\q'\cdot\r}v_{\q'}(\r)$, one arrives at
\begin{widetext}
\begin{subequations}
\begin{align}
\bar \Acal_{\alpha,ij}(\R) &\simeq
\frac{i}{N}\sum_{\b,\q} w_b b_\alpha
e^{-i(\q+\b/2)\cdot(\R+\btau_j-\btau_i)} \left[
\Wcal^\dagger(\q) {\mathbbm M}(\q,\q+\b) \Wcal(\q+\b)
\right]_{ij},\\
\bar \Bcal_{\alpha,ij}(\R) &\simeq
\frac{i}{N}\sum_{\b,\q} w_b b_\alpha
e^{-i(\q+\b/2)\cdot(\R+\btau_j-\btau_i)}
\left[
\Wcal^\dagger(\q) {\mathbbm H}(\q){\mathbbm M}(\q,\q+\b) \Wcal(\q+\b)
\right]_{ij},\\
\bar \Ccal_{\alpha\beta,ij}(\R) &\simeq
\frac{1}{N}\sum_{\b,\b',\q} w_b w_{b'} b_\alpha b'_\beta
e^{-i(\q+\b/2+\b'/2)\cdot(\R+\btau_j-\btau_i)}
\left[
\Wcal^\dagger(\q+\b) {\mathbbm M}(\q+\b,\q+\b') \Wcal(\q+\b')
\right]_{ij},\\
\bar \Dcal_{\alpha\beta,ij}(\R) &\simeq
\frac{1}{N}\sum_{\b,\b',\q} w_b w_{b'} b_\alpha b'_\beta
e^{-i(\q+\b/2+\b'/2)\cdot(\R+\btau_j-\btau_i)}
\left[
\Wcal^\dagger(\q+\b) {\mathbbm N}(\q+\b,\q+\b') \Wcal(\q+\b')
\right]_{ij},
\end{align}
\eqlab{recentered-finite-diff-new}
\end{subequations}
\end{widetext}
where $\mathbbm{H}, \mathbbm{M}$, and $\mathbbm{N}$ are the \textit{ab
  initio} matrices introduced in \eq{ab-initio-mat}, and $\Wcal$ is
the wannierization matrix of \eq{psi-W-q}.

\section{Numerical tests}
\seclab{num-tests}

\subsection{Symmetrization}
\seclab{symm}

As mentioned in \sref{wannier-constr}, the Wannier functions employed
in this work do not fully respect crystal symmetries. Here, we test
the two symmetrization procedures that we use to correct for this
slight symmetry breaking in the calculation of
$\sigma_{\alpha\beta\gamma}(\w)$. The first is symmetrization of the
real-space matrices defined in \sref{sigma-interp}, for which we
follow the approach first implemented in Ref.~\cite{gresch-prm2018}
for the Hamiltonian matrix, and later generalized to other matrices as
well~\cite{xiaoxiong-phdthesis2023}.  The second, performed in
reciprocal space, uses point-group symmetry to reduce the $\k$
summations to the irreducible wedge of the BZ~\cite{tsirkin-npjcm21}.

\begin{figure}[t]
\centering\includegraphics[width=\columnwidth]{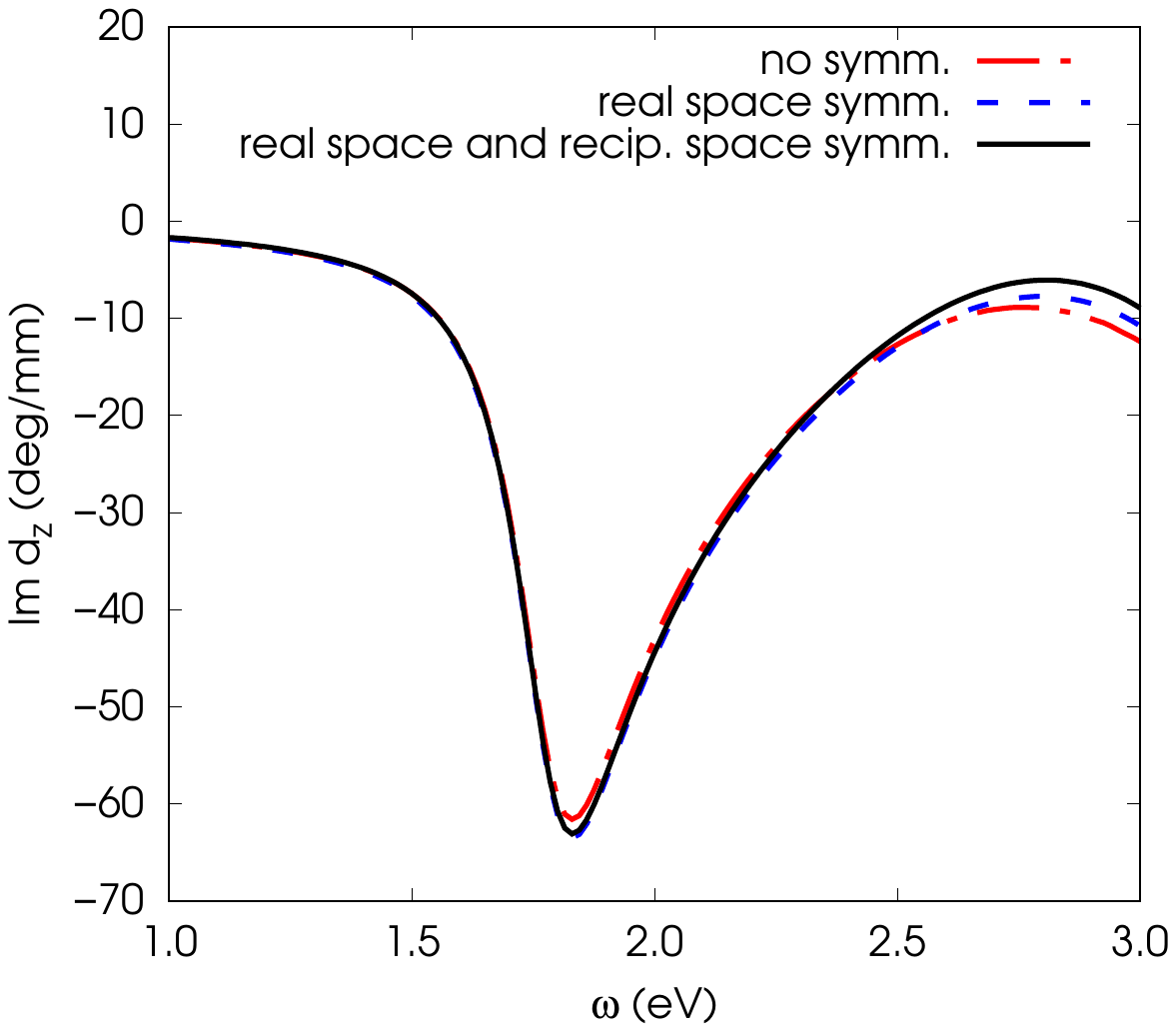}
\caption{Absorptive part of the polar optical activity spectrum of GaN
  obtained with different levels of symmetrization: no symmetrization
  (red dashed-dotted line), symmetrization of the real-space matrices
  (blue dashed line), symmetrization both in real space and in
  reciprocal space (black solid line). The latter is the same as the
  black solid line in \fref{fig4}(a).}
\figlab{fig7}
\end{figure}

To check that the above symmetrization procedures are properly
implemented, we calculate in three different ways for GaN the quantity
$\Im\, d_z(\w)$ displayed in \fref{fig4}: (i) no symmetrization, (ii)
symmetrization for the real-space matrix elements but no
symmetrization in reciprocal space, and (iii) symmetrization in both
real and reciprocal space. The results are shown in \fref{fig7}. If
the Wannier functions respected all point-group symmetries, the three
curves would be identical; the small deviations among them suggest
that the symmetry breaking is weak.

Symmetry breaking in Wannier interpolation can be avoided altogether
by using symmetry-adapted Wannier
functions~\cite{sakuma-sawf}. However, their current implementation in
\textsc{Wannier90} has two limitations: it does not allow for a frozen
window in the disentanglement procedure, and it does not take SOC into
account. These limations prevented us from using symmetry-adapted
Wannier functions in the present work.\footnote{During the preparation
  of this manuscript, the construction of symmetry-adapted Wannier
  functions with frozen energy windows and with SOC was implemented in
  \textsc{WannierBerri}. However, the present work did not make use of
  this development.}

\subsection{Convergence of rotatory power vs. wannierized conduction states}
\seclab{trunc-error}

As discussed in \sref{optical}, in our Wannier interpolation scheme
the reactive part of $\sigma_{\alpha\beta\gamma}(\w)$ is affected by a
band-truncation error. To assess its magnitude, here we recalculate
the rotatory power of Se using augmented sets of Wannier functions.

\begin{figure}[b]
\centering\includegraphics[width=\columnwidth]{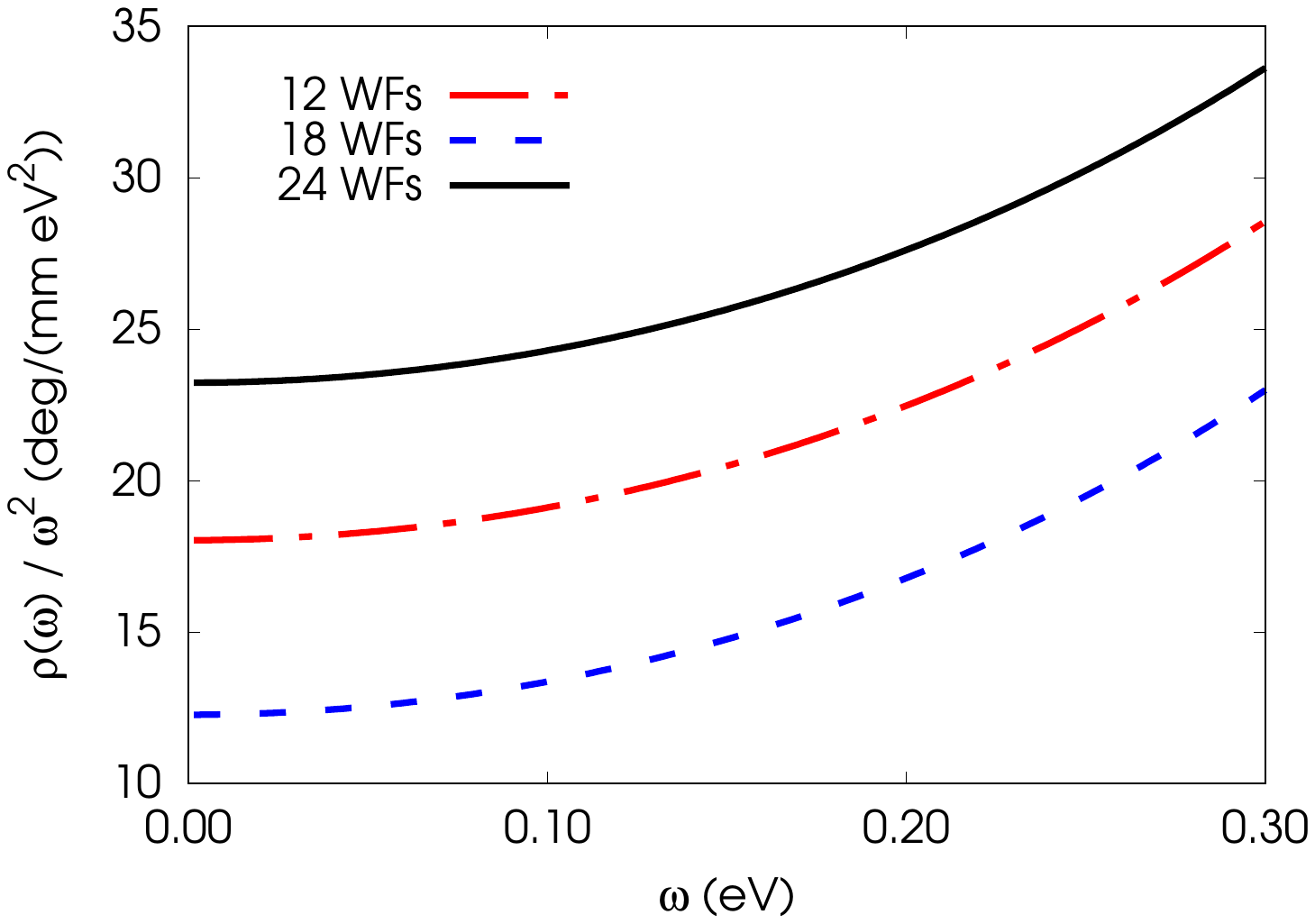}
\caption{Dependence on the size of the Wannier-function (WF) basis of
  the calculated rotatory power of trigonal Se divided by $\w^2$.}
\figlab{fig8}
\end{figure}

The results for Se reported in \sref{se-quartz} were obtained with 12
Wannier functions per primitive cell, constructed from atom-centered
$s$ and $p$ trial orbitals. This Wannier basis covers the nine valence
bands and the three lowest conduction bands. To describe higher
conduction bands, we use additional $s$-like Wannier functions
centered at interstitial sites, identified using the SCDM method (see
Supplementary Material for more details). In this way, we are able to
describe a total of 18 (9 valence and 9 conduction) and 24 (9 valence
and 15 conduction) bands.

In \fref{fig8} we display the quantity $\bar{\rho}(\w)=\rho(\w)/\w^2$
defined in \eq{rho-bar}, computed using the three Wannier basis sets;
the curve corresponding to 12 Wannier functions is the same as in the
inset of \fref{fig6}.  The convergence with respect to the number of
Wannier functions is somewhat slow and nonmonotonic, with
$\bar{\rho}(\w)$ changing by roughly a factor of two when going from
18 to 24 orbitals.

\bibliography{SDOC}

\begin{thebibliography}{60}%
\makeatletter
\providecommand \@ifxundefined [1]{%
 \@ifx{#1\undefined}
}%
\providecommand \@ifnum [1]{%
 \ifnum #1\expandafter \@firstoftwo
 \else \expandafter \@secondoftwo
 \fi
}%
\providecommand \@ifx [1]{%
 \ifx #1\expandafter \@firstoftwo
 \else \expandafter \@secondoftwo
 \fi
}%
\providecommand \natexlab [1]{#1}%
\providecommand \enquote  [1]{``#1''}%
\providecommand \bibnamefont  [1]{#1}%
\providecommand \bibfnamefont [1]{#1}%
\providecommand \citenamefont [1]{#1}%
\providecommand \href@noop [0]{\@secondoftwo}%
\providecommand \href [0]{\begingroup \@sanitize@url \@href}%
\providecommand \@href[1]{\@@startlink{#1}\@@href}%
\providecommand \@@href[1]{\endgroup#1\@@endlink}%
\providecommand \@sanitize@url [0]{\catcode `\\12\catcode `\$12\catcode `\&12\catcode `\#12\catcode `\^12\catcode `\_12\catcode `\%12\relax}%
\providecommand \@@startlink[1]{}%
\providecommand \@@endlink[0]{}%
\providecommand \url  [0]{\begingroup\@sanitize@url \@url }%
\providecommand \@url [1]{\endgroup\@href {#1}{\urlprefix }}%
\providecommand \urlprefix  [0]{URL }%
\providecommand \Eprint [0]{\href }%
\providecommand \doibase [0]{https://doi.org/}%
\providecommand \selectlanguage [0]{\@gobble}%
\providecommand \bibinfo  [0]{\@secondoftwo}%
\providecommand \bibfield  [0]{\@secondoftwo}%
\providecommand \translation [1]{[#1]}%
\providecommand \BibitemOpen [0]{}%
\providecommand \bibitemStop [0]{}%
\providecommand \bibitemNoStop [0]{.\EOS\space}%
\providecommand \EOS [0]{\spacefactor3000\relax}%
\providecommand \BibitemShut  [1]{\csname bibitem#1\endcsname}%
\let\auto@bib@innerbib\@empty
\bibitem [{\citenamefont {Landau}\ and\ \citenamefont {Lifshitz}(1984)}]{landau-84}%
  \BibitemOpen
  \bibfield  {author} {\bibinfo {author} {\bibfnamefont {L.~D.}\ \bibnamefont {Landau}}\ and\ \bibinfo {author} {\bibfnamefont {E.~M.}\ \bibnamefont {Lifshitz}},\ }\href@noop {} {\emph {\bibinfo {title} {Electrodynamics of continuous media}}}\ (\bibinfo  {publisher} {Elsevier, Amsterdam, Netherlands},\ \bibinfo {year} {1984})\BibitemShut {NoStop}%
\bibitem [{\citenamefont {Barron}(2004)}]{barron-04}%
  \BibitemOpen
  \bibfield  {author} {\bibinfo {author} {\bibfnamefont {L.~D.}\ \bibnamefont {Barron}},\ }\href {https://doi.org/https://doi.org/10.1017/CBO9780511535468} {\emph {\bibinfo {title} {Molecular light scattering and optical activity}}}\ (\bibinfo  {publisher} {Cambridge University Press, Cambridge, UK},\ \bibinfo {year} {2004})\BibitemShut {NoStop}%
\bibitem [{\citenamefont {Newnham}(2005)}]{newnham-book05}%
  \BibitemOpen
  \bibfield  {author} {\bibinfo {author} {\bibfnamefont {R.~E.}\ \bibnamefont {Newnham}},\ }\href {https://doi.org/10.1093/oso/9780198520757.001.0001} {\emph {\bibinfo {title} {{Properties of Materials}}}}\ (\bibinfo  {publisher} {Oxford University Press},\ \bibinfo {year} {2005})\BibitemShut {NoStop}%
\bibitem [{\citenamefont {Malgrange}\ \emph {et~al.}(2014)\citenamefont {Malgrange}, \citenamefont {Ricolleau},\ and\ \citenamefont {Schlenker}}]{malgrange-14}%
  \BibitemOpen
  \bibfield  {author} {\bibinfo {author} {\bibfnamefont {C.}~\bibnamefont {Malgrange}}, \bibinfo {author} {\bibfnamefont {C.}~\bibnamefont {Ricolleau}},\ and\ \bibinfo {author} {\bibfnamefont {M.}~\bibnamefont {Schlenker}},\ }\href {https://doi.org/https://doi.org/10.1007/978-94-017-8993-6} {\emph {\bibinfo {title} {Symmetry and physical properties of crystals}}}\ (\bibinfo  {publisher} {Springer, Dordrecht},\ \bibinfo {year} {2014})\BibitemShut {NoStop}%
\bibitem [{\citenamefont {Raab}\ and\ \citenamefont {De~Lange}(2005)}]{raab-05}%
  \BibitemOpen
  \bibfield  {author} {\bibinfo {author} {\bibfnamefont {R.~E.}\ \bibnamefont {Raab}}\ and\ \bibinfo {author} {\bibfnamefont {O.~L.}\ \bibnamefont {De~Lange}},\ }\href@noop {} {\emph {\bibinfo {title} {Multipole Theory in Electromagnetism}}}\ (\bibinfo  {publisher} {Oxford University Press, Oxford, UK},\ \bibinfo {year} {2005})\BibitemShut {NoStop}%
\bibitem [{\citenamefont {Brown}\ \emph {et~al.}(1963)\citenamefont {Brown}, \citenamefont {Shtrikman},\ and\ \citenamefont {Treves}}]{brown-jap63}%
  \BibitemOpen
  \bibfield  {author} {\bibinfo {author} {\bibfnamefont {J.}~\bibnamefont {Brown}, \bibfnamefont {W.~F.}}, \bibinfo {author} {\bibfnamefont {S.}~\bibnamefont {Shtrikman}},\ and\ \bibinfo {author} {\bibfnamefont {D.}~\bibnamefont {Treves}},\ }\bibfield  {title} {\bibinfo {title} {{Possibility of Visual Observation of Antiferromagnetic Domains}},\ }\href {https://doi.org/10.1063/1.1729451} {\bibfield  {journal} {\bibinfo  {journal} {J. Appl. Phys.}\ }\textbf {\bibinfo {volume} {34}},\ \bibinfo {pages} {1233} (\bibinfo {year} {1963})}\BibitemShut {NoStop}%
\bibitem [{\citenamefont {Hornreich}(1968)}]{hornreich-jap68}%
  \BibitemOpen
  \bibfield  {author} {\bibinfo {author} {\bibfnamefont {R.~M.}\ \bibnamefont {Hornreich}},\ }\bibfield  {title} {\bibinfo {title} {{Gyrotropic Birefringence~—~Phenomenological Theory}},\ }\href {https://doi.org/10.1063/1.2163466} {\bibfield  {journal} {\bibinfo  {journal} {J. Appl. Phys.}\ }\textbf {\bibinfo {volume} {39}},\ \bibinfo {pages} {432} (\bibinfo {year} {1968})}\BibitemShut {NoStop}%
\bibitem [{\citenamefont {Hornreich}\ and\ \citenamefont {Shtrikman}(1968)}]{hornreich-pr68}%
  \BibitemOpen
  \bibfield  {author} {\bibinfo {author} {\bibfnamefont {R.~M.}\ \bibnamefont {Hornreich}}\ and\ \bibinfo {author} {\bibfnamefont {S.}~\bibnamefont {Shtrikman}},\ }\bibfield  {title} {\bibinfo {title} {Theory of gyrotropic birefringence},\ }\href {https://doi.org/10.1103/PhysRev.171.1065} {\bibfield  {journal} {\bibinfo  {journal} {Phys. Rev.}\ }\textbf {\bibinfo {volume} {171}},\ \bibinfo {pages} {1065} (\bibinfo {year} {1968})}\BibitemShut {NoStop}%
\bibitem [{\citenamefont {Kimura}\ and\ \citenamefont {Kimura}(2024)}]{kimura-prl24}%
  \BibitemOpen
  \bibfield  {author} {\bibinfo {author} {\bibfnamefont {K.}~\bibnamefont {Kimura}}\ and\ \bibinfo {author} {\bibfnamefont {T.}~\bibnamefont {Kimura}},\ }\bibfield  {title} {\bibinfo {title} {Nonvolatile switching of large nonreciprocal optical absorption at shortwave infrared wavelengths},\ }\href {https://doi.org/10.1103/PhysRevLett.132.036901} {\bibfield  {journal} {\bibinfo  {journal} {Phys. Rev. Lett.}\ }\textbf {\bibinfo {volume} {132}},\ \bibinfo {pages} {036901} (\bibinfo {year} {2024})}\BibitemShut {NoStop}%
\bibitem [{\citenamefont {{Bondo Pedersen}}\ and\ \citenamefont {{Hansen}}(1995)}]{pedersen-cpl95}%
  \BibitemOpen
  \bibfield  {author} {\bibinfo {author} {\bibfnamefont {T.}~\bibnamefont {{Bondo Pedersen}}}\ and\ \bibinfo {author} {\bibfnamefont {A.~E.}\ \bibnamefont {{Hansen}}},\ }\bibfield  {title} {\bibinfo {title} {{Ab initio calculation and display of the rotatory strength tensor in the random phase approximation. Method and model studies}},\ }\href {https://doi.org/10.1016/0009-2614(95)01036-9} {\bibfield  {journal} {\bibinfo  {journal} {Chem. Phys. Lett.}\ }\textbf {\bibinfo {volume} {246}},\ \bibinfo {pages} {1} (\bibinfo {year} {1995})}\BibitemShut {NoStop}%
\bibitem [{\citenamefont {{Polavarapu}}(1997)}]{polavarapu-molphys97}%
  \BibitemOpen
  \bibfield  {author} {\bibinfo {author} {\bibfnamefont {P.~L.}\ \bibnamefont {{Polavarapu}}},\ }\bibfield  {title} {\bibinfo {title} {{Ab initio molecular optical rotations and absolute configurations}},\ }\href {https://doi.org/10.1080/00268979709482744} {\bibfield  {journal} {\bibinfo  {journal} {Mol. Phys.}\ }\textbf {\bibinfo {volume} {91}},\ \bibinfo {pages} {551} (\bibinfo {year} {1997})}\BibitemShut {NoStop}%
\bibitem [{\citenamefont {Yabana}\ and\ \citenamefont {Bertsch}(1999)}]{yabana-pra99}%
  \BibitemOpen
  \bibfield  {author} {\bibinfo {author} {\bibfnamefont {K.}~\bibnamefont {Yabana}}\ and\ \bibinfo {author} {\bibfnamefont {G.~F.}\ \bibnamefont {Bertsch}},\ }\bibfield  {title} {\bibinfo {title} {Application of the time-dependent local density approximation to optical activity},\ }\href {https://doi.org/10.1103/PhysRevA.60.1271} {\bibfield  {journal} {\bibinfo  {journal} {Phys. Rev. A}\ }\textbf {\bibinfo {volume} {60}},\ \bibinfo {pages} {1271} (\bibinfo {year} {1999})}\BibitemShut {NoStop}%
\bibitem [{\citenamefont {Varsano}\ \emph {et~al.}(2009)\citenamefont {Varsano}, \citenamefont {Espinosa-Leal}, \citenamefont {Andrade}, \citenamefont {Marques}, \citenamefont {di~Felice},\ and\ \citenamefont {Rubio}}]{varsano-pccp09}%
  \BibitemOpen
  \bibfield  {author} {\bibinfo {author} {\bibfnamefont {D.}~\bibnamefont {Varsano}}, \bibinfo {author} {\bibfnamefont {L.~A.}\ \bibnamefont {Espinosa-Leal}}, \bibinfo {author} {\bibfnamefont {X.}~\bibnamefont {Andrade}}, \bibinfo {author} {\bibfnamefont {M.~A.~L.}\ \bibnamefont {Marques}}, \bibinfo {author} {\bibfnamefont {R.}~\bibnamefont {di~Felice}},\ and\ \bibinfo {author} {\bibfnamefont {A.}~\bibnamefont {Rubio}},\ }\bibfield  {title} {\bibinfo {title} {Towards a gauge invariant method for molecular chiroptical properties in {TDDFT}},\ }\href {https://doi.org/10.1039/B903200B} {\bibfield  {journal} {\bibinfo  {journal} {Phys. Chem. Chem. Phys.}\ }\textbf {\bibinfo {volume} {11}},\ \bibinfo {pages} {4481} (\bibinfo {year} {2009})}\BibitemShut {NoStop}%
\bibitem [{\citenamefont {Polavarapu}(2002)}]{polavarapu-chirality02}%
  \BibitemOpen
  \bibfield  {author} {\bibinfo {author} {\bibfnamefont {P.~L.}\ \bibnamefont {Polavarapu}},\ }\bibfield  {title} {\bibinfo {title} {Optical rotation: Recent advances in determining the absolute configuration},\ }\href {https://doi.org/https://doi.org/10.1002/chir.10145} {\bibfield  {journal} {\bibinfo  {journal} {Chirality}\ }\textbf {\bibinfo {volume} {14}},\ \bibinfo {pages} {768} (\bibinfo {year} {2002})}\BibitemShut {NoStop}%
\bibitem [{\citenamefont {Autschbach}\ \emph {et~al.}(2010)\citenamefont {Autschbach}, \citenamefont {Nitsch-Velasquez},\ and\ \citenamefont {Rudolph}}]{autschbach-10}%
  \BibitemOpen
  \bibfield  {author} {\bibinfo {author} {\bibfnamefont {J.}~\bibnamefont {Autschbach}}, \bibinfo {author} {\bibfnamefont {L.}~\bibnamefont {Nitsch-Velasquez}},\ and\ \bibinfo {author} {\bibfnamefont {M.}~\bibnamefont {Rudolph}},\ }\href {https://doi.org/https://doi.org/10.1007/128_2010_72} {\emph {\bibinfo {title} {Time-dependent density functional response theory for electronic chiroptical properties of chiral molecules}}}\ (\bibinfo  {publisher} {Springer, Berlin, Germany},\ \bibinfo {year} {2010})\BibitemShut {NoStop}%
\bibitem [{\citenamefont {Mattiat}\ and\ \citenamefont {Luber}(2021)}]{mattiat-helvchim21}%
  \BibitemOpen
  \bibfield  {author} {\bibinfo {author} {\bibfnamefont {J.}~\bibnamefont {Mattiat}}\ and\ \bibinfo {author} {\bibfnamefont {S.}~\bibnamefont {Luber}},\ }\bibfield  {title} {\bibinfo {title} {Recent progress in the simulation of chiral systems with real time propagation methods},\ }\href {https://doi.org/https://doi.org/10.1002/hlca.202100154} {\bibfield  {journal} {\bibinfo  {journal} {Helv. Chim. Acta}\ }\textbf {\bibinfo {volume} {104}},\ \bibinfo {pages} {12} (\bibinfo {year} {2021})}\BibitemShut {NoStop}%
\bibitem [{\citenamefont {Natori}(1975)}]{natori-jpsj75}%
  \BibitemOpen
  \bibfield  {author} {\bibinfo {author} {\bibfnamefont {K.}~\bibnamefont {Natori}},\ }\bibfield  {title} {\bibinfo {title} {Band theory of the optical activity of crystals},\ }\href {https://doi.org/10.1143/JPSJ.39.1013} {\bibfield  {journal} {\bibinfo  {journal} {J. Phys. Soc. Jpn.}\ }\textbf {\bibinfo {volume} {39}},\ \bibinfo {pages} {1013} (\bibinfo {year} {1975})}\BibitemShut {NoStop}%
\bibitem [{\citenamefont {Zhong}\ \emph {et~al.}(1993)\citenamefont {Zhong}, \citenamefont {Levine}, \citenamefont {Allan},\ and\ \citenamefont {Wilkins}}]{zhong-prb93}%
  \BibitemOpen
  \bibfield  {author} {\bibinfo {author} {\bibfnamefont {H.}~\bibnamefont {Zhong}}, \bibinfo {author} {\bibfnamefont {Z.~H.}\ \bibnamefont {Levine}}, \bibinfo {author} {\bibfnamefont {D.~C.}\ \bibnamefont {Allan}},\ and\ \bibinfo {author} {\bibfnamefont {J.~W.}\ \bibnamefont {Wilkins}},\ }\bibfield  {title} {\bibinfo {title} {Band-theoretic calculations of the optical-activity tensor of \ensuremath{\alpha}-quartz and trigonal {Se}},\ }\href {https://doi.org/10.1103/PhysRevB.48.1384} {\bibfield  {journal} {\bibinfo  {journal} {Phys. Rev. B}\ }\textbf {\bibinfo {volume} {48}},\ \bibinfo {pages} {1384} (\bibinfo {year} {1993})}\BibitemShut {NoStop}%
\bibitem [{\citenamefont {Malashevich}\ and\ \citenamefont {Souza}(2010)}]{malashevich-prb10}%
  \BibitemOpen
  \bibfield  {author} {\bibinfo {author} {\bibfnamefont {A.}~\bibnamefont {Malashevich}}\ and\ \bibinfo {author} {\bibfnamefont {I.}~\bibnamefont {Souza}},\ }\bibfield  {title} {\bibinfo {title} {Band theory of spatial dispersion in magnetoelectrics},\ }\href {https://doi.org/10.1103/PhysRevB.82.245118} {\bibfield  {journal} {\bibinfo  {journal} {Phys. Rev. B}\ }\textbf {\bibinfo {volume} {82}},\ \bibinfo {pages} {245118} (\bibinfo {year} {2010})}\BibitemShut {NoStop}%
\bibitem [{\citenamefont {Wang}\ and\ \citenamefont {Yan}(2023)}]{wang-prb23}%
  \BibitemOpen
  \bibfield  {author} {\bibinfo {author} {\bibfnamefont {X.}~\bibnamefont {Wang}}\ and\ \bibinfo {author} {\bibfnamefont {Y.}~\bibnamefont {Yan}},\ }\bibfield  {title} {\bibinfo {title} {Optical activity of solids from first principles},\ }\href {https://doi.org/10.1103/PhysRevB.107.045201} {\bibfield  {journal} {\bibinfo  {journal} {Phys. Rev. B}\ }\textbf {\bibinfo {volume} {107}},\ \bibinfo {pages} {045201} (\bibinfo {year} {2023})}\BibitemShut {NoStop}%
\bibitem [{\citenamefont {Ocaña}\ and\ \citenamefont {Souza}(2023)}]{pozo-scipost23}%
  \BibitemOpen
  \bibfield  {author} {\bibinfo {author} {\bibfnamefont {O.~P.}\ \bibnamefont {Ocaña}}\ and\ \bibinfo {author} {\bibfnamefont {I.}~\bibnamefont {Souza}},\ }\bibfield  {title} {\bibinfo {title} {{Multipole theory of optical spatial dispersion in crystals}},\ }\href {https://doi.org/10.21468/SciPostPhys.14.5.118} {\bibfield  {journal} {\bibinfo  {journal} {SciPost Phys.}\ }\textbf {\bibinfo {volume} {14}},\ \bibinfo {pages} {118} (\bibinfo {year} {2023})}\BibitemShut {NoStop}%
\bibitem [{\citenamefont {Zabalo}\ and\ \citenamefont {Stengel}(2023)}]{zabalo-prl23}%
  \BibitemOpen
  \bibfield  {author} {\bibinfo {author} {\bibfnamefont {A.}~\bibnamefont {Zabalo}}\ and\ \bibinfo {author} {\bibfnamefont {M.}~\bibnamefont {Stengel}},\ }\bibfield  {title} {\bibinfo {title} {Natural optical activity from density-functional perturbation theory},\ }\href {https://doi.org/10.1103/PhysRevLett.131.086902} {\bibfield  {journal} {\bibinfo  {journal} {Phys. Rev. Lett.}\ }\textbf {\bibinfo {volume} {131}},\ \bibinfo {pages} {086902} (\bibinfo {year} {2023})}\BibitemShut {NoStop}%
\bibitem [{\citenamefont {J\"onsson}\ \emph {et~al.}(1996)\citenamefont {J\"onsson}, \citenamefont {Levine},\ and\ \citenamefont {Wilkins}}]{jonsson-prl96}%
  \BibitemOpen
  \bibfield  {author} {\bibinfo {author} {\bibfnamefont {L.}~\bibnamefont {J\"onsson}}, \bibinfo {author} {\bibfnamefont {Z.~H.}\ \bibnamefont {Levine}},\ and\ \bibinfo {author} {\bibfnamefont {J.~W.}\ \bibnamefont {Wilkins}},\ }\bibfield  {title} {\bibinfo {title} {Large local-field corrections in optical rotatory power of quartz and selenium},\ }\href {https://doi.org/10.1103/PhysRevLett.76.1372} {\bibfield  {journal} {\bibinfo  {journal} {Phys. Rev. Lett.}\ }\textbf {\bibinfo {volume} {76}},\ \bibinfo {pages} {1372} (\bibinfo {year} {1996})}\BibitemShut {NoStop}%
\bibitem [{\citenamefont {Wang}\ \emph {et~al.}(2006)\citenamefont {Wang}, \citenamefont {Yates}, \citenamefont {Souza},\ and\ \citenamefont {Vanderbilt}}]{wang-prb06}%
  \BibitemOpen
  \bibfield  {author} {\bibinfo {author} {\bibfnamefont {X.}~\bibnamefont {Wang}}, \bibinfo {author} {\bibfnamefont {J.~R.}\ \bibnamefont {Yates}}, \bibinfo {author} {\bibfnamefont {I.}~\bibnamefont {Souza}},\ and\ \bibinfo {author} {\bibfnamefont {D.}~\bibnamefont {Vanderbilt}},\ }\bibfield  {title} {\bibinfo {title} {{{\it Ab initio} calculation of the anomalous Hall conductivity by Wannier interpolation}},\ }\href {https://doi.org/10.1103/PhysRevB.74.195118} {\bibfield  {journal} {\bibinfo  {journal} {Phys. Rev. B}\ }\textbf {\bibinfo {volume} {74}},\ \bibinfo {pages} {195118} (\bibinfo {year} {2006})}\BibitemShut {NoStop}%
\bibitem [{\citenamefont {Yates}\ \emph {et~al.}(2007)\citenamefont {Yates}, \citenamefont {Wang}, \citenamefont {Vanderbilt},\ and\ \citenamefont {Souza}}]{yates-prb07}%
  \BibitemOpen
  \bibfield  {author} {\bibinfo {author} {\bibfnamefont {J.~R.}\ \bibnamefont {Yates}}, \bibinfo {author} {\bibfnamefont {X.}~\bibnamefont {Wang}}, \bibinfo {author} {\bibfnamefont {D.}~\bibnamefont {Vanderbilt}},\ and\ \bibinfo {author} {\bibfnamefont {I.}~\bibnamefont {Souza}},\ }\bibfield  {title} {\bibinfo {title} {{Spectral and Fermi surface properties from Wannier interpolation}},\ }\href {https://doi.org/10.1103/PhysRevB.75.195121} {\bibfield  {journal} {\bibinfo  {journal} {Phys. Rev. B}\ }\textbf {\bibinfo {volume} {75}},\ \bibinfo {pages} {195121} (\bibinfo {year} {2007})}\BibitemShut {NoStop}%
\bibitem [{\citenamefont {Lopez}\ \emph {et~al.}(2012)\citenamefont {Lopez}, \citenamefont {Vanderbilt}, \citenamefont {Thonhauser},\ and\ \citenamefont {Souza}}]{lopez-prb12}%
  \BibitemOpen
  \bibfield  {author} {\bibinfo {author} {\bibfnamefont {M.~G.}\ \bibnamefont {Lopez}}, \bibinfo {author} {\bibfnamefont {D.}~\bibnamefont {Vanderbilt}}, \bibinfo {author} {\bibfnamefont {T.}~\bibnamefont {Thonhauser}},\ and\ \bibinfo {author} {\bibfnamefont {I.}~\bibnamefont {Souza}},\ }\bibfield  {title} {\bibinfo {title} {Wannier-based calculation of the orbital magnetization in crystals},\ }\href {https://doi.org/10.1103/PhysRevB.85.014435} {\bibfield  {journal} {\bibinfo  {journal} {Phys. Rev. B}\ }\textbf {\bibinfo {volume} {85}},\ \bibinfo {pages} {014435} (\bibinfo {year} {2012})}\BibitemShut {NoStop}%
\bibitem [{\citenamefont {Ades}\ and\ \citenamefont {Champness}(1975)}]{ades-josa75}%
  \BibitemOpen
  \bibfield  {author} {\bibinfo {author} {\bibfnamefont {S.}~\bibnamefont {Ades}}\ and\ \bibinfo {author} {\bibfnamefont {C.~H.}\ \bibnamefont {Champness}},\ }\bibfield  {title} {\bibinfo {title} {Optical activity of tellurium to 20 {\textmu}m},\ }\href {https://doi.org/10.1364/JOSA.65.000217} {\bibfield  {journal} {\bibinfo  {journal} {J. Opt. Soc. Am.}\ }\textbf {\bibinfo {volume} {65}},\ \bibinfo {pages} {217} (\bibinfo {year} {1975})}\BibitemShut {NoStop}%
\bibitem [{\citenamefont {Fukuda}\ \emph {et~al.}(1975)\citenamefont {Fukuda}, \citenamefont {Shiosaki},\ and\ \citenamefont {Kawabata}}]{fukuda-pssb75}%
  \BibitemOpen
  \bibfield  {author} {\bibinfo {author} {\bibfnamefont {S.}~\bibnamefont {Fukuda}}, \bibinfo {author} {\bibfnamefont {T.}~\bibnamefont {Shiosaki}},\ and\ \bibinfo {author} {\bibfnamefont {A.}~\bibnamefont {Kawabata}},\ }\bibfield  {title} {\bibinfo {title} {Infrared optical activity in tellurium},\ }\href {https://doi.org/https://doi.org/10.1002/pssb.2220680247} {\bibfield  {journal} {\bibinfo  {journal} {Phys. Status Solidi B}\ }\textbf {\bibinfo {volume} {68}},\ \bibinfo {pages} {K107} (\bibinfo {year} {1975})}\BibitemShut {NoStop}%
\bibitem [{\citenamefont {Jerphagnon}\ and\ \citenamefont {Chemla}(1976)}]{jerphagnon-jchemphys76}%
  \BibitemOpen
  \bibfield  {author} {\bibinfo {author} {\bibfnamefont {J.}~\bibnamefont {Jerphagnon}}\ and\ \bibinfo {author} {\bibfnamefont {D.~S.}\ \bibnamefont {Chemla}},\ }\bibfield  {title} {\bibinfo {title} {Optical activity of crystals},\ }\href {https://doi.org/10.1063/1.433207} {\bibfield  {journal} {\bibinfo  {journal} {J. Chem. Phys.}\ }\textbf {\bibinfo {volume} {65}},\ \bibinfo {pages} {1522} (\bibinfo {year} {1976})}\BibitemShut {NoStop}%
\bibitem [{\citenamefont {Halasyamani}\ and\ \citenamefont {Poeppelmeier}(1998)}]{halasyamani-chemmat98}%
  \BibitemOpen
  \bibfield  {author} {\bibinfo {author} {\bibfnamefont {P.~S.}\ \bibnamefont {Halasyamani}}\ and\ \bibinfo {author} {\bibfnamefont {K.~R.}\ \bibnamefont {Poeppelmeier}},\ }\bibfield  {title} {\bibinfo {title} {Noncentrosymmetric oxides},\ }\href {https://doi.org/10.1021/cm980140w} {\bibfield  {journal} {\bibinfo  {journal} {Chem. Mater.}\ }\textbf {\bibinfo {volume} {10}},\ \bibinfo {pages} {2753} (\bibinfo {year} {1998})}\BibitemShut {NoStop}%
\bibitem [{\citenamefont {Provost}\ and\ \citenamefont {Vallee}(1980)}]{provost80}%
  \BibitemOpen
  \bibfield  {author} {\bibinfo {author} {\bibfnamefont {J.~P.}\ \bibnamefont {Provost}}\ and\ \bibinfo {author} {\bibfnamefont {G.}~\bibnamefont {Vallee}},\ }\bibfield  {title} {\bibinfo {title} {Riemannian structure on manifolds of quantum states},\ }\href {https://doi.org/10.1007/BF02193559} {\bibfield  {journal} {\bibinfo  {journal} {Commun. Math. Phys.}\ }\textbf {\bibinfo {volume} {76}},\ \bibinfo {pages} {289} (\bibinfo {year} {1980})}\BibitemShut {NoStop}%
\bibitem [{\citenamefont {Marzari}\ and\ \citenamefont {Vanderbilt}(1997)}]{marzari-prb97}%
  \BibitemOpen
  \bibfield  {author} {\bibinfo {author} {\bibfnamefont {N.}~\bibnamefont {Marzari}}\ and\ \bibinfo {author} {\bibfnamefont {D.}~\bibnamefont {Vanderbilt}},\ }\bibfield  {title} {\bibinfo {title} {{Maximally localized generalized Wannier functions for composite energy bands}},\ }\href {https://doi.org/10.1103/PhysRevB.56.12847} {\bibfield  {journal} {\bibinfo  {journal} {Phys. Rev. B}\ }\textbf {\bibinfo {volume} {56}},\ \bibinfo {pages} {12847} (\bibinfo {year} {1997})}\BibitemShut {NoStop}%
\bibitem [{\citenamefont {Souza}\ \emph {et~al.}(2001)\citenamefont {Souza}, \citenamefont {Marzari},\ and\ \citenamefont {Vanderbilt}}]{souza-prb01}%
  \BibitemOpen
  \bibfield  {author} {\bibinfo {author} {\bibfnamefont {I.}~\bibnamefont {Souza}}, \bibinfo {author} {\bibfnamefont {N.}~\bibnamefont {Marzari}},\ and\ \bibinfo {author} {\bibfnamefont {D.}~\bibnamefont {Vanderbilt}},\ }\bibfield  {title} {\bibinfo {title} {Maximally localized wannier functions for entangled energy bands},\ }\href {https://doi.org/10.1103/PhysRevB.65.035109} {\bibfield  {journal} {\bibinfo  {journal} {Phys. Rev. B}\ }\textbf {\bibinfo {volume} {65}},\ \bibinfo {pages} {035109} (\bibinfo {year} {2001})}\BibitemShut {NoStop}%
\bibitem [{\citenamefont {Pizzi}\ \emph {et~al.}(2020)\citenamefont {Pizzi}, \citenamefont {Vitale}, \citenamefont {Arita}, \citenamefont {Blügel}, \citenamefont {Freimuth}, \citenamefont {Géranton}, \citenamefont {Gibertini}, \citenamefont {Gresch}, \citenamefont {Johnson}, \citenamefont {Koretsune}, \citenamefont {Ibañez-Azpiroz}, \citenamefont {Lee}, \citenamefont {Lihm}, \citenamefont {Marchand}, \citenamefont {Marrazzo}, \citenamefont {Mokrousov}, \citenamefont {Mustafa}, \citenamefont {Nohara}, \citenamefont {Nomura}, \citenamefont {Paulatto}, \citenamefont {Poncé}, \citenamefont {Ponweiser}, \citenamefont {Qiao}, \citenamefont {Thöle}, \citenamefont {Tsirkin}, \citenamefont {Wierzbowska}, \citenamefont {Marzari}, \citenamefont {Vanderbilt}, \citenamefont {Souza}, \citenamefont {Mostofi},\ and\ \citenamefont {Yates}}]{pizzi-jpcm20}%
  \BibitemOpen
  \bibfield  {author} {\bibinfo {author} {\bibfnamefont {G.}~\bibnamefont {Pizzi}}, \bibinfo {author} {\bibfnamefont {V.}~\bibnamefont {Vitale}}, \bibinfo {author} {\bibfnamefont {R.}~\bibnamefont {Arita}}, \bibinfo {author} {\bibfnamefont {S.}~\bibnamefont {Blügel}}, \bibinfo {author} {\bibfnamefont {F.}~\bibnamefont {Freimuth}}, \bibinfo {author} {\bibfnamefont {G.}~\bibnamefont {Géranton}}, \bibinfo {author} {\bibfnamefont {M.}~\bibnamefont {Gibertini}}, \bibinfo {author} {\bibfnamefont {D.}~\bibnamefont {Gresch}}, \bibinfo {author} {\bibfnamefont {C.}~\bibnamefont {Johnson}}, \bibinfo {author} {\bibfnamefont {T.}~\bibnamefont {Koretsune}}, \bibinfo {author} {\bibfnamefont {J.}~\bibnamefont {Ibañez-Azpiroz}}, \bibinfo {author} {\bibfnamefont {H.}~\bibnamefont {Lee}}, \bibinfo {author} {\bibfnamefont {J.-M.}\ \bibnamefont {Lihm}}, \bibinfo {author} {\bibfnamefont {D.}~\bibnamefont {Marchand}}, \bibinfo {author} {\bibfnamefont {A.}~\bibnamefont {Marrazzo}}, \bibinfo {author} {\bibfnamefont
  {Y.}~\bibnamefont {Mokrousov}}, \bibinfo {author} {\bibfnamefont {J.~I.}\ \bibnamefont {Mustafa}}, \bibinfo {author} {\bibfnamefont {Y.}~\bibnamefont {Nohara}}, \bibinfo {author} {\bibfnamefont {Y.}~\bibnamefont {Nomura}}, \bibinfo {author} {\bibfnamefont {L.}~\bibnamefont {Paulatto}}, \bibinfo {author} {\bibfnamefont {S.}~\bibnamefont {Poncé}}, \bibinfo {author} {\bibfnamefont {T.}~\bibnamefont {Ponweiser}}, \bibinfo {author} {\bibfnamefont {J.}~\bibnamefont {Qiao}}, \bibinfo {author} {\bibfnamefont {F.}~\bibnamefont {Thöle}}, \bibinfo {author} {\bibfnamefont {S.~S.}\ \bibnamefont {Tsirkin}}, \bibinfo {author} {\bibfnamefont {M.}~\bibnamefont {Wierzbowska}}, \bibinfo {author} {\bibfnamefont {N.}~\bibnamefont {Marzari}}, \bibinfo {author} {\bibfnamefont {D.}~\bibnamefont {Vanderbilt}}, \bibinfo {author} {\bibfnamefont {I.}~\bibnamefont {Souza}}, \bibinfo {author} {\bibfnamefont {A.~A.}\ \bibnamefont {Mostofi}},\ and\ \bibinfo {author} {\bibfnamefont {J.~R.}\ \bibnamefont {Yates}},\ }\bibfield  {title}
  {\bibinfo {title} {Wannier90 as a community code: new features and applications},\ }\href {https://doi.org/10.1088/1361-648X/ab51ff} {\bibfield  {journal} {\bibinfo  {journal} {J. Phys.: Condens. Matter}\ }\textbf {\bibinfo {volume} {32}},\ \bibinfo {pages} {165902} (\bibinfo {year} {2020})}\BibitemShut {NoStop}%
\bibitem [{\citenamefont {Tsirkin}(2021)}]{tsirkin-npjcm21}%
  \BibitemOpen
  \bibfield  {author} {\bibinfo {author} {\bibfnamefont {S.~S.}\ \bibnamefont {Tsirkin}},\ }\bibfield  {title} {\bibinfo {title} {High performance {Wannier} interpolation of {Berry} curvature and related quantities with wannierberri code},\ }\href {https://doi.org/10.1038/s41524-021-00498-5} {\bibfield  {journal} {\bibinfo  {journal} {npj Comput. Mater.}\ }\textbf {\bibinfo {volume} {7}},\ \bibinfo {pages} {33} (\bibinfo {year} {2021})}\BibitemShut {NoStop}%
\bibitem [{\citenamefont {Giannozzi}\ \emph {et~al.}(2009)\citenamefont {Giannozzi}, \citenamefont {Baroni}, \citenamefont {Bonini}, \citenamefont {Calandra}, \citenamefont {Car}, \citenamefont {Cavazzoni}, \citenamefont {Ceresoli}, \citenamefont {Chiarotti}, \citenamefont {Cococcioni}, \citenamefont {Dabo} \emph {et~al.}}]{giannozzi-jpcm09}%
  \BibitemOpen
  \bibfield  {author} {\bibinfo {author} {\bibfnamefont {P.}~\bibnamefont {Giannozzi}}, \bibinfo {author} {\bibfnamefont {S.}~\bibnamefont {Baroni}}, \bibinfo {author} {\bibfnamefont {N.}~\bibnamefont {Bonini}}, \bibinfo {author} {\bibfnamefont {M.}~\bibnamefont {Calandra}}, \bibinfo {author} {\bibfnamefont {R.}~\bibnamefont {Car}}, \bibinfo {author} {\bibfnamefont {C.}~\bibnamefont {Cavazzoni}}, \bibinfo {author} {\bibfnamefont {D.}~\bibnamefont {Ceresoli}}, \bibinfo {author} {\bibfnamefont {G.~L.}\ \bibnamefont {Chiarotti}}, \bibinfo {author} {\bibfnamefont {M.}~\bibnamefont {Cococcioni}}, \bibinfo {author} {\bibfnamefont {I.}~\bibnamefont {Dabo}}, \emph {et~al.},\ }\bibfield  {title} {\bibinfo {title} {{QUANTUM ESPRESSO: A modular and open-source software project for quantum simulations of materials}},\ }\href {https://doi.org/10.1088/0953-8984/21/39/395502} {\bibfield  {journal} {\bibinfo  {journal} {J. Phys.: Condens. Matter}\ }\textbf {\bibinfo {volume} {21}},\ \bibinfo {pages} {395502} (\bibinfo {year}
  {2009})}\BibitemShut {NoStop}%
\bibitem [{\citenamefont {Giannozzi}\ \emph {et~al.}(2017)\citenamefont {Giannozzi}, \citenamefont {Andreussi}, \citenamefont {Brumme}, \citenamefont {Bunau}, \citenamefont {Buongiorno~Nardelli}, \citenamefont {Calandra}, \citenamefont {Car}, \citenamefont {Cavazzoni}, \citenamefont {Ceresoli}, \citenamefont {Cococcioni} \emph {et~al.}}]{giannozzi-jpcm17}%
  \BibitemOpen
  \bibfield  {author} {\bibinfo {author} {\bibfnamefont {P.}~\bibnamefont {Giannozzi}}, \bibinfo {author} {\bibfnamefont {O.}~\bibnamefont {Andreussi}}, \bibinfo {author} {\bibfnamefont {T.}~\bibnamefont {Brumme}}, \bibinfo {author} {\bibfnamefont {O.}~\bibnamefont {Bunau}}, \bibinfo {author} {\bibfnamefont {M.}~\bibnamefont {Buongiorno~Nardelli}}, \bibinfo {author} {\bibfnamefont {M.}~\bibnamefont {Calandra}}, \bibinfo {author} {\bibfnamefont {R.}~\bibnamefont {Car}}, \bibinfo {author} {\bibfnamefont {C.}~\bibnamefont {Cavazzoni}}, \bibinfo {author} {\bibfnamefont {D.}~\bibnamefont {Ceresoli}}, \bibinfo {author} {\bibfnamefont {M.}~\bibnamefont {Cococcioni}}, \emph {et~al.},\ }\bibfield  {title} {\bibinfo {title} {{Advanced capabilities for materials modelling with Quantum ESPRESSO}},\ }\href {https://doi.org/10.1088/1361-648X/aa8f79} {\bibfield  {journal} {\bibinfo  {journal} {J. Phys.: Condens. Matter}\ }\textbf {\bibinfo {volume} {29}},\ \bibinfo {pages} {465901} (\bibinfo {year} {2017})}\BibitemShut
  {NoStop}%
\bibitem [{\citenamefont {Lihm}\ \emph {et~al.}()\citenamefont {Lihm}, \citenamefont {Ghim}, \citenamefont {Hong},\ and\ \citenamefont {Park}}]{lihm-unpublished}%
  \BibitemOpen
  \bibfield  {author} {\bibinfo {author} {\bibfnamefont {J.-M.}\ \bibnamefont {Lihm}}, \bibinfo {author} {\bibfnamefont {M.}~\bibnamefont {Ghim}}, \bibinfo {author} {\bibfnamefont {S.-J.}\ \bibnamefont {Hong}},\ and\ \bibinfo {author} {\bibfnamefont {C.-H.}\ \bibnamefont {Park}},\ }\bibfield  {title} {\bibinfo {title} {Accurate calculation of {Wannier} centers, spreads, and position-related matrix elements.},\ }\bibinfo {note} {unpublished}\BibitemShut {NoStop}%
\bibitem [{\citenamefont {Perdew}\ and\ \citenamefont {Zunger}(1981)}]{perdew-prb81}%
  \BibitemOpen
  \bibfield  {author} {\bibinfo {author} {\bibfnamefont {J.~P.}\ \bibnamefont {Perdew}}\ and\ \bibinfo {author} {\bibfnamefont {A.}~\bibnamefont {Zunger}},\ }\bibfield  {title} {\bibinfo {title} {Self-interaction correction to density-functional approximations for many-electron systems},\ }\href {https://doi.org/10.1103/PhysRevB.23.5048} {\bibfield  {journal} {\bibinfo  {journal} {Phys. Rev. B}\ }\textbf {\bibinfo {volume} {23}},\ \bibinfo {pages} {5048} (\bibinfo {year} {1981})}\BibitemShut {NoStop}%
\bibitem [{\citenamefont {Perdew}\ \emph {et~al.}(1996)\citenamefont {Perdew}, \citenamefont {Burke},\ and\ \citenamefont {Ernzerhof}}]{perdew-prl96}%
  \BibitemOpen
  \bibfield  {author} {\bibinfo {author} {\bibfnamefont {J.~P.}\ \bibnamefont {Perdew}}, \bibinfo {author} {\bibfnamefont {K.}~\bibnamefont {Burke}},\ and\ \bibinfo {author} {\bibfnamefont {M.}~\bibnamefont {Ernzerhof}},\ }\bibfield  {title} {\bibinfo {title} {Generalized gradient approximation made simple},\ }\href {https://doi.org/10.1103/PhysRevLett.77.3865} {\bibfield  {journal} {\bibinfo  {journal} {Phys. Rev. Lett.}\ }\textbf {\bibinfo {volume} {77}},\ \bibinfo {pages} {3865} (\bibinfo {year} {1996})}\BibitemShut {NoStop}%
\bibitem [{\citenamefont {Heyd}\ \emph {et~al.}(2003)\citenamefont {Heyd}, \citenamefont {Scuseria},\ and\ \citenamefont {Ernzerhof}}]{heyd-jcp03}%
  \BibitemOpen
  \bibfield  {author} {\bibinfo {author} {\bibfnamefont {J.}~\bibnamefont {Heyd}}, \bibinfo {author} {\bibfnamefont {G.~E.}\ \bibnamefont {Scuseria}},\ and\ \bibinfo {author} {\bibfnamefont {M.}~\bibnamefont {Ernzerhof}},\ }\bibfield  {title} {\bibinfo {title} {{Hybrid functionals based on a screened Coulomb potential}},\ }\href {https://doi.org/10.1063/1.1564060} {\bibfield  {journal} {\bibinfo  {journal} {J. Chem. Phys.}\ }\textbf {\bibinfo {volume} {118}},\ \bibinfo {pages} {8207} (\bibinfo {year} {2003})}\BibitemShut {NoStop}%
\bibitem [{\citenamefont {Hamann}(2013)}]{hamann-prb13}%
  \BibitemOpen
  \bibfield  {author} {\bibinfo {author} {\bibfnamefont {D.~R.}\ \bibnamefont {Hamann}},\ }\bibfield  {title} {\bibinfo {title} {{Optimized norm-conserving Vanderbilt pseudopotentials}},\ }\href {https://doi.org/10.1103/PhysRevB.88.085117} {\bibfield  {journal} {\bibinfo  {journal} {Phys. Rev. B}\ }\textbf {\bibinfo {volume} {88}},\ \bibinfo {pages} {085117} (\bibinfo {year} {2013})}\BibitemShut {NoStop}%
\bibitem [{\citenamefont {Monkhorst}\ and\ \citenamefont {Pack}(1976)}]{monkhorst-prb76}%
  \BibitemOpen
  \bibfield  {author} {\bibinfo {author} {\bibfnamefont {H.~J.}\ \bibnamefont {Monkhorst}}\ and\ \bibinfo {author} {\bibfnamefont {J.~D.}\ \bibnamefont {Pack}},\ }\bibfield  {title} {\bibinfo {title} {{Special points for Brillouin-zone integrations}},\ }\href {https://doi.org/10.1103/PhysRevB.13.5188} {\bibfield  {journal} {\bibinfo  {journal} {Phys. Rev. B}\ }\textbf {\bibinfo {volume} {13}},\ \bibinfo {pages} {5188} (\bibinfo {year} {1976})}\BibitemShut {NoStop}%
\bibitem [{\citenamefont {Damle}\ \emph {et~al.}(2015)\citenamefont {Damle}, \citenamefont {Lin},\ and\ \citenamefont {Ying}}]{damle-jctc15}%
  \BibitemOpen
  \bibfield  {author} {\bibinfo {author} {\bibfnamefont {A.}~\bibnamefont {Damle}}, \bibinfo {author} {\bibfnamefont {L.}~\bibnamefont {Lin}},\ and\ \bibinfo {author} {\bibfnamefont {L.}~\bibnamefont {Ying}},\ }\bibfield  {title} {\bibinfo {title} {Compressed representation of {Kohn–Sham} orbitals via selected columns of the density matrix},\ }\href {https://doi.org/10.1021/ct500985f} {\bibfield  {journal} {\bibinfo  {journal} {J. Chem. Theory Comput.}\ }\textbf {\bibinfo {volume} {11}},\ \bibinfo {pages} {1463} (\bibinfo {year} {2015})}\BibitemShut {NoStop}%
\bibitem [{\citenamefont {Damle}\ and\ \citenamefont {Lin}(2018)}]{damle-mms18}%
  \BibitemOpen
  \bibfield  {author} {\bibinfo {author} {\bibfnamefont {A.}~\bibnamefont {Damle}}\ and\ \bibinfo {author} {\bibfnamefont {L.}~\bibnamefont {Lin}},\ }\bibfield  {title} {\bibinfo {title} {Disentanglement via entanglement: A unified method for {Wannier} localization},\ }\href {https://doi.org/10.1137/17M1129696} {\bibfield  {journal} {\bibinfo  {journal} {Multiscale Model. Simul.}\ }\textbf {\bibinfo {volume} {16}},\ \bibinfo {pages} {1392} (\bibinfo {year} {2018})}\BibitemShut {NoStop}%
\bibitem [{\citenamefont {Vitale}\ \emph {et~al.}(2020)\citenamefont {Vitale}, \citenamefont {Pizzi}, \citenamefont {Marrazzo}, \citenamefont {Yates}, \citenamefont {Marzari},\ and\ \citenamefont {Mostofi}}]{vitale-npjcm20}%
  \BibitemOpen
  \bibfield  {author} {\bibinfo {author} {\bibfnamefont {V.}~\bibnamefont {Vitale}}, \bibinfo {author} {\bibfnamefont {G.}~\bibnamefont {Pizzi}}, \bibinfo {author} {\bibfnamefont {A.}~\bibnamefont {Marrazzo}}, \bibinfo {author} {\bibfnamefont {J.~R.}\ \bibnamefont {Yates}}, \bibinfo {author} {\bibfnamefont {N.}~\bibnamefont {Marzari}},\ and\ \bibinfo {author} {\bibfnamefont {A.~A.}\ \bibnamefont {Mostofi}},\ }\bibfield  {title} {\bibinfo {title} {Automated high-throughput wannierisation},\ }\href {https://doi.org/10.1038/s41524-020-0312-y} {\bibfield  {journal} {\bibinfo  {journal} {npj Comput. Mater.}\ }\textbf {\bibinfo {volume} {6}},\ \bibinfo {pages} {66} (\bibinfo {year} {2020})}\BibitemShut {NoStop}%
\bibitem [{\citenamefont {Anzin}\ \emph {et~al.}(1977)\citenamefont {Anzin}, \citenamefont {Eremets}, \citenamefont {Kosichkin}, \citenamefont {Nadezhdinskii},\ and\ \citenamefont {Shirokov}}]{anzin-pssa77}%
  \BibitemOpen
  \bibfield  {author} {\bibinfo {author} {\bibfnamefont {V.~B.}\ \bibnamefont {Anzin}}, \bibinfo {author} {\bibfnamefont {M.~I.}\ \bibnamefont {Eremets}}, \bibinfo {author} {\bibfnamefont {Y.~V.}\ \bibnamefont {Kosichkin}}, \bibinfo {author} {\bibfnamefont {A.~I.}\ \bibnamefont {Nadezhdinskii}},\ and\ \bibinfo {author} {\bibfnamefont {A.~M.}\ \bibnamefont {Shirokov}},\ }\bibfield  {title} {\bibinfo {title} {Measurement of the energy gap in tellurium under pressure},\ }\href {https://doi.org/https://doi.org/10.1002/pssa.2210420143} {\bibfield  {journal} {\bibinfo  {journal} {Phys. Status Solidi A}\ }\textbf {\bibinfo {volume} {42}},\ \bibinfo {pages} {385} (\bibinfo {year} {1977})}\BibitemShut {NoStop}%
\bibitem [{\citenamefont {Assmann}\ \emph {et~al.}(2016)\citenamefont {Assmann}, \citenamefont {Wissgott}, \citenamefont {Kuneš}, \citenamefont {Toschi}, \citenamefont {Blaha},\ and\ \citenamefont {Held}}]{assmann-cpc2016}%
  \BibitemOpen
  \bibfield  {author} {\bibinfo {author} {\bibfnamefont {E.}~\bibnamefont {Assmann}}, \bibinfo {author} {\bibfnamefont {P.}~\bibnamefont {Wissgott}}, \bibinfo {author} {\bibfnamefont {J.}~\bibnamefont {Kuneš}}, \bibinfo {author} {\bibfnamefont {A.}~\bibnamefont {Toschi}}, \bibinfo {author} {\bibfnamefont {P.}~\bibnamefont {Blaha}},\ and\ \bibinfo {author} {\bibfnamefont {K.}~\bibnamefont {Held}},\ }\bibfield  {title} {\bibinfo {title} {{woptic: Optical conductivity with Wannier functions and adaptive k-mesh refinement}},\ }\href {https://doi.org/https://doi.org/10.1016/j.cpc.2015.12.010} {\bibfield  {journal} {\bibinfo  {journal} {Comput. Phys. Commun.}\ }\textbf {\bibinfo {volume} {202}},\ \bibinfo {pages} {1} (\bibinfo {year} {2016})}\BibitemShut {NoStop}%
\bibitem [{\citenamefont {Hanke}\ \emph {et~al.}(2016)\citenamefont {Hanke}, \citenamefont {Freimuth}, \citenamefont {Nandy}, \citenamefont {Zhang}, \citenamefont {Bl\"ugel},\ and\ \citenamefont {Mokrousov}}]{hanke-prb16}%
  \BibitemOpen
  \bibfield  {author} {\bibinfo {author} {\bibfnamefont {J.-P.}\ \bibnamefont {Hanke}}, \bibinfo {author} {\bibfnamefont {F.}~\bibnamefont {Freimuth}}, \bibinfo {author} {\bibfnamefont {A.~K.}\ \bibnamefont {Nandy}}, \bibinfo {author} {\bibfnamefont {H.}~\bibnamefont {Zhang}}, \bibinfo {author} {\bibfnamefont {S.}~\bibnamefont {Bl\"ugel}},\ and\ \bibinfo {author} {\bibfnamefont {Y.}~\bibnamefont {Mokrousov}},\ }\bibfield  {title} {\bibinfo {title} {{Role of Berry phase theory for describing orbital magnetism: From magnetic heterostructures to topological orbital ferromagnets}},\ }\href {https://doi.org/10.1103/PhysRevB.94.121114} {\bibfield  {journal} {\bibinfo  {journal} {Phys. Rev. B}\ }\textbf {\bibinfo {volume} {94}},\ \bibinfo {pages} {121114} (\bibinfo {year} {2016})}\BibitemShut {NoStop}%
\bibitem [{\citenamefont {Marrazzo}\ \emph {et~al.}(2024)\citenamefont {Marrazzo}, \citenamefont {Beck}, \citenamefont {Margine}, \citenamefont {Marzari}, \citenamefont {Mostofi}, \citenamefont {Qiao}, \citenamefont {Souza}, \citenamefont {Tsirkin}, \citenamefont {Yates},\ and\ \citenamefont {Pizzi}}]{marrazzo-rmp24}%
  \BibitemOpen
  \bibfield  {author} {\bibinfo {author} {\bibfnamefont {A.}~\bibnamefont {Marrazzo}}, \bibinfo {author} {\bibfnamefont {S.}~\bibnamefont {Beck}}, \bibinfo {author} {\bibfnamefont {E.~R.}\ \bibnamefont {Margine}}, \bibinfo {author} {\bibfnamefont {N.}~\bibnamefont {Marzari}}, \bibinfo {author} {\bibfnamefont {A.~A.}\ \bibnamefont {Mostofi}}, \bibinfo {author} {\bibfnamefont {J.}~\bibnamefont {Qiao}}, \bibinfo {author} {\bibfnamefont {I.}~\bibnamefont {Souza}}, \bibinfo {author} {\bibfnamefont {S.~S.}\ \bibnamefont {Tsirkin}}, \bibinfo {author} {\bibfnamefont {J.~R.}\ \bibnamefont {Yates}},\ and\ \bibinfo {author} {\bibfnamefont {G.}~\bibnamefont {Pizzi}},\ }\bibfield  {title} {\bibinfo {title} {Wannier-function software ecosystem for materials simulations},\ }\href {https://doi.org/10.1103/RevModPhys.96.045008} {\bibfield  {journal} {\bibinfo  {journal} {Rev. Mod. Phys.}\ }\textbf {\bibinfo {volume} {96}},\ \bibinfo {pages} {045008} (\bibinfo {year} {2024})}\BibitemShut {NoStop}%
\bibitem [{\citenamefont {Urru}\ \emph {et~al.}(2025)\citenamefont {Urru}, \citenamefont {Souza}, \citenamefont {Tsirkin}, \citenamefont {Pozo~Ocaña},\ and\ \citenamefont {Vanderbilt}}]{urru_2025_15236403}%
  \BibitemOpen
  \bibfield  {author} {\bibinfo {author} {\bibfnamefont {A.}~\bibnamefont {Urru}}, \bibinfo {author} {\bibfnamefont {I.}~\bibnamefont {Souza}}, \bibinfo {author} {\bibfnamefont {S.~S.}\ \bibnamefont {Tsirkin}}, \bibinfo {author} {\bibfnamefont {O.}~\bibnamefont {Pozo~Ocaña}},\ and\ \bibinfo {author} {\bibfnamefont {D.}~\bibnamefont {Vanderbilt}},\ }\bibfield  {title} {\bibinfo {title} {Data relative to manuscript ``{Optical} spatial dispersion via {Wannier} interpolation''},\ }\href {https://doi.org/10.5281/zenodo.15236403} {10.5281/zenodo.15236403} (\bibinfo {year} {2025})\BibitemShut {NoStop}%
\bibitem [{\citenamefont {Melrose}\ and\ \citenamefont {McPhedran}(1991)}]{melrose-91}%
  \BibitemOpen
  \bibfield  {author} {\bibinfo {author} {\bibfnamefont {D.~B.}\ \bibnamefont {Melrose}}\ and\ \bibinfo {author} {\bibfnamefont {R.~C.}\ \bibnamefont {McPhedran}},\ }\href@noop {} {\emph {\bibinfo {title} {Electromagnetic processes in dispersive media}}}\ (\bibinfo  {publisher} {Cambridge University Press, Cambridge, UK},\ \bibinfo {year} {1991})\BibitemShut {NoStop}%
\bibitem [{\citenamefont {Pezo}\ \emph {et~al.}(2022)\citenamefont {Pezo}, \citenamefont {Garc\'{\i}a~Ovalle},\ and\ \citenamefont {Manchon}}]{pezo-prb22}%
  \BibitemOpen
  \bibfield  {author} {\bibinfo {author} {\bibfnamefont {A.}~\bibnamefont {Pezo}}, \bibinfo {author} {\bibfnamefont {D.}~\bibnamefont {Garc\'{\i}a~Ovalle}},\ and\ \bibinfo {author} {\bibfnamefont {A.}~\bibnamefont {Manchon}},\ }\bibfield  {title} {\bibinfo {title} {{Orbital Hall effect in crystals: Interatomic versus intra-atomic contributions}},\ }\href {https://doi.org/10.1103/PhysRevB.106.104414} {\bibfield  {journal} {\bibinfo  {journal} {Phys. Rev. B}\ }\textbf {\bibinfo {volume} {106}},\ \bibinfo {pages} {104414} (\bibinfo {year} {2022})}\BibitemShut {NoStop}%
\bibitem [{\citenamefont {G\"obel}\ and\ \citenamefont {Mertig}(2024)}]{gobel-prl24}%
  \BibitemOpen
  \bibfield  {author} {\bibinfo {author} {\bibfnamefont {B.}~\bibnamefont {G\"obel}}\ and\ \bibinfo {author} {\bibfnamefont {I.}~\bibnamefont {Mertig}},\ }\bibfield  {title} {\bibinfo {title} {Orbital {Hall} effect accompanying quantum hall effect: Landau levels cause orbital polarized edge currents},\ }\href {https://doi.org/10.1103/PhysRevLett.133.146301} {\bibfield  {journal} {\bibinfo  {journal} {Phys. Rev. Lett.}\ }\textbf {\bibinfo {volume} {133}},\ \bibinfo {pages} {146301} (\bibinfo {year} {2024})}\BibitemShut {NoStop}%
\bibitem [{\citenamefont {Gao}\ and\ \citenamefont {Xiao}(2019)}]{gao-prl19}%
  \BibitemOpen
  \bibfield  {author} {\bibinfo {author} {\bibfnamefont {Y.}~\bibnamefont {Gao}}\ and\ \bibinfo {author} {\bibfnamefont {D.}~\bibnamefont {Xiao}},\ }\bibfield  {title} {\bibinfo {title} {{Nonreciprocal Directional Dichroism Induced by the Quantum Metric Dipole}},\ }\href {https://doi.org/10.1103/PhysRevLett.122.227402} {\bibfield  {journal} {\bibinfo  {journal} {Phys. Rev. Lett.}\ }\textbf {\bibinfo {volume} {122}},\ \bibinfo {pages} {227402} (\bibinfo {year} {2019})}\BibitemShut {NoStop}%
\bibitem [{\citenamefont {Xiao}\ \emph {et~al.}(2005)\citenamefont {Xiao}, \citenamefont {Shi},\ and\ \citenamefont {Niu}}]{xiao-prl05}%
  \BibitemOpen
  \bibfield  {author} {\bibinfo {author} {\bibfnamefont {D.}~\bibnamefont {Xiao}}, \bibinfo {author} {\bibfnamefont {J.}~\bibnamefont {Shi}},\ and\ \bibinfo {author} {\bibfnamefont {Q.}~\bibnamefont {Niu}},\ }\bibfield  {title} {\bibinfo {title} {Berry phase correction to electron density of states in solids},\ }\href {https://doi.org/10.1103/PhysRevLett.95.137204} {\bibfield  {journal} {\bibinfo  {journal} {Phys. Rev. Lett.}\ }\textbf {\bibinfo {volume} {95}},\ \bibinfo {pages} {137204} (\bibinfo {year} {2005})}\BibitemShut {NoStop}%
\bibitem [{\citenamefont {Stengel}\ and\ \citenamefont {Spaldin}(2006)}]{stengel-prb06}%
  \BibitemOpen
  \bibfield  {author} {\bibinfo {author} {\bibfnamefont {M.}~\bibnamefont {Stengel}}\ and\ \bibinfo {author} {\bibfnamefont {N.~A.}\ \bibnamefont {Spaldin}},\ }\bibfield  {title} {\bibinfo {title} {{Accurate polarization within a unified Wannier function formalism}},\ }\href {https://doi.org/10.1103/PhysRevB.73.075121} {\bibfield  {journal} {\bibinfo  {journal} {Phys. Rev. B}\ }\textbf {\bibinfo {volume} {73}},\ \bibinfo {pages} {075121} (\bibinfo {year} {2006})}\BibitemShut {NoStop}%
\bibitem [{\citenamefont {Gresch}\ \emph {et~al.}(2018)\citenamefont {Gresch}, \citenamefont {Wu}, \citenamefont {Winkler}, \citenamefont {H\"auselmann}, \citenamefont {Troyer},\ and\ \citenamefont {Soluyanov}}]{gresch-prm2018}%
  \BibitemOpen
  \bibfield  {author} {\bibinfo {author} {\bibfnamefont {D.}~\bibnamefont {Gresch}}, \bibinfo {author} {\bibfnamefont {Q.}~\bibnamefont {Wu}}, \bibinfo {author} {\bibfnamefont {G.~W.}\ \bibnamefont {Winkler}}, \bibinfo {author} {\bibfnamefont {R.}~\bibnamefont {H\"auselmann}}, \bibinfo {author} {\bibfnamefont {M.}~\bibnamefont {Troyer}},\ and\ \bibinfo {author} {\bibfnamefont {A.~A.}\ \bibnamefont {Soluyanov}},\ }\bibfield  {title} {\bibinfo {title} {Automated construction of symmetrized {Wannier}-like tight-binding models from ab initio calculations},\ }\href {https://doi.org/10.1103/PhysRevMaterials.2.103805} {\bibfield  {journal} {\bibinfo  {journal} {Phys. Rev. Mater.}\ }\textbf {\bibinfo {volume} {2}},\ \bibinfo {pages} {103805} (\bibinfo {year} {2018})}\BibitemShut {NoStop}%
\bibitem [{\citenamefont {Liu}(2023)}]{xiaoxiong-phdthesis2023}%
  \BibitemOpen
  \bibfield  {author} {\bibinfo {author} {\bibfnamefont {X.}~\bibnamefont {Liu}},\ }\emph {\bibinfo {title} {Simulation of Nonlinear Electronic Transport Using Wannier Interpolation}},\ \href {https://www.zora.uzh.ch/id/eprint/253142/} {Ph.D. thesis},\ \bibinfo  {school} {University of Z\"urich} (\bibinfo {year} {2023})\BibitemShut {NoStop}%
\bibitem [{\citenamefont {Sakuma}(2013)}]{sakuma-sawf}%
  \BibitemOpen
  \bibfield  {author} {\bibinfo {author} {\bibfnamefont {R.}~\bibnamefont {Sakuma}},\ }\bibfield  {title} {\bibinfo {title} {Symmetry-adapted {Wannier} functions in the maximal localization procedure},\ }\href {https://doi.org/10.1103/PhysRevB.87.235109} {\bibfield  {journal} {\bibinfo  {journal} {Phys. Rev. B}\ }\textbf {\bibinfo {volume} {87}},\ \bibinfo {pages} {235109} (\bibinfo {year} {2013})}\BibitemShut {NoStop}%
\end{thebibliography}%

\end{document}